\documentclass[twocolumn]{aastex7}

\received{May 26, 2025}
\revised{Oct 15, 2025}
\accepted{Oct 22, 2025}
\published{Dec 5,2025}
\submitjournal{AJ}
\usepackage{siunitx}
\usepackage{hyperref}
\usepackage{amsmath}
\usepackage{amsfonts}

\usepackage{soul}
\begin{document}
\title{\texttt{iSTARMOD}: a Python Code to Quantify Chromospheric Activity by Using the Spectral Subtraction Technique}
\author[orcid=0000-0002-7143-0206,sname='Labarga']{Fernando Labarga}
\affiliation{Departamento de F\'{i}sica de la Tierra y Astrof\'{i}sica--Facultad de Ciencias F\'{i}sicas, Universidad Complutense de Madrid, E-28040, Madrid, Spain} 
\email[show]{flabarga@ucm.es}  

\author[orcid=0000-0002-7779-238X,gname=David, sname='Montes']{David Montes} 
\affiliation{Departamento de F\'{i}sica de la Tierra y Astrof\'{i}sica--Facultad de Ciencias F\'{i}sicas, Universidad Complutense de Madrid, E-28040, Madrid, Spain} 
\affiliation{IPARCOS--UCM 
(Instituto de F\'{i}sica de Part\'{i}culas y del Cosmos), Spain }
\email[show]{dmontes@ucm.es}

\begin{abstract}
The use of the spectral subtraction technique allows measurements of chromospheric activity in late-type stars across several activity indicators, such as H$\alpha$ and the other Balmer lines in the visible, He \textsc{i} D$_3$ and Na \textsc{i} D$_\text{1}$, D$_\text{2}$, Ca \textsc{ii} H and K, and Ca \textsc{ii} infrared triplet, as well as Paschen series and He \textsc{i} $\lambda 10830$ lines in the near infrared.
\texttt{iSTARMOD} is an updated and extended version of the original \texttt{STARMOD} code and its subsequent modifications. 
\texttt{iSTARMOD} is presented in this paper as a Python code developed to quantify chromospheric activity by using the spectral subtraction technique. \texttt{iSTARMOD} improves usability, modularity, and integration with modern data analysis workflows and is publicly available, including several examples that help one learn how to use and test the code. The  \texttt{iSTARMOD} code is accompanied here with a series of calibrations of $\chi$-functions, to transform the excess emission equivalent widths measured through \texttt{iSTARMOD} into absolute surface fluxes. The method provided with this code and the corresponding flux calibrations allows for the automatic characterization of the chromospheric activity of a large number of spectra or a large number of stars and is also very useful for mitigating the effect of activity on radial velocities in the search for exoplanets.
\end{abstract}

\keywords{--- \uat{Stellar astronomy}{1583};\uat{Stellar Activity}{1580};\uat{Astronomy Software}{1855};\uat{Late-type stars}{909};\uat{Stellar spectral lines}{1630} }

\section{Introduction \label{sect:intro}}
The spectral subtraction technique \citep{barden1, montes1995b, montes1995a, montes2000} is one of the most effective methods for quantifying the chromospheric contribution to the radiative outflow in active late-type stars across several activity indicators \citep{montes2000, 2007A&A...472..587G, 2009AJ....137.3965G, martinezarnaiz2011, marvinonTeff}. 
This contribution is typically measured as excess emission --relative to a quiescent reference--in equivalent width (EW) units for specific optical and near-infrared (NIR) spectral lines, which are well-established indicators of chromospheric activity in FGKM-type stars.
The excess emission needs to be measured against the photospheric contribution to the radiative outflow: 
\begin{equation}
	\mathcal{F}'_{\text{line}}= \mathcal{F}_{\text{line}} - \mathcal{F}_{\text{line,phot}}
	\label{eq:fluxsubtraction}
\end{equation}
where $\mathcal{F}'_{\text{line}}$ is the chromospheric excess emission-line flux, $\mathcal{F}_{\text{line}}$ is the total flux emitted along this line, and $\mathcal{F}_{\text{line,phot}}$ is the photospheric contribution to the outflow along this line.
So spectral subtraction requires recovering the continuum through synthesizing a quiescent version of the spectrum from a reference star, and with the same spectral type, radial velocity, and rotational state.
Finally, the EW measurement obtained through this method is converted to line flux or, equivalently, line luminosity normalized to bolometric luminosity.
For this purpose, the $\chi$-factor method \cite{walkowicz1} applies. This paper describes both spectral subtraction and a series of $\chi$-factor function calibrations.
Given the above, the spectral subtraction technique allows us to study magnetically active cool stars, analyze in detail its chromosphere using the information provided for several optical spectroscopic features that are formed at different heights, and discriminate between structures: plagues, prominences, flares, and micro-flares. 
This paper describes methods and tools, by means of \texttt{iSTARMOD}, to analyze these features with chromospheric activity indicators, such as the extensively studied H$\alpha$ along with the rest of Balmer lines in the visible, together with He~\textsc{i}~$D_3$ and Na~\textsc{i}~D$_\text{1}$, D$_\text{2}$, Ca~\textsc{ii}~H and K, and the less studied Ca~\textsc{ii} infrared triplet (IRT) lines in the NIR (see \citealt{montes2000}).
Further along NIR spectral range can be analyzed Paschen series (Pa$\delta$, Pa$\gamma$ and Pa$\beta$), together with He \textsc{i} $\lambda 10830$ \citep{schofer2019}.
All of these lines are included in most of the high-resolution echelle spectra.

Spectral subtraction has been routinely applied to measure chromospheric fluxes of active stars. 
\cite{herbig1985onHalphaemission} did so for analyzing H$\alpha$ emission in F8--G3 stars, and \cite{FrascaCatalanoHalphaforbinaries} applied it in the study of close binaries. 
\cite{huenemoerderonTiO} explored TiO bands to model the occurrence of spots in the surface of RS CVn stars.

Continuing the work of \cite{FrascaCatalanoHalphaforbinaries}, \texttt{ROTFIT}, a robust tool for automated spectral classification and activity analysis in late-type stars, was developed.
\texttt{ROTFIT} is a code written in IDL (interactive data language) designed to determine stellar parameters (effective temperature, surface gravity, and metallicity) by comparing observed spectra with a library of standard-star spectra
\citep{frascaetalonrotfit2003, frascaetalonrotfit2006}. 
The tool was used alongside spectral subtraction techniques to isolate chromospheric emission features and characterize the activity levels
\citep{guilloutonfurtherrotfit}. \cite{frascaetal2018} further used \texttt{ROTFIT} to derive stellar parameters ($T_\text{eff}$, log$\,g$, [Fe/H], and $v\text{sin}i$) by fitting observed spectra with a grid of reference star spectra, accounting for rotational broadening.
These works show the role of the spectral subtraction technique in enabling the discovery of numerous young, active stars and candidate binaries, with the characterization of their activity based on lithium abundance and chromospheric activity indicators.

Previous to development of \texttt{iSTARMOD}, as well as after the significant contribution to early models of extended matter and eclipse behavior in RS CVn binaries, that layed the groundwork for later observational and theoretical efforts \citep{1992AJ....104.1942H, 1994AJ....107.1149H}, its predecessor, \texttt{STARMOD}, have been extensively used in several studies. Among these studies we can mention the observation an analysis of contact binaries, focusing on low mass ratio systems \citep{2022AJ....164..202L,2024MNRAS.527.6406L,2025MNRAS.tmp..723L,2025ApJ...979...69W,2025AJ....169..139X,2025AJ....169...85X}; the detailed photometric and spectroscopic studies of W UMa-type binaries \citep{2021AJ....161..221P, 2022ApJ...927...12P}; studies of magnetic and chromospheric activity in RS CVn-type stars: \citep{2012A&A...538A.130C, 2015MNRAS.449.1380C,  2024ApJ...963...13C, 2025AJ....169..198C, 2023MNRAS.523.4146C}; studies of extended matter and active binaries, exploring the evolution of magnetic activity and orbital period variations \citep{2014AJ....147...50P, 2014NewA...27...81Z, 2014NewA...32....1Z, 2018Ap&SS.363..174Z}; studies in transmission spectroscopy, to measure or discard activity levels in the studied object \citep{alonsofloriano2019}; the study of molecular bands to probe starspots properties \citep{1997AJ....113.1129O, 1998ApJ...501L..73O, 2001AJ....122.1954O}; and finally, studies from our group on activity, kinematics, and age in Single and Binary Late-type Stars \citep{montes2000,2004LNEA....1..119M, 2002A&A...389..524G,2006Ap&SS.304...59G, 2007A&A...472..587G}, rotational modulation of activity \citep{2003A&A...411..489L, lopezsantiago2010}, and flux-flux relationships \citep{martinezarnaiz2011}.

This paper describes and provides the iSTARMOD code for the application of the spectral subtraction technique to all possible activity indicators available in the spectral range being analyzed. Its application is of great scientific interest to characterize the chromospheric activity of FGKM stars, both isolated and binary; to study in detail the temporal variability of especially active stars; and contribute to understanding the different magnetic phenomena that occur in them. Calibrations are also provided to convert the excess chromospheric emission to surface fluxes, facilitating the comparison between different activity indicators measured at different wavelengths that form at different heights in the chromosphere. This allows us to study flux--flux relationships and how they depend on stellar parameters. The method provided with this code and the corresponding flux calibrations allow for the automatic characterization of chromospheric activity in a large number of spectra of a single star, as occurs in spectroscopic surveys such as CARMENES \citep{quirrenbach1,quirrenbach2, datarelease1}, or in a large number of stars belonging to different clusters or moving groups in large spectroscopic surveys such as the Gaia ESO Survey \citealt{2022A&A...666A.120G}). Adequate characterization of chromospheric activity is also of great importance in studies dedicated to the search for exoplanets, in order to mitigate the effect of activity on radial velocities and to rule out false positives.

This paper is structured as follows: Section \ref{sect:section2SST} provides a presentation of the spectral subtraction technique and its basic assumptions. Section \ref{sect:section3code} presents the particular implementation of the spectral subtraction technique in \texttt{iSTARMOD} with its workflows, inputs, and improvements. 
Section \ref{sect:section4chi} deals with the $\chi$-factor methodology, to obtain fluxes from the EWs obtained with \texttt{iSTARMOD}, and the calibrations of the $\chi$-function for several chromospheric activity indicators are put in the annexes. 
Finally, Section \ref{sect:section5examples} presents some results of the execution of \texttt{iSTARMOD}.

\section{Spectral Subtraction Technique} \label{sect:section2SST}
The method involves subtracting the spectrum of an active star (the \textit{input} spectrum) from a synthetically constructed \textit{reference} spectrum. 
This reference is generated by applying artificial rotational broadening and radial velocity shifts to a weighted sum of spectra from inactive \textit{reference} stars that closely match the spectral type and luminosity class of the target star \citep{montes1995b, montes1995a}. 
A key feature of the technique is that subtraction is carried out either order by order in the echelle spectrum or over a predefined wavelength range, allowing for high spectral fidelity.

A major advantage of this method lies in its ability to circumvent the challenge of continuum placement in late-type, cool stars, which exhibit dense forests of absorption lines. 
By assuming that both the active and reference spectra--being nearly identical in spectral type--share the same pseudo-continuum, the subtraction process effectively isolates the chromospheric emission. 
This approach greatly reduces systematic uncertainties and simplifies EW measurement of emission features. 
The core algorithm is implemented as an iterative least-squares optimization to determine the best-fit synthetic spectrum: the one that minimizes the rms of the subtraction residuals. The main stages of the fitting process are as follows:

\begin{itemize}
    \item \textit{Doppler shifting} to align the spectral lines of the input and synthetic spectra.
    \item \textit{Rotational broadening} using a Gray profile \citep{graybible} to simulate the effects of stellar rotation on the reference spectrum.
    \item \textit{Spectral weighting} of multiple reference stars, where applicable, such as in the case of unresolved binary (SBx) or multiple systems, showing activity.
\end{itemize}

\subsection{Radial Velocity Determination \label{subsect:rv}}
The spectral subtraction technique requires a high-precision spectra alignment of both reference star and active star. This is achieved through Doppler shifting of the synthetic spectra. Observed wavelength becomes:
\begin{equation}
	\lambda_{\text{obs}} = \lambda_0 \left(1 + \frac{v_r}{c} \right)
\end{equation}
where \( \lambda_0 \) is the rest-frame (laboratory) wavelength of the spectral line, \( \lambda_{\text{obs}} \) is the observed wavelength, \( v_r \) is the supposed radial velocity of the star, and \( c \): speed of light.

Here the flux data array of the fits spectrum is shifted by a noninteger amount through performing both an integer and a residual noninteger shift. The combined process allows for subsample shifting of an array, as is commonly used in signal processing where subsample alignment is required. This approach allows the function to be translated smoothly without the discretization artifacts that would arise from integer-only shifts. Conceptually, the operation approximates the continuous transformation 
$f(x) \longrightarrow f(x - \Delta x)$, where $\Delta x$ corresponds to the imposed \texttt{shift}, thus enabling precise real-space translation of sampled data without resorting to Fourier-space manipulation. 

The integer part is handled by direct reindexing, and the fractional part is handled via local Lagrange interpolation. The input \texttt{shift} value is determined knowing the step in wavelength value between each pair of consecutive elements in the array, supposed constant throughout the whole spectral order, and transforming this wavelength step into velocity. Then this input value is decomposed into its integer and fractional components. The integer part is
\begin{equation}
	\texttt{ishft} = \lfloor \texttt{shft} \rfloor, \quad
\end{equation}
where $\lfloor \cdot \rfloor$ is the floor function, which returns the largest integer not greater than its argument. Then, the fractional part is simply
\begin{equation}
	\texttt{fshft} = \texttt{shft} - \texttt{ishft}, \quad
\end{equation}

The integer part of the process reindexes the array with bounds clamping:
\begin{equation*}
	y[i] = x[\max(0, \min(n - 1, i - \texttt{ishft}))]
\end{equation*}

If the noninteger or fractional part $\texttt{fshft} \neq 0$, the residual shift is applied through interpolation. For each index $i$, an interpolated value is computed at
\begin{equation*}
	x_{\text{in}} = i - \texttt{fshft}
\end{equation*}
using nearby values $\{(x_j, y_j)\}$ in a local window. The interpolated value is calculated with a Lagrange polynomial:
\begin{equation}
	y_{\text{out}} = \sum_{j=0}^{n-1} y_j \cdot \prod_{\substack{i=0 \\ i \ne j}}^{n-1} \frac{x_{\text{in}} - x_i}{x_j - x_i}
\end{equation}
In this implementation, a polynomial of degree 1 (two terms, linear interpolation) is typically used to ensure stability and performance.
This technique is well suited for high-precision alignment of an active star and the reference star performed order by order. The alignment requires a least-squares fit that minimizes the residuals between the input spectrum and the shifted reference star spectrum.
\begin{figure}
	\centering
	\includegraphics[width=0.85\linewidth]{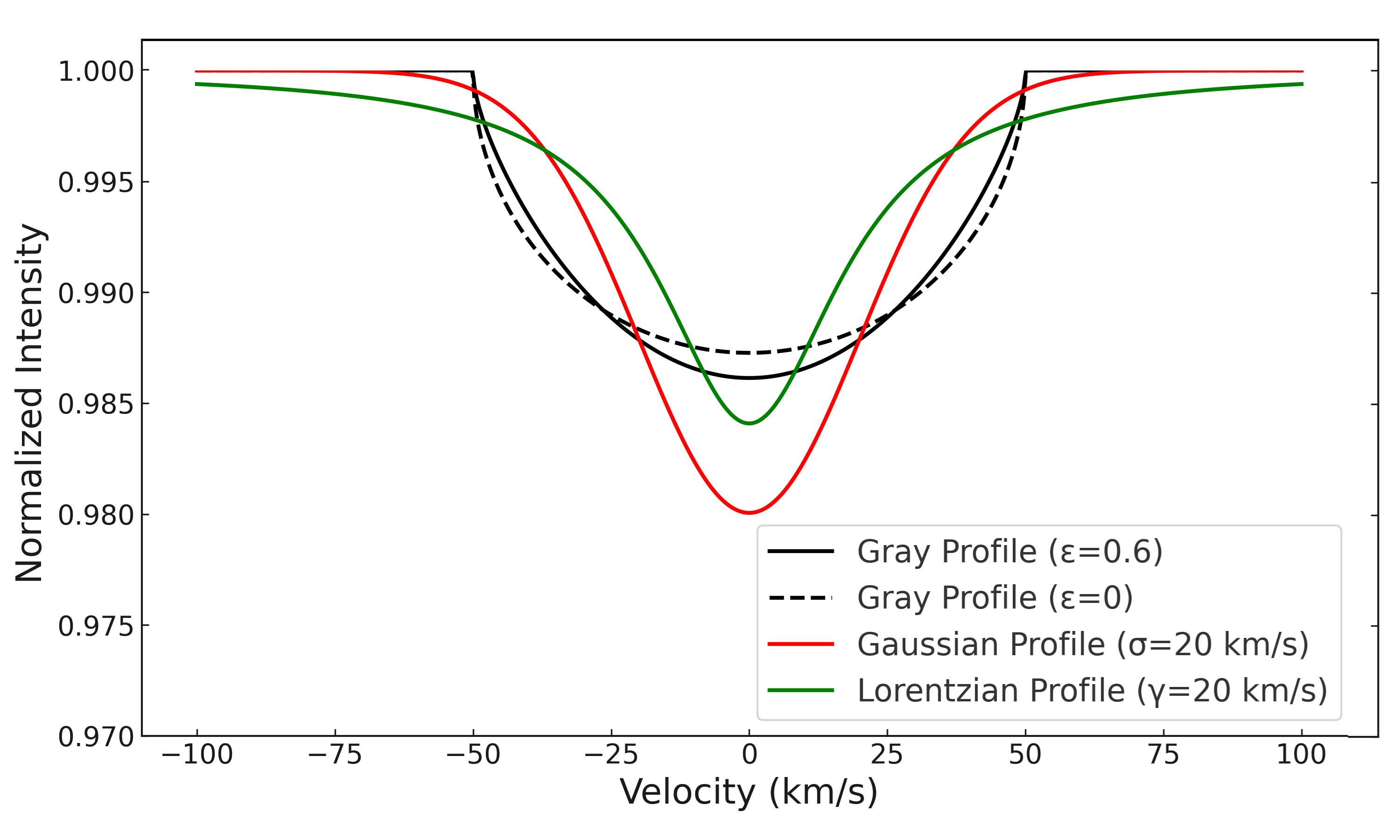}
	\caption{Comparison between Gray, Gaussian, and Lorentzian broadening profiles}
	\label{fig:grayprofile}
\end{figure}
\subsection{Rotational Broadening and the Gray Profile \label{subsect:starot}}
In stellar spectroscopy, rotational broadening refers to the perturbation of spectral lines caused by Doppler shifts resulting from a star's rotation. Different regions of the stellar surface move toward or away from the observer at different velocities, leading to a spread of the observed wavelengths.

A description of the emitted intensity across the stellar disk is needed to model the effect of stellar rotation on spectral lines. This is where the \textit{Gray profile} (or \textit{Gray rotation profile}) becomes essential. In the theoretical deduction of Gray profile, three main assumptions are made: \emph{solid-body rotation}, that is to say, the entire surface rotates at the same angular velocity, a \emph{simple linear limb-darkening law}, and \emph{neglect of macroturbulence}, implying that large-scale turbulent motions, producing additional patterns of Doppler shifts, are excluded from this theoretical deduction process, because they are considered already present in the reference star spectrum.
Instrumental and macroturbulence effects can also mimic the effect of rotation, but using reference stars' spectra taken with the same instrument, and with the same spectral type, luminosity class, and metallicity, these undesired side effects are avoided.
The normalized rotational broadening function \( G(v) \) as a function of velocity \( v \) (relative to line center) is \citep{graybible}
\begin{equation}
G(v) = \frac{2(1 - \epsilon) \sqrt{1 - \left( \frac{v}{v_{\text{max}}} \right)^2} + \frac{\pi \epsilon}{2} \left( 1 - \left( \frac{v}{v_{\text{max}}} \right)^2 \right)}{\pi v_{\text{max}} (1 - \epsilon/3)}
\label{eq:grayprofile}
\end{equation}
where \( v_{\text{max}} = v \sin i \) is the maximum projected rotational velocity, \( i \) is the inclination angle between the rotation axis and the line of sight, and \( \epsilon \) is the linear limb-darkening coefficient ($0\,\leq \epsilon\,\leq\,1$). 
The domain of the function, as is evident by Equation \ref{eq:grayprofile}, is in the interval $[0,\, v_{\text{max}}]$, and $G(v)\,=\,0$ outside it ($|v|\,>\,v_{\text{max}}$).
In the case of no limb darkening (\( \epsilon = 0 \)), the Gray profile reduces to:
\begin{align}
G(v) =
\begin{cases}
\displaystyle \frac{2}{\pi v_{\text{max}}} \sqrt{1 - \left( \frac{v}{v_{\text{max}}} \right)^2}, & |v| \leq v_{\text{max}} \\[10pt]
0, & |v| > v_{\text{max}}
\end{cases}
\end{align}
This is often referred to as the \textit{classical Gray rotation profile}. However, in \texttt{iSTARMOD}, the Gray profile employed is one defined as in equation \ref{eq:grayprofile} with \(\epsilon = 0.6\) and then this is the Gray parameter for a linear limb-darkening law:
\begin{equation*}
	\frac{I(\mu=cos\theta)}{I(\mu=0)} = 1-\epsilon(1-\mu)
\end{equation*}
where $I$ is the specific intensity. It is worth mentioning that this value for $\epsilon$ is the one typical for solar-type stars. It can be different for other types of stars and wavelength ranges \citep{graybible}.

Other options for formulating rotational broadening may involve the use of Gaussian or Lorentzian profiles. A comparison of the three formulations can be seen in Figure \ref{fig:grayprofile}.
\begin{figure}[t]
	\centering
	\includegraphics[width=1.10\linewidth]{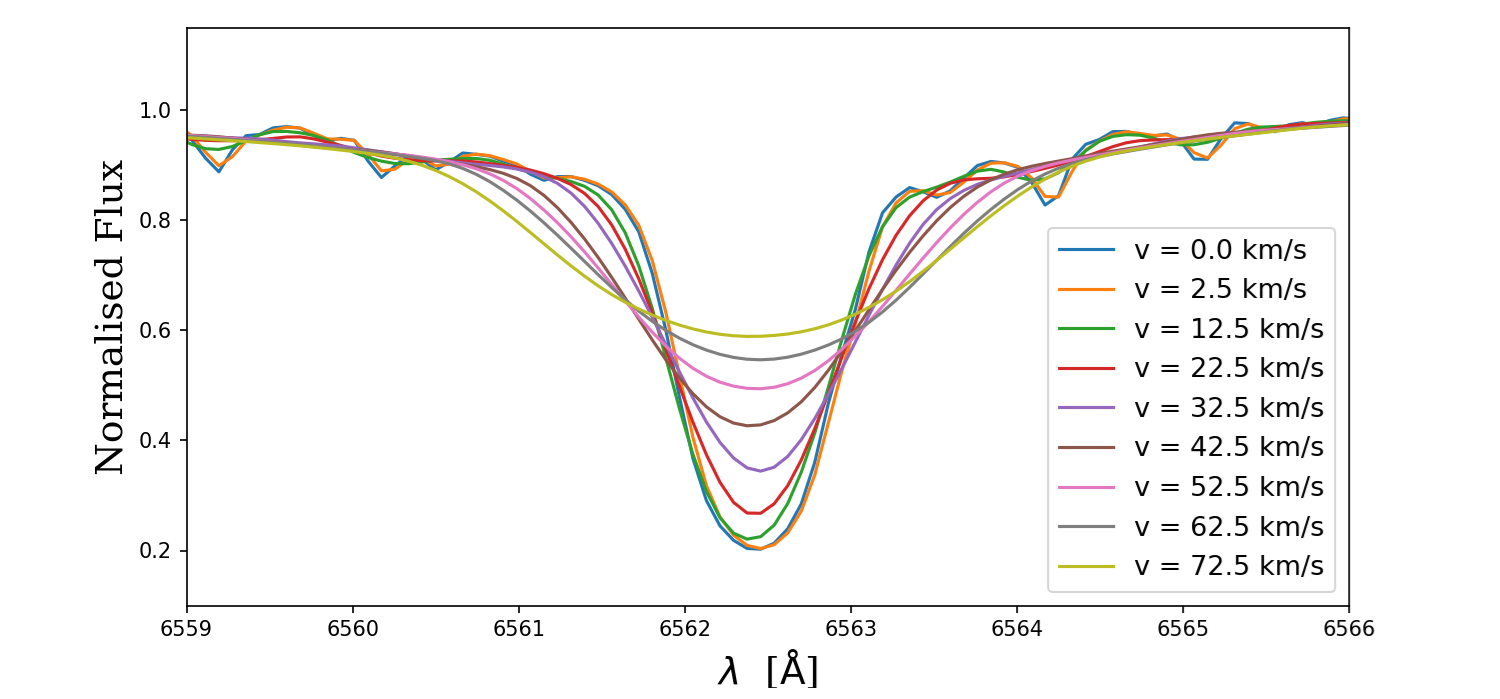}
	\caption{Synthetic line profiles with varying rotational velocities ($v\text{ sin}i$). The \texttt{iSTARMOD} code has been applied to an example H$\alpha$ absorption line, taken from a quiescent K2V star. Note that minor and noisy features of the intrinsic profile are erased for $v\,>\,10 km s^{-1}$}
	\label{fig:rotationalbroadening}
\end{figure}

The observed stellar line profile \( F_{\text{obs}}(v) \) is obtained by the convolution of the intrinsic stellar line profile \( F_{\text{int}}(v) \) with the rotational broadening function \( G(v) \):

\begin{equation}
F_{\text{obs}}(v) = \int_{-\infty}^{+\infty} F_{\text{int}}(v') \, G(v - v') \, dv'.
\end{equation}

Thus, the effect of rotation broadens and redistributes the intrinsic spectral line shape according to the profiles defined here.

The Gray profile describes the distribution of Doppler shifts across the visible stellar disk, where the limb regions exhibit the highest projected velocities \( \pm v_{\text{max}} \), while the central regions (near the rotation axis) contribute around \( v \approx 0 \).
Without limb darkening, the contribution depends purely on geometric projection, where limb darkening (\( \epsilon > 0 \)) reduces the flux contribution near the edges. Figure ~\ref{fig:rotationalbroadening} summarizes all these effects, when this convolution is applied with several projected rotational velocities ($v \text{sin} i$).

Macroturbulence refers to large--scale velocity fields in the stellar atmosphere, larger than the photon mean free path, that broaden absorption line profiles beyond the contributions of instrumental, intrinsic and thermal broadening. \emph{It is known that macroturbulence can also mimic the effect of rotation}, as described in \citet{graybible}. Here, the convolution is performed using a kernel Gaussian profile: 
\begin{equation}
	M_{\zeta}(\Delta \lambda) \;=\; \frac{1}{\sqrt{\pi} \, \zeta} 
	\exp\!\left[-\left(\frac{\Delta v}{\zeta}\right)^2 \right]
\end{equation}
with:
\begin{equation}
	\int M_\zeta \, d(\Delta v) = 1
\end{equation} 
The velocity fields are defined by $\zeta$, a set of empirical relationships (\citealt{Gray1984, Gray2010empiricaldec, Doyle2014, Tsantaki2018}) dependent on $T_\text{eff}$ and surface gravity log$ g$:
\begin{equation}
	\zeta = a + B_\text{n}(T_\text{eff}- T_\text{s})^n - c(log g - log g_\text{s})
	\label{zetaDoyle}
\end{equation}
The normalization condition implies that macroturbulence does not change the EW of absorption-line profiles, only their shape, broadening and reducing their depth. But in spectral subtraction, a small mismatch $\delta T$ between input and reference star spectrum temperatures and gravity surface therefore could produce residuals dominated by the core depression and wing enhancement pattern. To quantify this, we will treat $\delta T$ as a perturbation of the subtracting spectrum. Then, using equation \ref{zetaDoyle}, as derived in \cite{Doyle2014} to perform the perturbation calculations, the reference star spectrum $F_{\rm orig}$ is convolved with the macroturbulence kernel $M_{\zeta}$. The first-order perturbation in this convolved spectrum is
\begin{equation}
	\delta\,(F_{\rm orig} \ast M_{\zeta})
	\;=\;
	F_{\rm orig}
	\;\ast\;
	\left(
	\frac{\partial M}{\partial \zeta}\,
	\frac{\text{d}\zeta}{\text{d}T_{\text{eff}}}\,\delta T
	\right),
\end{equation}
so the sensitivity kernel is precisely $\partial M/\partial  \zeta$ scaled by
$d\zeta/dT_{\text{eff}}$. Forcing the validity of equation \ref{zetaDoyle} outside its effective temperature range ($[5200, 6400]$) in \cite{Doyle2014} and putting numbers for the examples of Section \ref{sect:section5examples} (e.g., V1216 Sgr with respect to the reference of Barnard's Star, two M3.5V stars), this scale is $9\times 10^{-4} \,\delta T$ and the relative perturbation $\delta \zeta/ \zeta$ is lower than $8\%$. Given that an additional convolution with the Gray profile is cumulative, the scale and induced relative error are even lower. This highlights the key point of the choice of a suitable reference star spectrum with comparable $T_\text{eff}$, luminosity class (then log $g$) and metallicity.

There are additional features of stars that can be confidently ignored for the case of cool ones: oblateness and the gravity darkening derived from it. 
\textit{Gravity darkening} is due to the variation in $T_\text{eff}$ and brightness over the surface of a rotating star due to differences in effective gravity (von Zeipel theorem, see \citealt{lucyongravitydarkening1967}), in turn due to oblateness of the body, as
\begin{equation*}
	T_\text{eff} (\theta) \simeq \sqrt[4]{g_\text{eff}(\theta)}
\end{equation*}
where $\theta$ is the latitude.

This is important only for very fast rotators \citep{solaronvsinigravitydarkening}, as Be stars, or with intense gravitational fields.
Then, for late-type stars, except perhaps the youngest and consequently most active ones, this effect can be safely neglected.
To what extent is it negligible? In the majority of cool dwarfs, the rotational velocities are low enough that centrifugal distortion and gravity darkening are insignificant.
The oblateness \( f \) of a star can be approximated by
\begin{equation}
	f \equiv \frac{R_{\text{eq}} - R_{\text{pol}}}{R_{\text{eq}}} \approx \frac{v_{\text{eq}}^2 R}{2 G M},
	\label{eq:oblateness}
\end{equation}
where $R_\text{eq}$ is the equatorial radius, $R_\text{pol}$ is the polar radius, and $R$ is the mean one. Oblateness is typically much less than 1\% for stars with $v_{\text{eq}} \lesssim 10 \text{ km s}^{-1}$.
Putting the numbers in Equation \ref{eq:oblateness} with the parameters of, e.g, EV Lac, a particularly active M3.5Ve flare star, with $v_\text{eq sin} i \approx 5\text{ km s}^{-1}$, results in an oblateness parameter of $f \approx 0.009\%$.
Similar stars in mass and radius, such as V1274 Her, a BY Dra variable with  $v_\text{eq} \text{sin} i \approx 50 \text{ km s}^{-1}$ yield values for $f$ of around 0.6~\%. In such cases, both the deviation from spherical symmetry and the latitudinal temperature variation due to gravity darkening can be safely ignored.

Consequently, for cool stars the use of standard limb-darkened rotation profiles (e.g., Gray profile) assuming a spherical stellar surface is well justified for modeling the rotational broadening of the spectral lines of reference stars.

\subsection{Weighted Spectra for Binary Systems}
The case for binaries requires the third set of parameters specified in Section \ref{sect:section2SST}: weight of the component stars participating in building the synthetic reference star. Since the spectra are normalized, these weights are applied directly by forming a weighted sum:
\begin{equation}
	\label{eq:weighting}
	F_\text{syn}(\lambda) = w_1*F_\text{prim}(\lambda) + w_2*F_\text{sec}(\lambda)
\end{equation}
subject to the normalization condition
\begin{equation}
	\label{eq:weightcond}
	w_1 + w_2 = 1.0 
\end{equation}
The contribution of each component to the total continuum can be obtained from the luminosity ratio in the H$\alpha$ region of the spectra \citep{montes1995a}, given by
\begin{equation*}
	\frac{L_1}{L_2} = \left(\frac{R_1}{R_2}\right)^2 \left(\frac{B_{\lambda=\text{H}\alpha} (T_\text{eff,1})}{B_{\lambda=\text{H}\alpha} (T_\text{eff,2})}\right)
\end{equation*}
where $B(\lambda,T_\text{eff})$ is the Planck function and $R_\text{1,2}$ are the corresponding radii of each star, where 1,2 refers to the hot and cool components of the binary.
Then, the different weights can be calculated as 
\begin{equation*}
	w_1 = \frac{\alpha}{1+\alpha}, \quad w_2 = \frac{1}{1+\alpha}
\end{equation*}
and $\alpha = L_1/L_2$. This normalization of the weights affects the calculation of EWs, as we will see in Section \ref{subsect:improvements} 
 
\subsection{Quality Tests \label{subsect:quality}}
Once the three processes mentioned above are executed and optimal radial velocities, $v \text{sin} i$ and spectral weights are determined, the synthetic spectra are finally built and subtracted to the input spectrum. Then, the rms of the residuals are calculated:
\begin{equation}
	\label{eq:rms}
	rms = \sum_{\substack{i=0 \\ i\neq excl}}^{npts} (F_\text{input}(\lambda)- F_\text{synth}(\lambda))^2
\end{equation}
where \(i\neq excl\) expresses the condition that points within excluded intervals are not summed.

This suggests an important quality check of this method in the features of the subtracted spectrum: a successful subtraction should yield a near-zero normalized flux level across the spectrum, resulting in a successful alignment of reference and input spectra, except in narrow regions around activity-sensitive lines (measured in \si{\angstrom}), which will display residual excess emission (see discussion in Section \ref{sect:section5examples}).
Then, in order to put a numerical condition for a successful subtraction performed by \texttt{iSTARMOD}, a limit can be stated where 
\begin{equation*}
	rms \ngeq 0.5
\end{equation*}
depending on the signal-to-noise ratio (S/N) of the spectrum taken.

This approach for spectral subtraction proves particularly valuable for stars with low levels of activity, where chromospheric emission does not significantly rise above the continuum. 
Hence, the method can be effectively applied across a wide range of stellar activity levels, not only to highly active stars.

\section{\texttt{iSTARMOD} Code} \label{sect:section3code}
\subsection{Overall Description}
\texttt{iSTARMOD} constitutes the actual implementation of the spectral subtraction technique for the study of the chromospheric activity in cool stars.
\texttt{iSTARMOD} \citep{istarmodatsea} is an updated and extended version of the original STARMOD code developed at Penn State University \citep{barden1}.
Subsequent modifications were introduced by \cite{montes2000} and further developed in the \texttt{JSTARMOD } variant by \cite{lopezsantiago2010}. 
In this work we present the migration of the code from \texttt{FORTRAN} to \texttt{Python} and implement additional enhancements to improve usability and precision. 
These upgrades include automated EW measurements (a feature previously not implemented by the code), error estimation routines, and the capability to process large time-series datasets or extensive stellar samples. This functionality is essential at performing, for example, analyses of flux--flux or activity--rotation relationships based on large numbers of spectra-—whether for individual stars or for many stars observed in large spectroscopic surveys, such as the CARMENES GTO sample of M stars (\citealt{quirrenbach2} \citealt{2023hsa..conf..272L}).

The migration of the original \texttt{FORTRAN} \texttt{STARMOD} code to Python was motivated by the need for improved usability, modularity, and integration with modern data analysis workflows. 
Python provides a versatile programming environment supported by a broad ecosystem of scientific libraries--such as \texttt{Astropy} \citep{astropy2018}, \texttt{NumPy} \citep{numpy2020}, and \texttt{SciPy} \citep{scipy2020}--that facilitate spectral data handling, numerical optimization, and reproducible research. 
The Python implementation of \texttt{iSTARMOD} allows for automated processing of large spectral datasets, integration into time-series analysis pipelines, and enhanced flexibility for spectral modeling.
The code is publicly available \citep{Labarga2025_iSTARMOD} under an open-source license and will be actively maintained by the authors at \url{https://github.com/flabarga/iSTARMOD}.

The subtraction is carried out order by order, as mentioned above (section \ref{sect:section2SST}), in the echelle spectrum to mitigate the effects of instrumental profile and variations across different orders and to optimize memory and computational efficiency. 

This order-by-order approach is important in high-resolution echelle spectroscopy owing to the wavelength-dependent sensitivity and instrumental profile that vary across different spectral orders. 
One of the primary challenges arises from the blaze function, an intrinsic feature of echelle gratings that causes a nonuniform intensity distribution within each order. 
Although standard reduction pipelines attempt to correct for the blaze response, residual effects often remain and can introduce low-frequency modulation or distortions in the continuum shape. 
By performing the subtraction within each spectral order independently, \texttt{iSTARMOD} minimizes the propagation of such instrumental artifacts into the final residual spectrum.
Furthermore, order-by-order processing allows for local optimization of continuum placement and velocity alignment, improving the accuracy of chromospheric excess emission measurements, especially for narrow spectral lines affected by order-edge distortions or varying S/N across the orders. Moreover, this local optimization of continuum placement makes \texttt{iSTARMOD} especially suitable for the study of chromospheric activity in the NIR channel of the spectrum.

\subsection{Improvements with Respect to \texttt{STARMOD} \label{subsect:improvements}}
Auxiliary steps are integrated in addition to primary subtraction and fitting procedures:
\begin{itemize}
    \item Rebinning of spectra that shows variable wavelength step sizes.
    \item Determination of continuum level using an adaptation of the \texttt{ARES} algorithm \citep{ares1, ares2}. This is, as mentioned above, particularly important for M dwarfs, where continuum placement is highly uncertain owing to profusion of absorption of atomic and molecular lines.
    \item Normalization of the spectrum after continuum placement.
    \item Calculation of the EW by numerical integration over predefined wavelength limits.
    \item Estimation of EW uncertainties using the Cayrel formula \citep{1988IAUS..132.....C, cayrel2004}, based on the full width at half maximum, S/N, and pixel step size ($\sigma_x$):
    \begin{equation}
         \sigma_W = \frac{1.06}{S/N} \sqrt{FWHM \cdot \sigma_x}
    \end{equation}
\end{itemize}
\begin{figure}
	\includegraphics[width=1.0\linewidth]{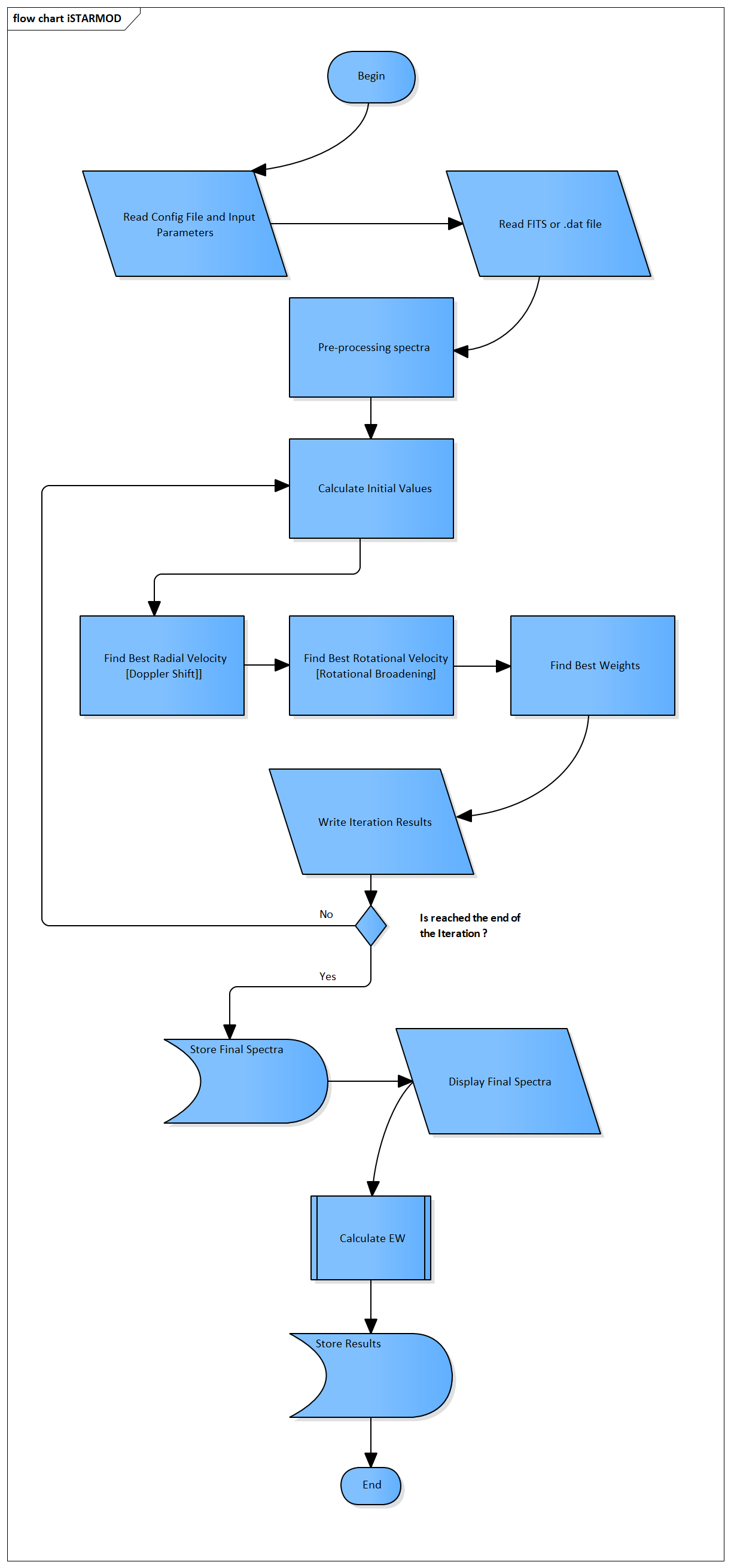}
    \caption{Flowchart of \texttt{iSTARMOD} process, showing the implementation of the algorithm of spectral subtraction technique, the steps followed and the information flow \label{fig:workflow}
    }
\end{figure}
All processing steps are encapsulated in a coherent iterative loop as shown in the workflow depicted in Figure \ref{fig:workflow}. 
The code supports multiple \textsc{FITS} formats produced by high-resolution spectrographs such as \textsc{FOCES}, \textsc{HERMES} and \textsc{CARMENES}, which present their own data structures and spectral characteristics, and the most simple .dat files, where the data arrays of wavelengths and fluxes are extracted, with a format \textit{wvl (space-space) flux}. 
Execution examples of iSTARMOD are shown in  Section 5. for several chromospherica activity indicators.

As previously noted, the algorithm is also suitable for spectroscopic binaries, where each stellar component can be represented by a separate reference spectrum. 
A composite synthetic spectrum can then be built by a weighted sum of each individually broadened and shifted reference spectra. 
Here the preliminary calculation of the EWs is performed through the deconvolution of the weighted sum of two Lorentzian profiles.
This deconvolution is performed by means of a least-squares fit, implemented via the \texttt{scipy:optimize:curve\_fit} function. 
Subsequently, a numerical integration is performed between two predefined wavelengths, resulting in two EWs values, which must be corrected as
\begin{equation*}
	EW_\text{n} = \left(\frac{1}{w_\text{n}}\right) EW_\text{n,0}
\end{equation*}
where $w_n$ are the weights, subject to condition eq. \ref{eq:weightcond}, and $n=\text{1,2}$ are the \textit{hot} and \textit{cool} components of the binary.
The suitability of this functionality implemented by previous versions of this code was demonstrated in \cite{montes1995b, montes1995a, montes2000} and most of the studies on spectroscopic binaries mentioned in the introduction, as the \texttt{STARMOD} code was made available for the scientific community. See examples of binaries in section \ref{subsect:examplesSB2}

\subsection{Input Parameters}
\texttt{iSTARMOD} performs, in the context of \textit{least-squares fitting}, an iterative minimization over spectral segments to best match a synthetic composite of reference spectra to an observed one. The process involves, on one hand, Iterative optimization (\texttt{N\_ITER}), where parameters like radial velocities, rotational broadening, and weights for primary and secondary components are adjusted to minimize the residuals between observed and synthetic spectra. On the other hand, there is a need to define exclusion zones (\texttt{PIX\_EXCL}) that allow ignoring regions contaminated by stellar activity or instrumental effects to prevent biasing the fit.

The \textit{least-squares criterion} is used to evaluate the sum of squared residuals within a defined spectral range (\texttt{PIX\_ZONE}), avoiding specified zones, according to equation \ref{eq:rms}.

Thus, the least-squares fit that the spectral subtraction technique depends on requires a large number of input and configuration parameters, settings for running the \texttt{iSTARMOD} tool. These parameters are specified in a configuration file (\texttt{inputParameters\_filename.sm}) using a \texttt{KEYWORD = value} syntax and are organized into several key sections:
\begin{enumerate}
	\item General Information--defines the input FITS spectrum location and filenames.
	\item Output Spectra--specifies output options for synthetic and subtracted spectra.
	\item Interpolation Parameters--includes the number of least-squares iterations (\texttt{N\_ITER}), fitting wavelength ranges, and exclusion zones to ignore during residual summation (e.g., due to stellar activity or edge effects).
	\item Primary and Secondary Star Parameters--specifies the Doppler shift (radial velocity), rotational broadening ($v \text{sin} i$), and weights for reference spectra. They will be the initial guesses for the fit process. They can be provided with the option to fix or vary these values during the whole iterative process.
	\item Spectra Format--selects the spectral order and line region to analyze, allowing the calculation of EWs.
	\item Algorithm and Visualization Parameters--controls display settings and tolerance for identifying emission peaks in subtracted spectra.
\end{enumerate}

In summary, and from the spectral processing perspective, \texttt{iSTARMOD} allows precise subtraction of reference spectra from observed ones, enabling detection of excess emission from chromospheric activity. This requires an accurate wavelength calibration and matching of instrumental profiles via Doppler shift and broadening adjustments. Finally, \texttt{iSTARMOD} provides post-processing analysis such as EW measurements, with a generation of synthetic (\texttt{SYN\_NAME}) and subtracted (\texttt{SUB\_NAME}) spectra to isolate features of interest (e.g., Ca \textsc{ii} IRT lines).

Together, these configurations ensure robust, reproducible spectral analysis tailored for stellar activity diagnostics or binary star component separation.

\section{Calibration of \texorpdfstring{$\chi$}{χ} Factor Functions for Different Activity Indicators}\label{sect:section4chi}
The ratio of the luminosity in H$\alpha$ to the bolometric luminosity ($L_{H_{\alpha}}/L_\text{Bol}$) is a measure of the strength of activity, as it is the main channel for chromospheric radiative loss of the magnetic energy in the surface of cool stars. 
So a method for calculating the transformation of EW of excess emission in H$\alpha$, measured as shown in the previous section, to absolute surface fluxes is required. 
The method is based on the $\chi$ factor, as was defined in \cite{walkowicz1}, and implemented as in \cite{reinersbasri}. 
It has been devised to derive a distance-independent method, free of systematic errors, for calculating $L_{H_{\alpha}}/L_\text{Bol}$. 
Alternate examples of application of this method can be found in \cite{douglasAltChi1} and \cite{douglasAltChi2}. More recently, \cite{otherchi} studied the $\chi$-factor methodology for the complete Balmer series.
In this study the method is generalized to obtain the flux in other usual chromospheric activity indicators. Along with the \texttt{iSTARMOD} code, the calibrations provided here will allow performing extensive studies of the chromospheric activity in single and binary cool stars.

Starting with the relation
\begin{equation}
        \\log \left(\frac{L_\text{line}}{L_\text{Bol}}\right) = log(\chi) + log( EW)
\end{equation}
we can express the specific line luminosities normalized to the bolometric luminosity as a function of EW. This ratio is clearly analogous to the ratio of specific line flux to bolometric flux: $\mathcal{F}_\text{line}/\mathcal{F}_\text{bol}$. So, the relationship between fluxes and EWs is established through the $\chi$ factor. 

A set of calibrations log$\chi = f(T_\text{eff})$ use synthetic spectra in the temperature range: [2200, 7000] K for most of the chromospheric activity indicators.
These temperature ranges cover spectral types from F to M or even early L, of class luminosity V.

A grid of synthetic spectra from the BT-Settl-CIFIST models \cite{baraffe2015}, especially suited for low-mass stars, based on \cite{allard2012} and \cite{husser2013} was used. The models assume solar metallicity [Fe/H] = 0,  visual extinction $A_V=0$ mag and $\text{log} g$ in the range [4.5, 5.5]. 
They can be found in the VOSA online repository\footnote{\hyperlink{http://svo2.cab.inta-csic.es/theory/newov2/index.php}{http://svo2.cab.inta-csic.es/theory/newov2/index.php}}.
The set of synthetic spectra have been degraded in resolution by convolving each spectrum with a Voigt profile (see \citealt{voigtfunc}) in order to match the spectral resolution of the synthetic spectra ($R \approx 500,000$) to the spectral resolution of, e.g., the CARMENES instrument ($R \approx 85,000$). This procedure accounts for the instrumental profile of CARMENES in its VIS and NIR channels. These steps were executed following the methods described in \cite{marfil2021}. After that, fluxes at wavelengths near the wavelength of the line under consideration for each $T_\text{eff}$, are measured.

Different fits have been obtained, and the coefficients ($C_1$, $\alpha$ and $\beta$) are shown in Table \ref{table:t1} in Appendix \ref{annexa}. The curves can be seen in Figure \ref{fig:FigCalibChiHalpha} for H$\alpha$ and Figure \ref{fig:FigCalibChi} in Appendix \ref{annexa} for Ca \textsc{ii} H \& K, He \textsc{i} D$_3$ + Na \textsc{i} D$_\text{1}, D_\text{2}$, Ca \textsc{ii} IRT, Pa$\delta$, Pa$\gamma$ + He \textsc{i} $\lambda 10830$ and Pa$\beta$.

The fits from these calibrations were obtained proposing a functional form for $\chi$:
\begin{equation}
        \label{eq:functionalform}
        \\ \\ \chi = C_0\left(\frac{T_\text{eff}}{T_s}\right)^\alpha   10^{P{_5(T}_\text{eff})}
\end{equation}
where T$\text{eff}$ is the effective temperature of the star. 
Equation \ref{eq:functionalform} resembles a $\text{log}_{10}$-Schechter-type or Yukawa functions. 
Taking logs, we end up with
\begin{equation}
        \\ \\ log\chi = C_1 + \alpha logT_\text{eff} + P{_5(T}_\text{eff})
\end{equation}
where
\begin{equation}
    \\    P{_5 (T}_\text{eff}) = \beta_1T_\text{eff} + \beta_2T_\text{eff}^2+\beta_3T_\text{eff}^3+ \beta_4T_\text{eff}^4+\beta_5T_\text{eff}^5
\end{equation}
and then
\begin{equation}
    \label{eq:independent}
    \\ \\ C_1 = \left(log { } C_0+\beta_0\right) - \alpha logT_s 
\end{equation}
The independent term for the polynomial, $\beta_0$, and the $T_s^\alpha$ have been incorporated into $C_1$, given that these coefficients cannot be fully determined by any fit. 

The first $\chi$ factor to be calculated is the one that abounds in the literature: the $\chi$ factor corresponding to H$\alpha$.
To test the limits and accuracy of this methodology, our calibration, spanning the range $[2000, 7000]$, was put together with the model from \cite{reinersbasri}, which covers the range $[1200, 4000]$. It is important to note that the values from \cite{reinersbasri} were originally intended for the study of ML-type stars.

The polynomial defined in \cite{reinersbasri}, being of fifth degree, exhibits oscillations beyond 4000 K, hence making it unsuitable for extending the calibration up to 7000 K. 
However both fits match almost perfectly in the range of $T_\text{eff}(K)$ of [2200, 2700] and remain consistently within [2700, 4000]. Beyond 4000 K the fit of our study extends smoothly.
The situation is summarized in Figure~\ref{fig:FigCalibChiHalpha}.
\begin{figure}[ht]
    \includegraphics[width=0.95\linewidth]{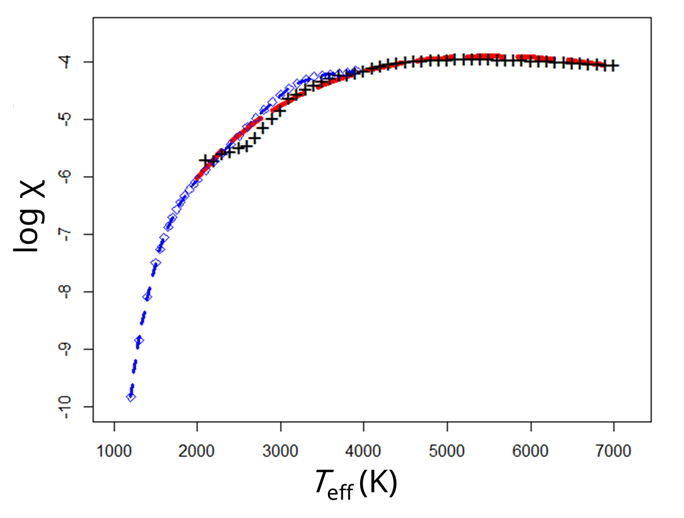}
    \caption{Comparison between the polynomial fit provided in \cite{reinersbasri} (blue) in the range [1200, 4000] and this paper calculations (red), in the range [2000, 7000]. The points obtained from the set of synthetic spectra are the black crosses.}
    \label{fig:FigCalibChiHalpha}
\end{figure}

The combined EWs of the Ca \textsc{ii} H and Ca \textsc{ii} K lines allows us to define one the most widely used activity indicators, $R_{H{K}}^{'}$ \citep{linskyetayres}. This parameter is defined as the ratio between Ca \textsc{ii} H and K emission and the stellar bolometric flux \citep{linskyetal},
\begin{equation}
   R_{H{K}}^{'} = \frac{\mathcal{F}'_H + \mathcal{F}'_K}{\sigma T_\text{eff}^4}\\ \\
\end{equation}
where the primes denote fluxes where the photospheric contribution has been subtracted as in Eq. \ref{eq:fluxsubtraction}. Considering that
\begin{equation*}
     \frac{\mathcal{F}_{\text{line}}}{\mathcal{F}_{\text{bol}}} = \frac{L_{\text{line}}}{L_{\text{bol}}} \\ \\
\end{equation*}
both luminosities normalized to bolometric, calculated by means of the $\chi$ factor, must be summed up in order to obtain $R`_\text{HK}$. Hence,
\begin{equation}
 \\   R_{H{K}}^{'} = \frac{L_\text{Ca \textsc{ii} H}}{L_\text{Bol}} + \frac{L_\text{Ca \textsc{ii} K}}{L_\text{Bol}} = \chi_\text{\textsc{hk}} (EW(H) + EW(K))
 \label{eq:RpHK}
\end{equation}
As demonstrated in \cite{2010A&A...520A..79M}, the $R_{H{K}}^{'}$ values obtained by means of the equation \ref{eq:RpHK} are equivalent to the ones obtained with the classical Mount Wilson S-index method \citep{vaughan1978}.

The lines He~\textsc{i}~D$_3$ and Na~\textsc{i}~D$_1$, D{$_2$} are clear chromospheric activity indicators. 
These lines, being sensitive to the chromospheric environment, often mirror the activity observed in H$\alpha$, though with differences due to their distinct formation mechanisms and the depth of the chromosphere layer \citep{kumaretfares2023}. 
The proximity between the He~\textsc{i}~D$_3$ and Na~\textsc{i}~D$_1$,D$_2$ lines allows for a single calibration for all three lines (see the examples in Section~\ref{sect:section5examples}), measuring the continuum emission halfway both groups of lines in the synthetic spectra. The values of their fit coefficients are shown in the second column of Table \ref{table:t1} in the Appendix \ref{annexa}

All lines of Ca \textsc{ii} IRT, Ca \textsc{ii} IRT-a $\lambda {8498}$, Ca \textsc{ii} IRT-b $\lambda {8542}$, and Ca \textsc{ii} IRT-c $\lambda {8664}$, are subject to the same calibration, summarized in top row of Figure ~\ref{fig:FigCalibChi} and Table \ref{table:t1}, both in Annex \ref{annexa}.
The Ca \textsc{ii} IRT is an alternative indicator in the IR of Ca \textsc{ii} H and K lines as one of the primary indicators of stellar activity \citep{hintzCaii}, and is specially suited for its study in M-type stars. 
\begin{figure*}[ht]
	\includegraphics[width=0.5\linewidth]{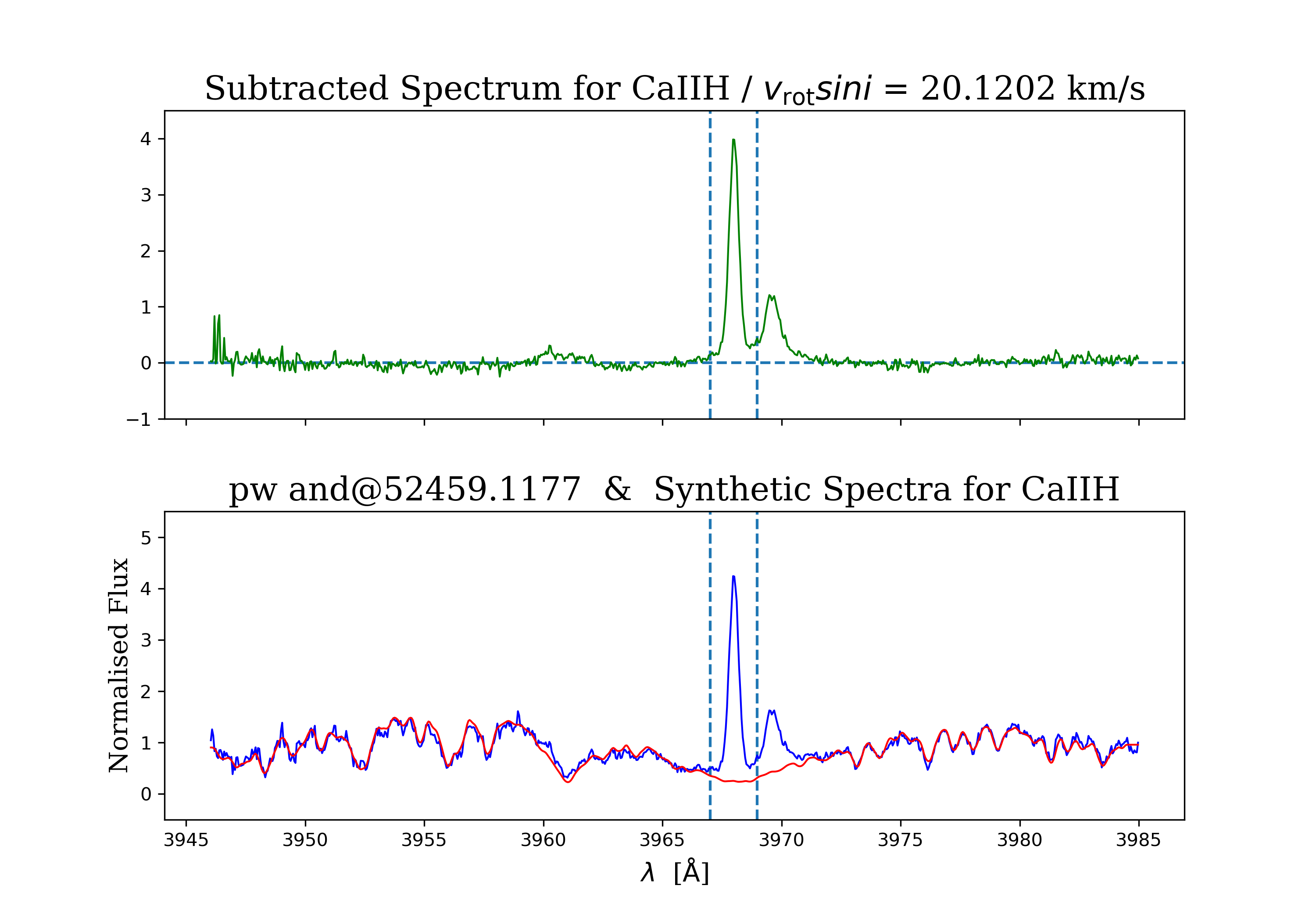}
	\includegraphics[width=0.5\linewidth]{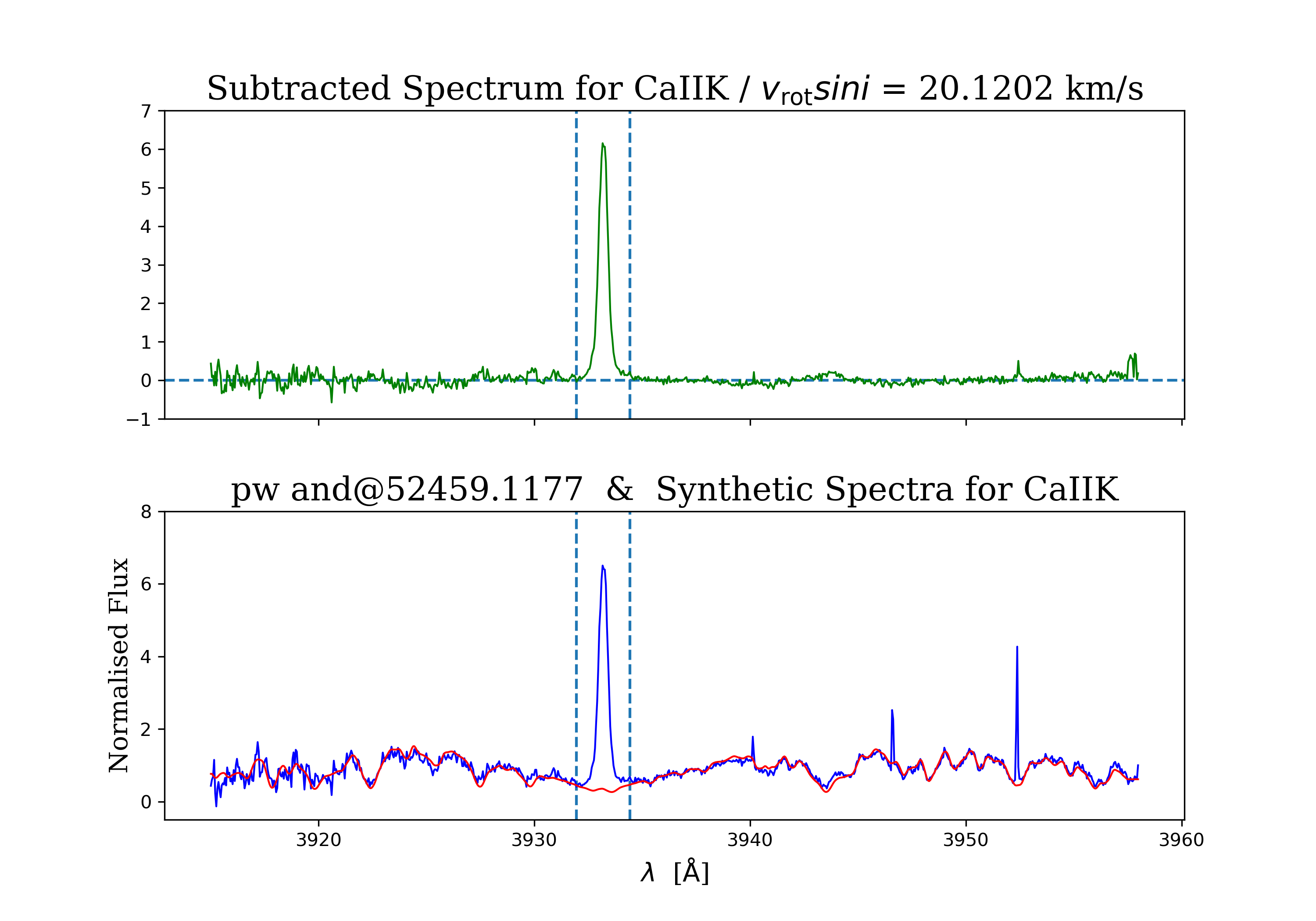}
	\includegraphics[width=0.5\linewidth]{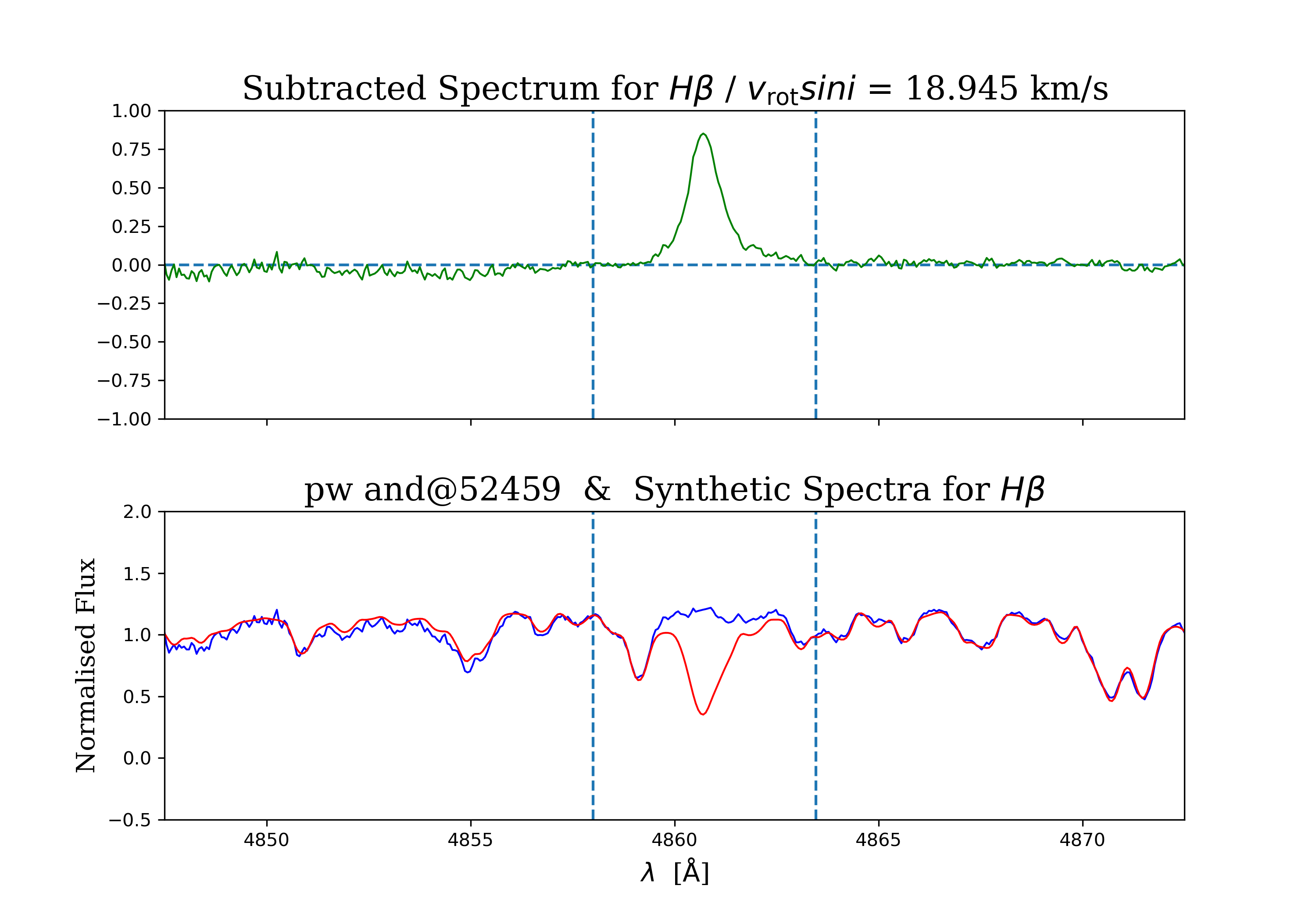}
	\includegraphics[width=0.5\linewidth]{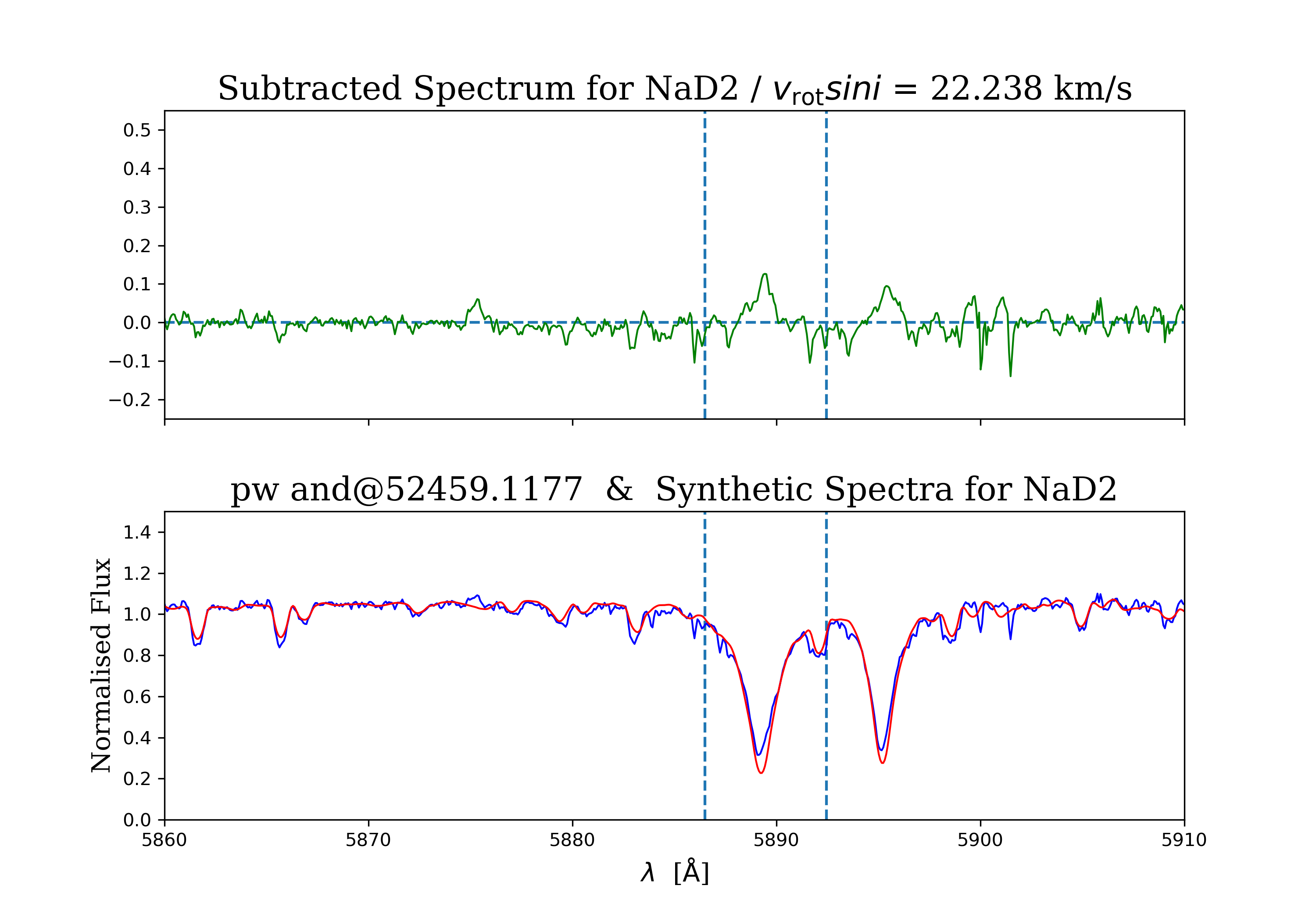}
	\includegraphics[width=0.5\linewidth]{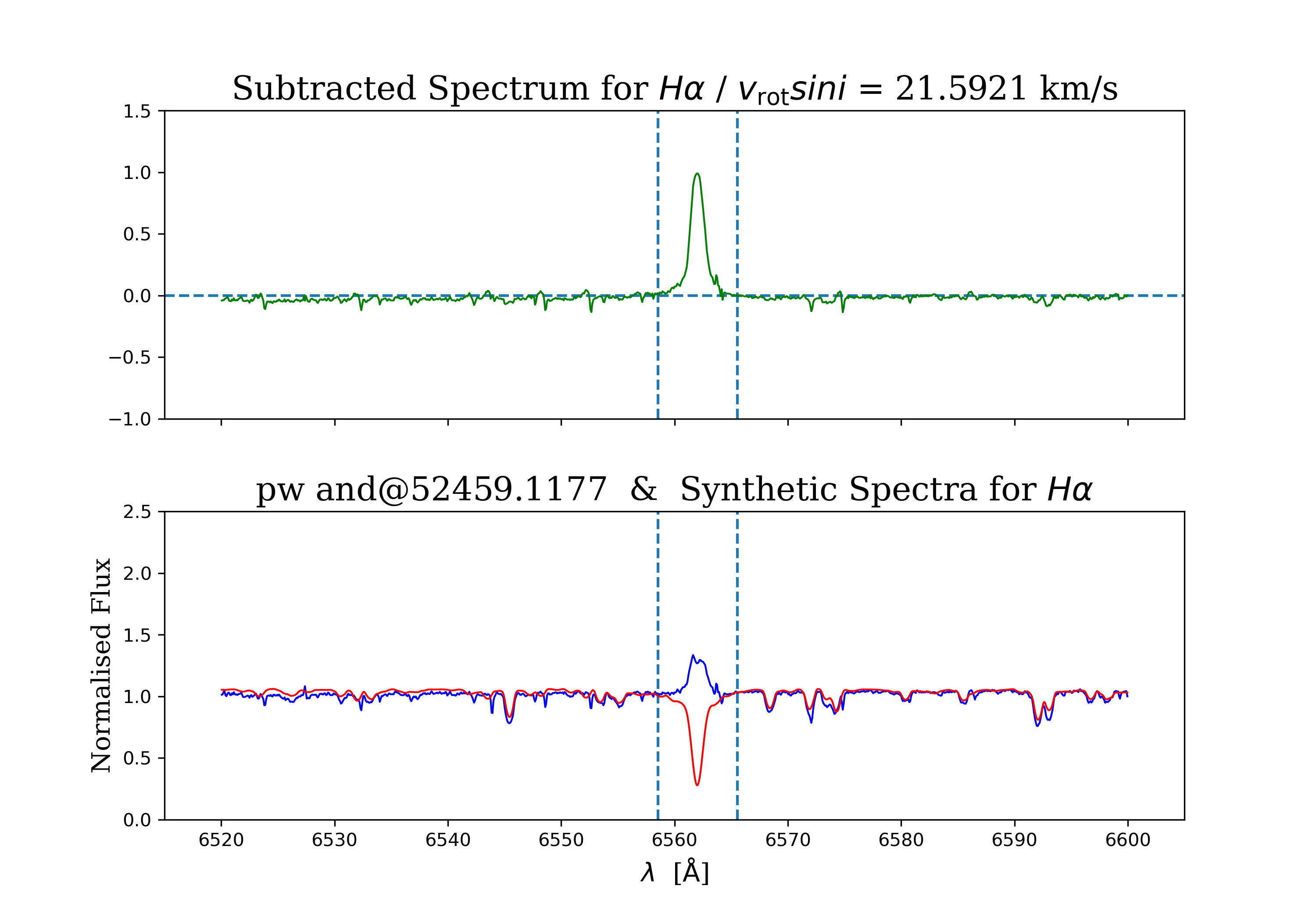}
	\includegraphics[width=0.5\linewidth]{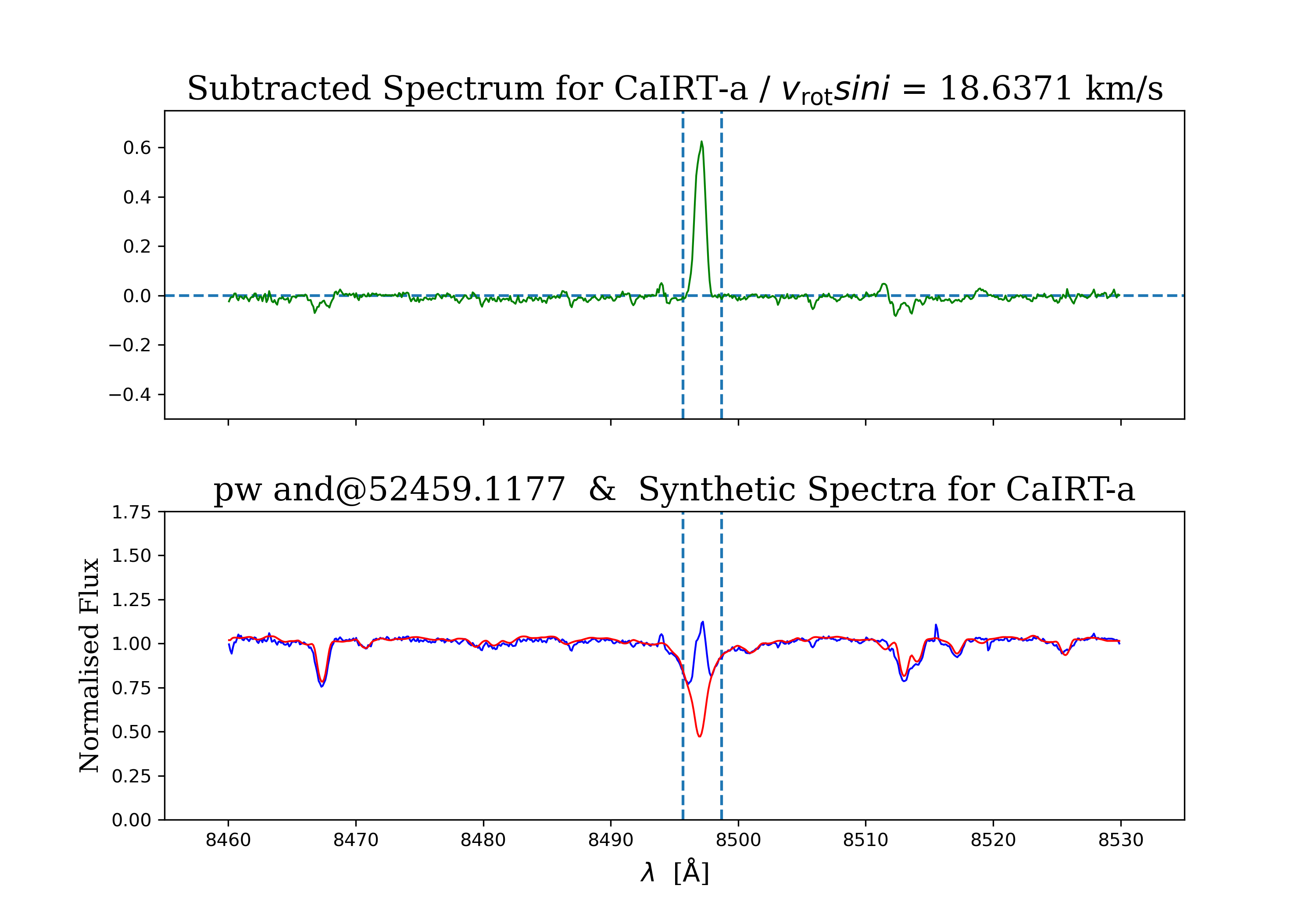}
	\caption{Spectral Subtraction Example from PW And, a K2V star. The different spectra shows chromospheric indicators in Ca \textsc{ii} H, Ca \textsc{ii} K (\textit{top panels}), H$\beta$ and Na \textsc{i} $D_2$ (\textit{middle panels}), and H$\alpha$ and Ca \textsc{ii} IRT-a lines (\textit{bottom panels}). Using \textsc{FOCES} provided spectra
	Lower panel: observed target spectrum (blue) and synthetic spectrum (red), obtained from a reference star spectrum.
	Upper panel: subtracted spectrum (green). 
	In both panels: the vertical blue dashed lines mark the integration limits for the chromosoheric excess emission EW determination.}
	\label{fig:PWAnd}
\end{figure*}

The fit for He \textsc{i} $\lambda10830$ will also apply to the Pa$\gamma$ line. Both lines are very sensitive activity indicators \citep{fuhrmeisterHeI10830}. Finally the last calibrations correspond to two lines of the Paschen series: Pa$\delta$ (bluer $\lambda{10052.14}$ than He \textsc{i} $\lambda{10830}$) and Pa$\beta$ (on the redder side). The calibrations performed in these spectral ranges, where the position of continuum is difficult to determine for late-type M-dwarfs, make them especially suitable for the study by spectral subtraction technique and \texttt{iSTARMOD}.

\section{Spectral Subtraction Examples \label{sect:section5examples}}
\begin{figure*}[ht]
	\includegraphics[width=0.5\linewidth]{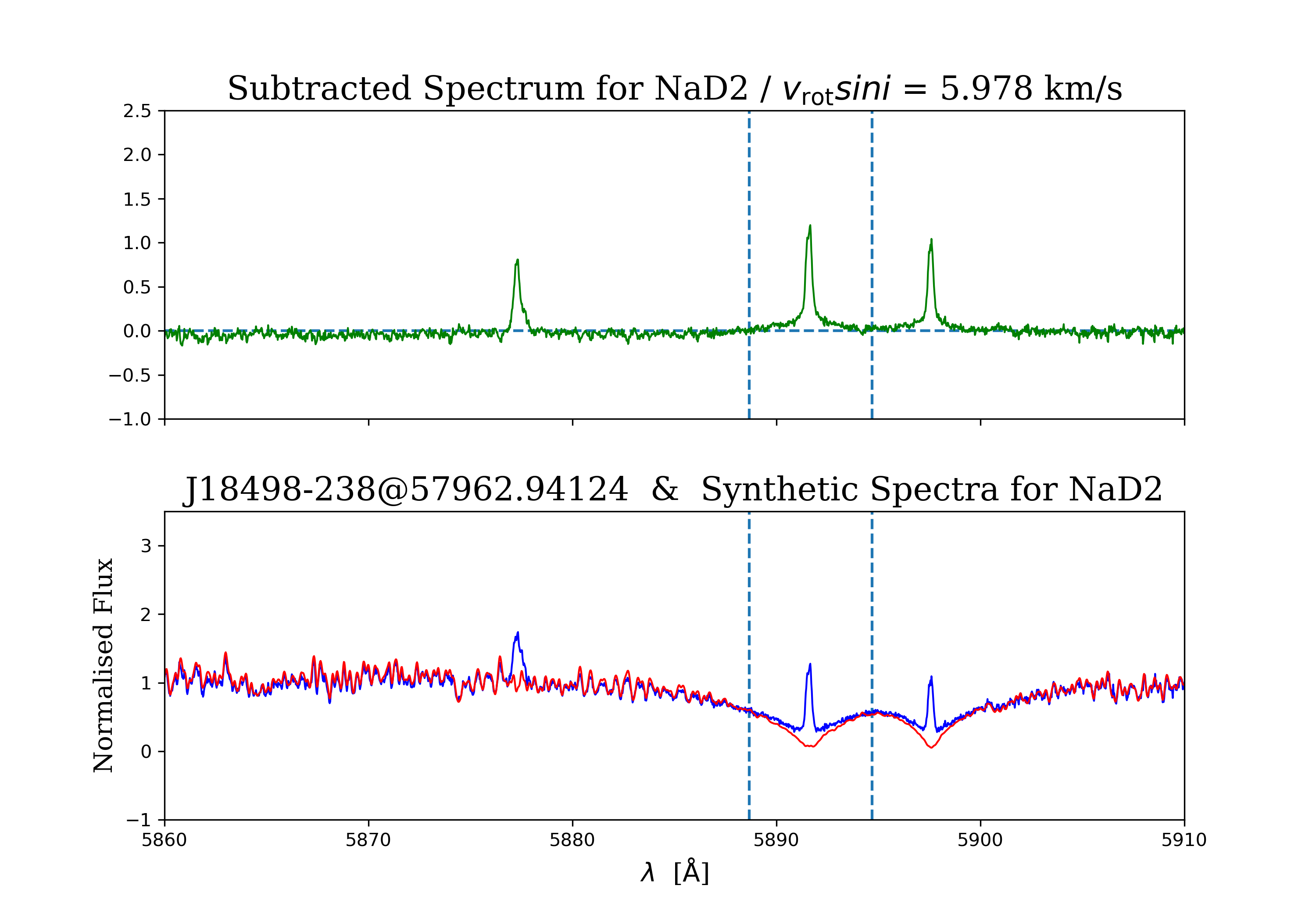}
	\includegraphics[width=0.5\linewidth]{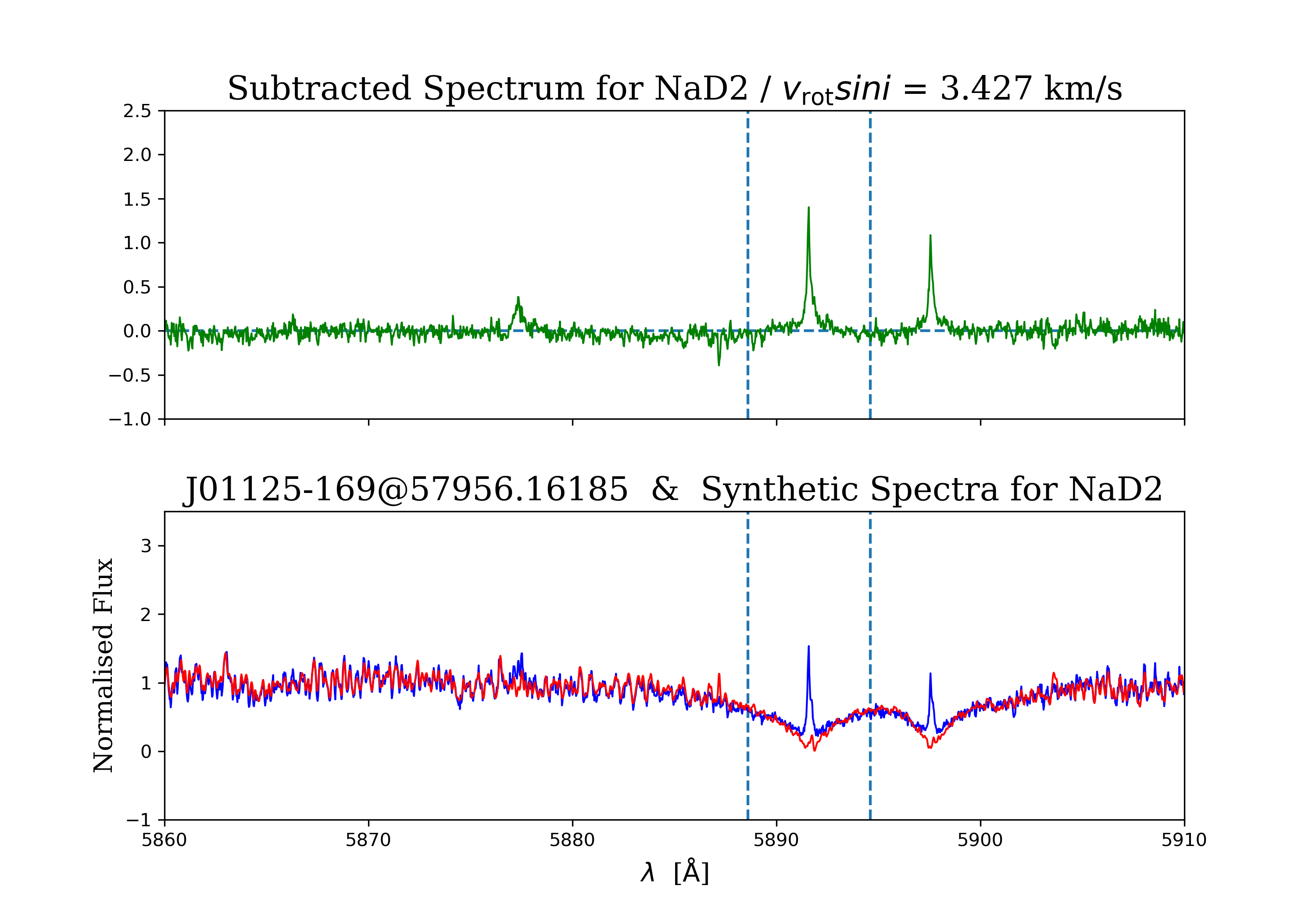}
	\includegraphics[width=0.5\linewidth]{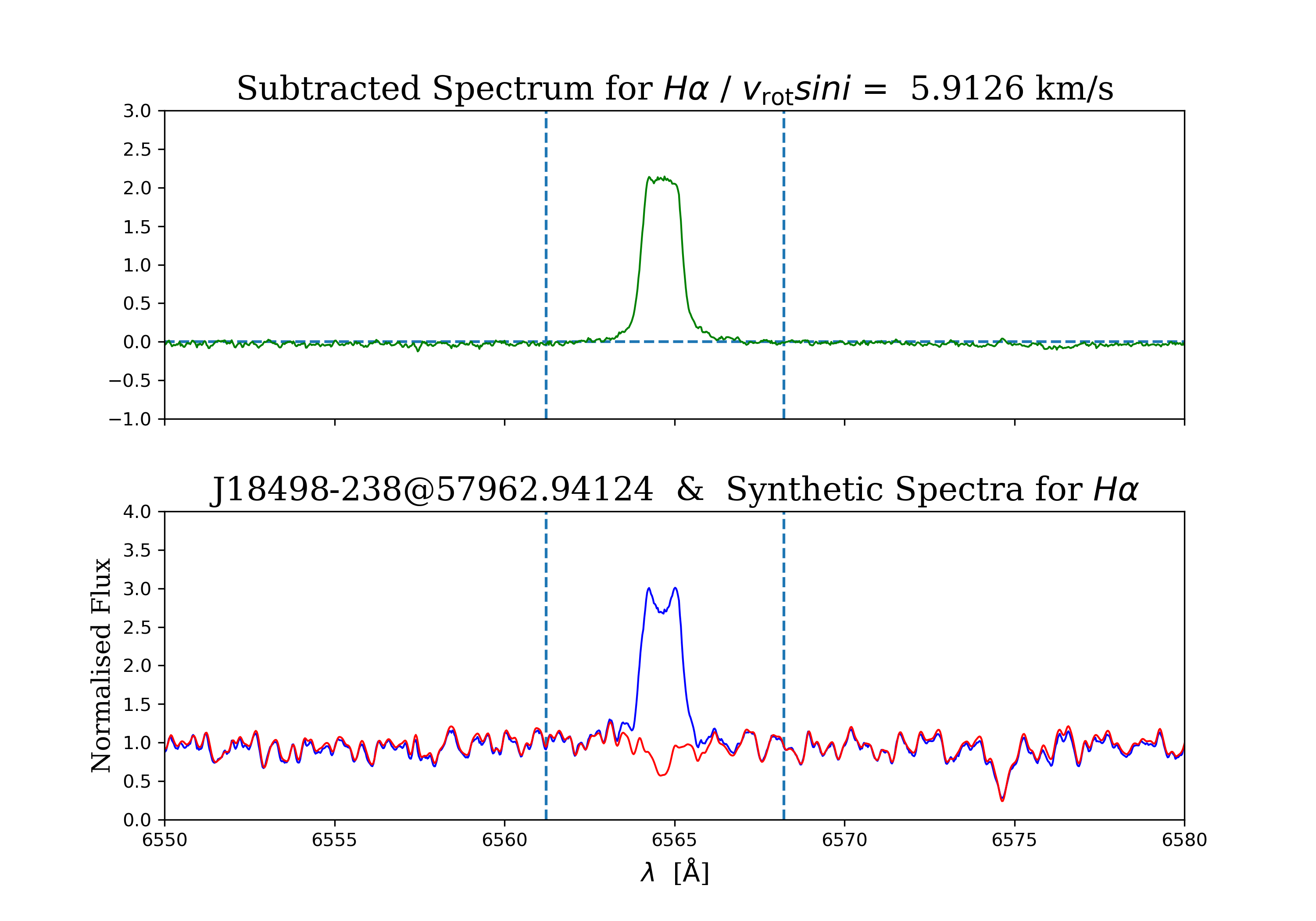}
	\includegraphics[width=0.5\linewidth]{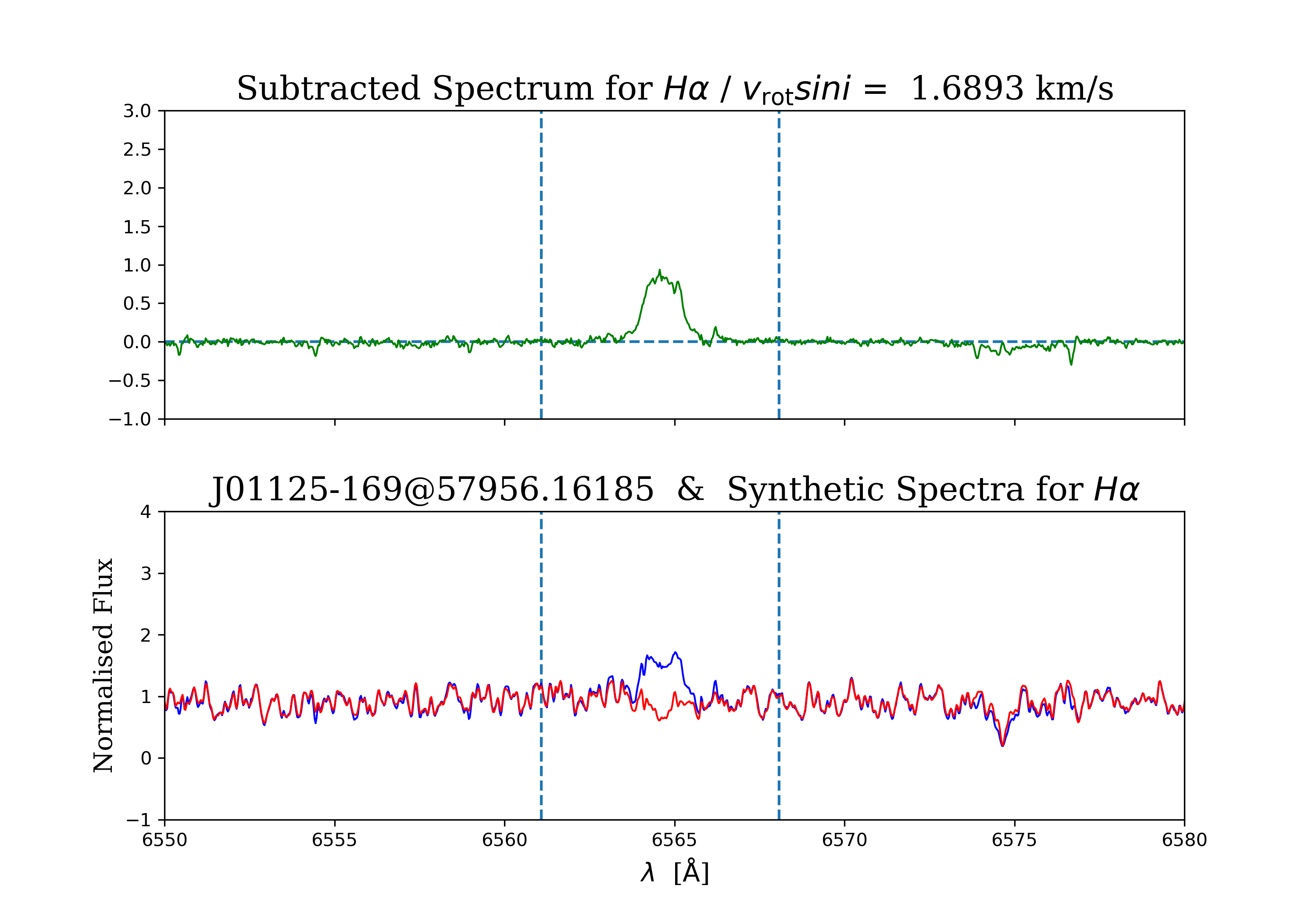}
	\includegraphics[width=0.5\linewidth]{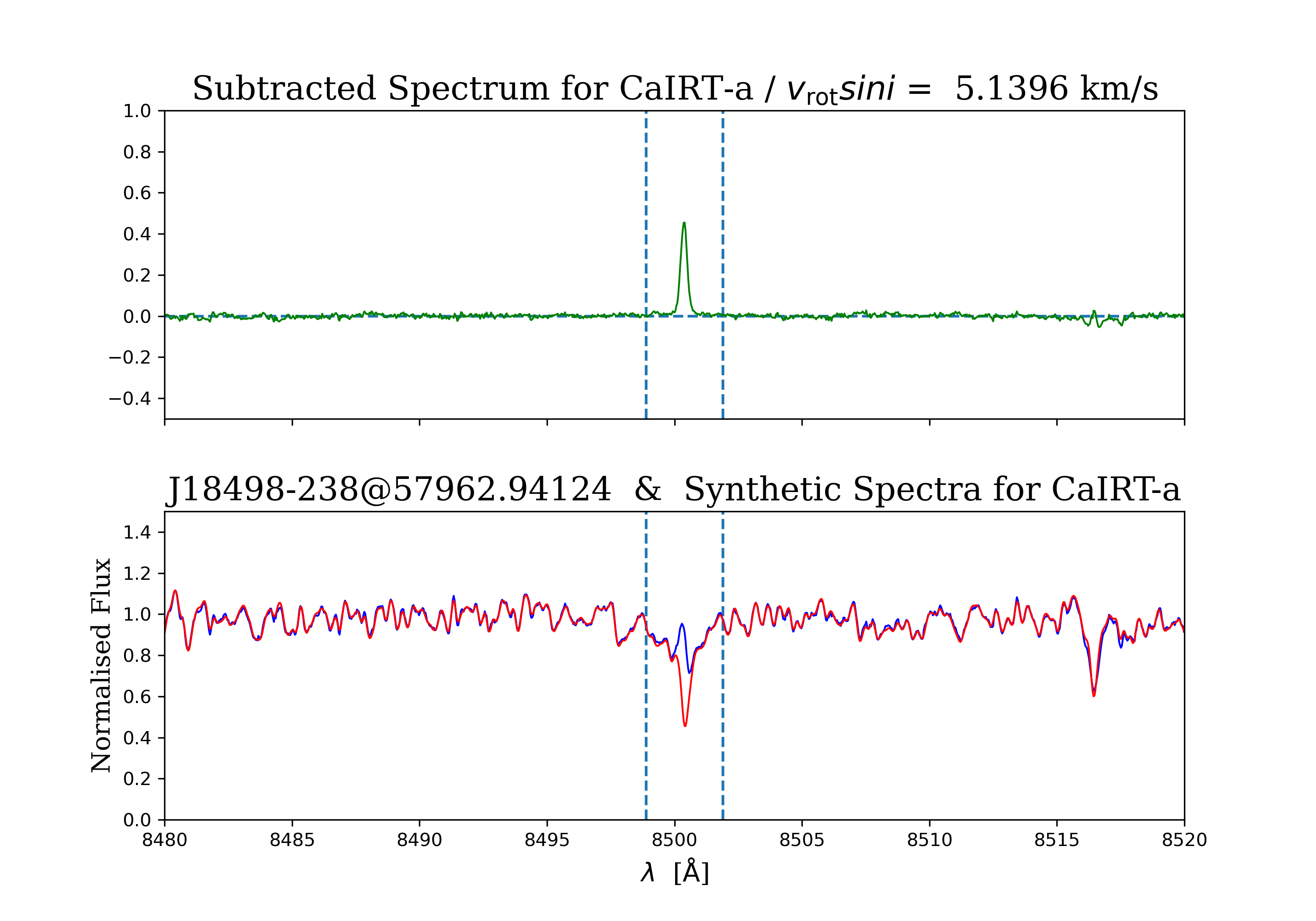}
	\includegraphics[width=0.5\linewidth]{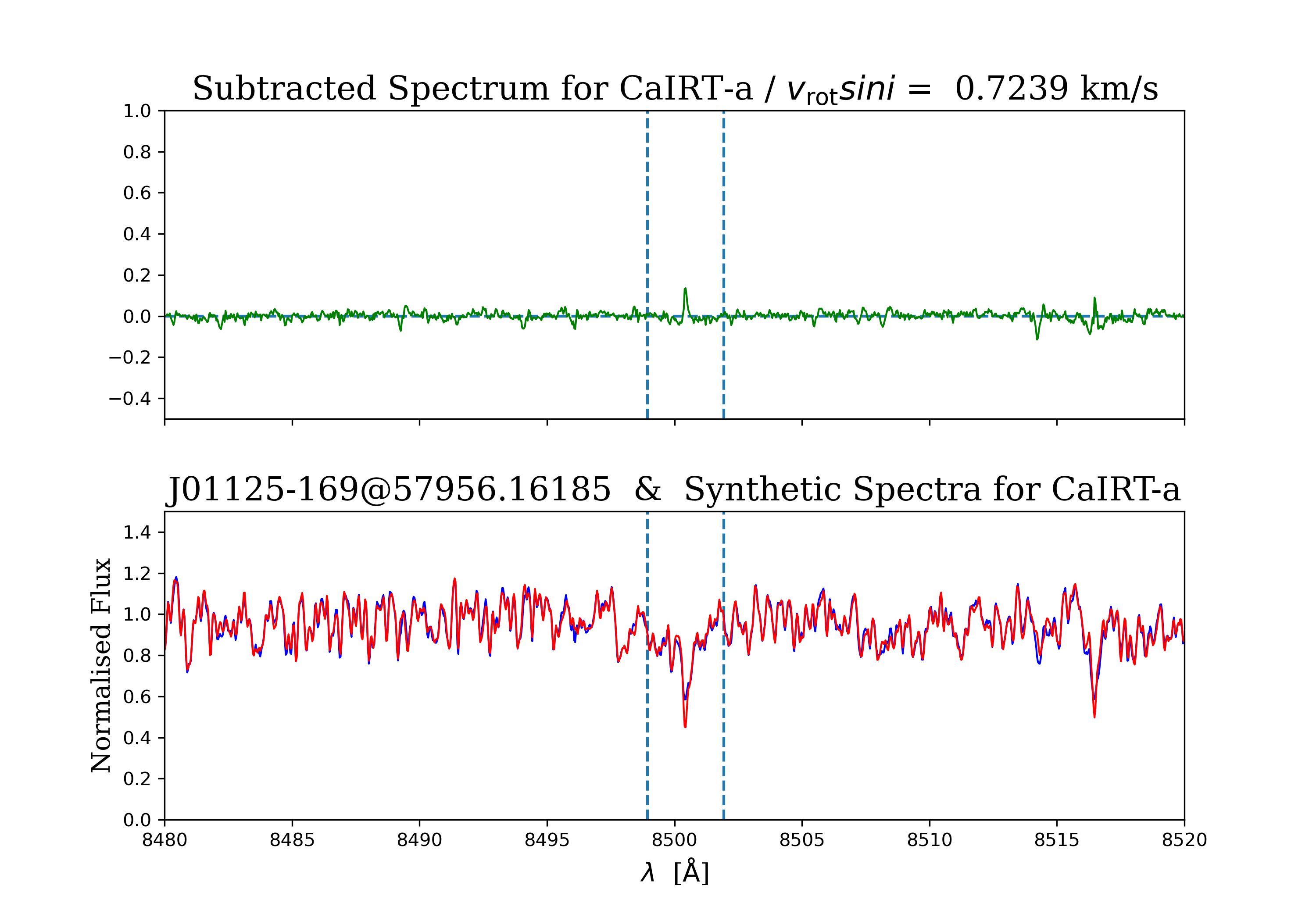}
	\caption{Left panels: Spectral Subtraction Example from V1216 Sgr (J18498--238), an M3.5 dwarf, performed with the VIS channel data of CARMENES spectrum, as provided in DR1 \citep{datarelease1}, showing: He \textsc{i} $D_3$, Na \textsc{i} D$_\text{1}$,D$_\text{2}$ (top), H$\alpha$ (middle) and Ca \textsc{ii} IRT-a $\lambda 8498$ (bottom).
	Right panels: Spectral Subtraction Example from YZ Cet (J01125-169), an M4.5 dwarf, performed with the VIS channel of a CARMENES spectrum as provided in DR1, showing also He \textsc{i} $D_3$, Na \textsc{i} D$_\text{1}$,D$_\text{2}$ (top), H$\alpha$ (middle) and Ca \textsc{ii} IRT-a $\lambda 8498$ (bottom).
	Lines and color codes as in Fig. \ref{fig:PWAnd}.}
	\label{fig:istarmod1}
\end{figure*}

To illustrate the capabilities of \texttt{iSTARMOD} as an actual implementation of the spectral subtraction technique, we present a selection of stellar spectra that have been processed using the methods described in Section \ref{sect:section2SST} and \ref{sect:section3code}. These examples demonstrate the software’s effectiveness in subtracting chromospheric emission and serve to validate its application to a range of disparate stellar types.

Each spectrum shown below has been selected to highlight specific aspects of the processing pipeline, such as continuum normalization, ability of processing spectra of different sources, and activity indicator extraction.

\subsection{Case: Spectral Subtraction of Single Stars} \label{subsect:examplessingles}
Figure \ref{fig:PWAnd} shows several orders of the same spectrum taken with spectrograph FOCES \citep{pfeifferetalonFOCES} at the 2.2 m telescope at Centro Astronómico Hispano en Andalucía (CAHA).
The spectrum was taken from PW And (HD 1405), a relatively bright star ($m_v = 8.6$) classified as a neighborhood Pleiades-age K2 dwarf with high Li \textsc{i} abundance and a member of the Local Association moving group \citep{montesonpwandLi}. 
It is a fast rotator with a photometric period $P_\text{phot} = 1.745$ days and $v \text{sin} i= 22.25$ km s$^{-1}$ \citep{2003A&A...411..489L, baharonPWAnd}.
It is apparent from the figures that the 'quality test' mentioned in Section \ref{sect:section2SST} is fulfilled: the rms of the residuals of the least-squares fit process performed are very low. For the case of Ca \textsc{ii} H, rms = 0.15, with maximum values $\simeq\pm0.3$. 
It is worth mentioning that this line can be resolved against H$\epsilon$, which leads to an error in determining EW of $\leq9$\%. 
The rms value for Ca \textsc{ii} K is slightly worse rms = 0.26, with maximum values of $\simeq\pm0.4$. The best case is achieved for H$\alpha$, where rms = 0.09. In all the above cases the reference, from which the synthetic spectrum have been built is HD166620, a nearby K2V very stable and quiescent star, with spectra also taken with spectrograph FOCES.

\begin{figure*}
	\includegraphics[width=0.5\linewidth]{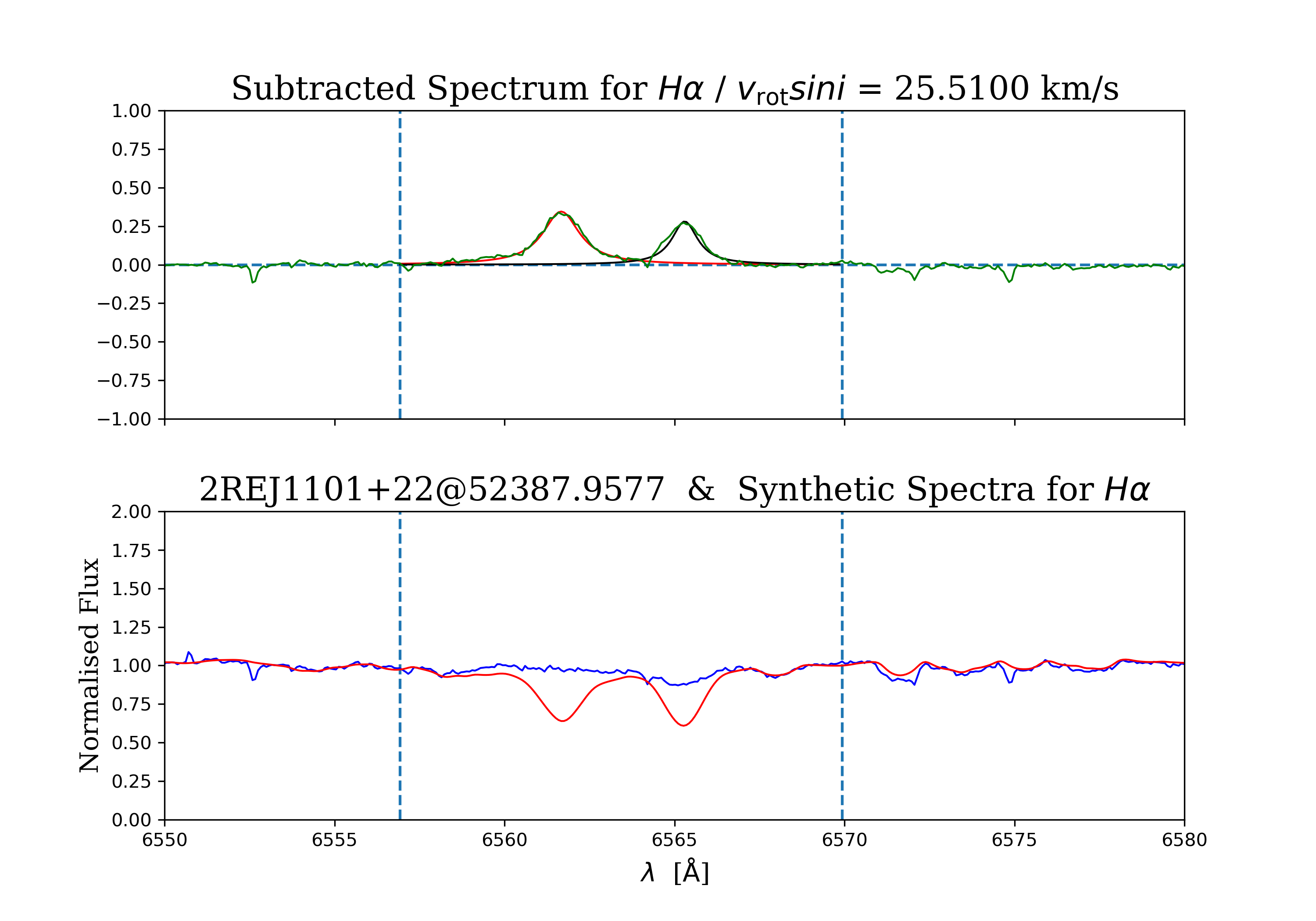}
	\includegraphics[width=0.5\linewidth]{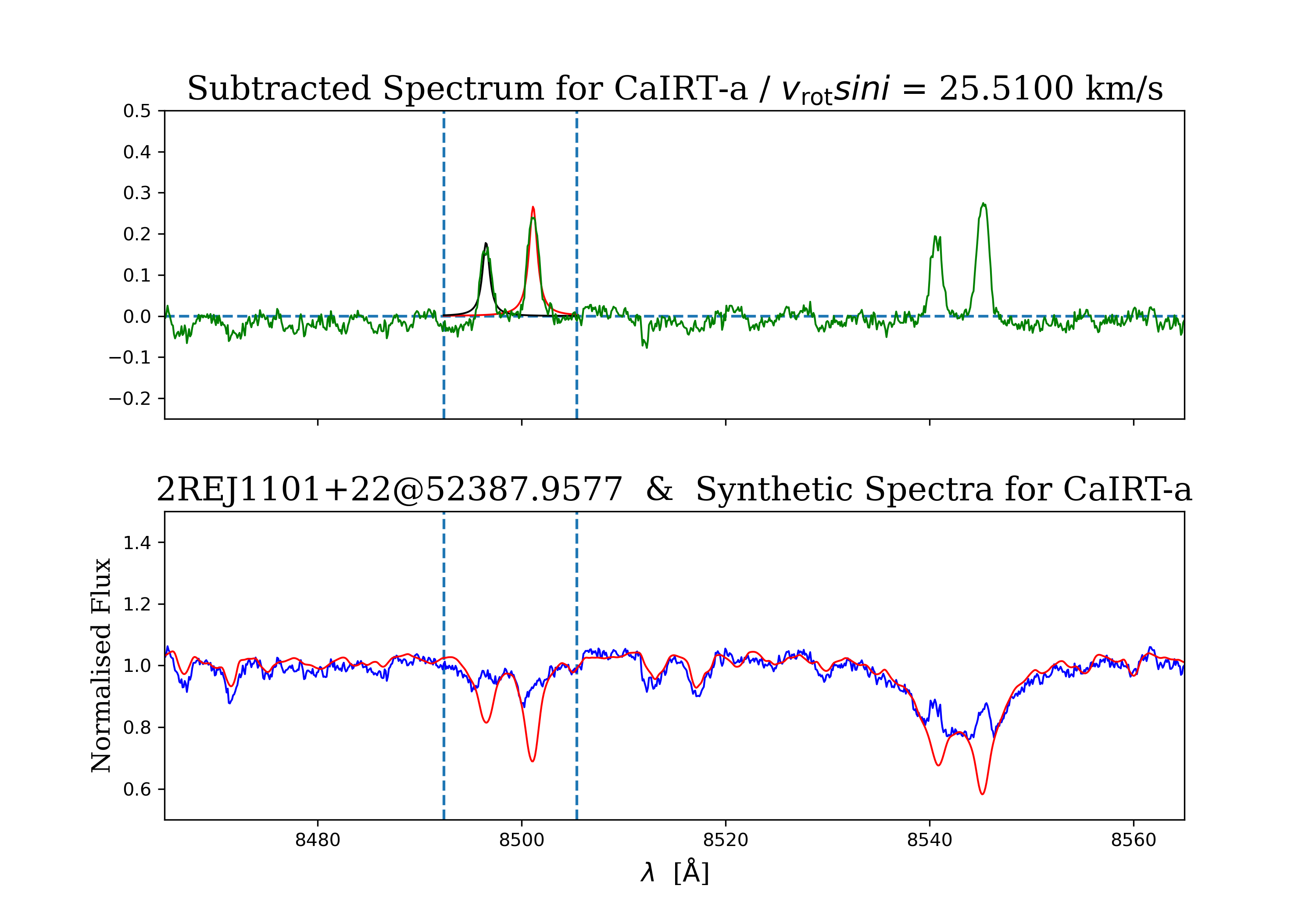}
	\includegraphics[width=0.5\linewidth]{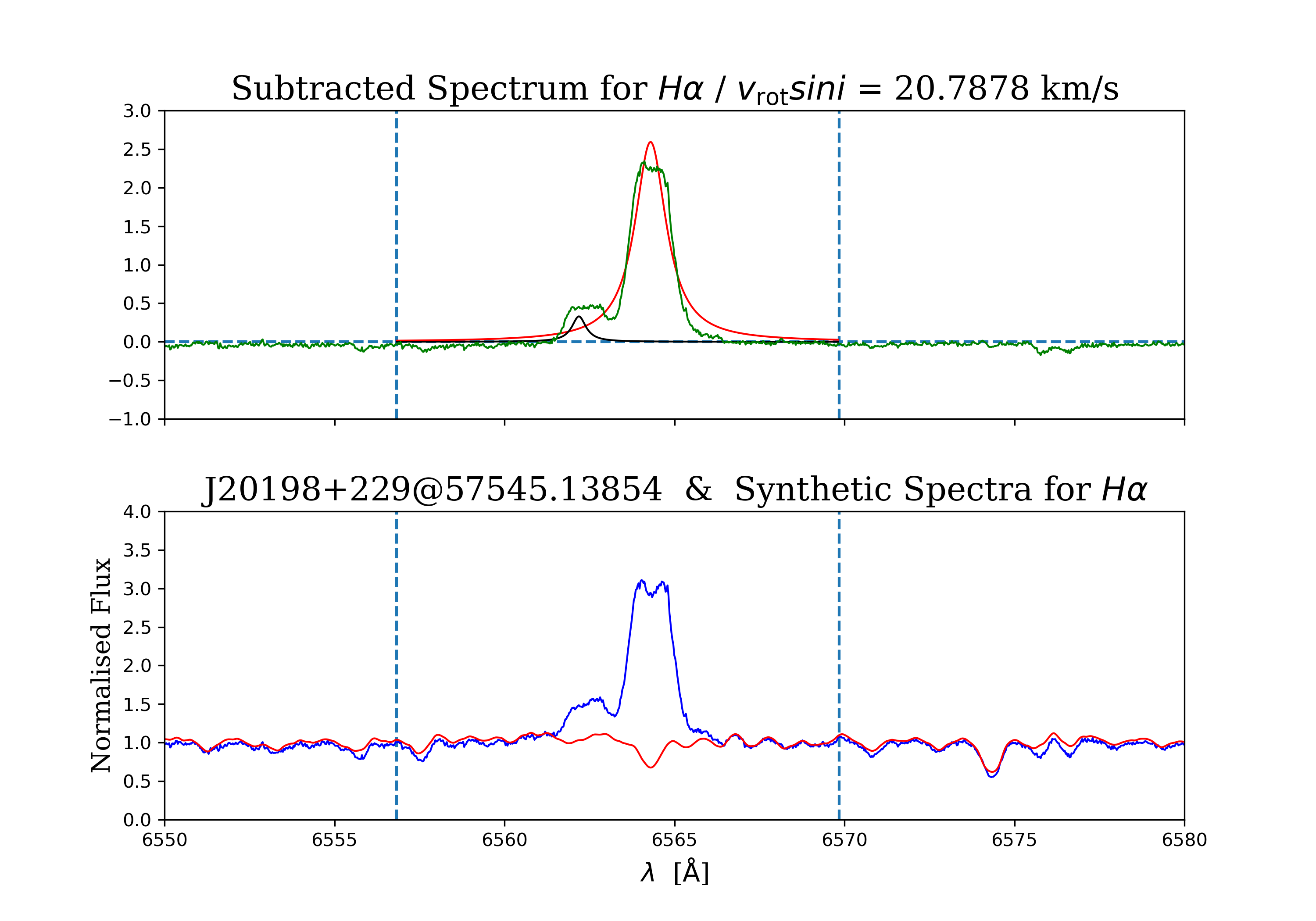}
	\includegraphics[width=0.5\linewidth]{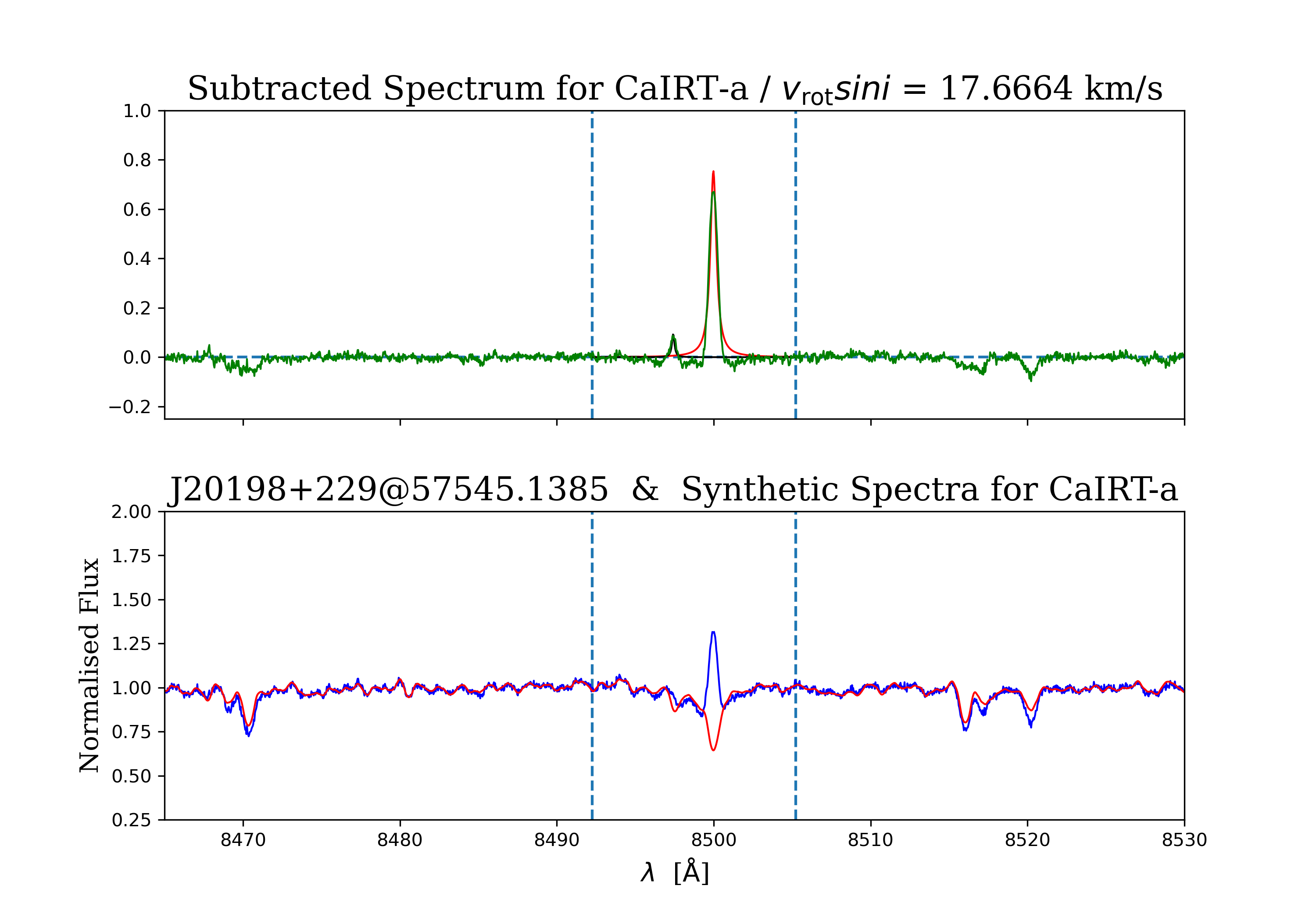}
	\caption{Spectral Subtraction examples of two SB2 systems, with $H_\alpha$ order in the left column and Ca \textsc{ii} IRT-a in the right one.	\textit{Upper Row}: GZ Leo (2RE J1101+223), a K1V+K1V SB2 star, using FOCES provided spectrum \citep{2009AJ....137.3965G}.
	\textit{Lower Row} LP 395-8 (J20198+229), a M3.0V+M3.5V SB2 system, using a spectrum taken by the CARMENES spectrograph, from CARMENES Data Release 1 \citep{datarelease1}. 
	Lines and color codes as in Fig. \ref{fig:PWAnd}.}
	\label{fig:binaries}
\end{figure*}

A selection of high-resolution spectra from CARMENES Data Release 1 (DR1;\citealt{datarelease1, caballerocarmenesinst}) has been used to illustrate the examples presented. CARMENES is a dual-channel spectrograph operating at the 3.5 m telescope of the Calar Alto Observatory, designed for high-precision radial velocity measurements of M dwarfs \citep{quirrenbach1, quirrenbach2}. 
Figure \ref{fig:istarmod1} shows subtracted spectra in several orders from spectra taken of two stars. The left panel of Figure \ref{fig:istarmod1} corresponds to V1216 Sgr (Ross 154 or J18498--238), a nearby red dwarf star, one of the closest stellar neighbors to our solar system. V1216 Sgr is a UV Ceti type flare star, M3.5V, with strong magnetic activity that is reflected in the figures. The He~\textsc{i}~D$_3$ and Na~\textsc{i}~D$_1$,D$_2$ lines are clearly visible in the first example analyzed. This example demonstrates the effectiveness of spectral subtraction when several \texttt{PIX\_EXCL} regions are defined as exclusion intervals--one for each of the three lines present--during the least-squares fit. Across all the spectra taken rms$\approx$0.16 in this order, meaning that outside the lines studied the normalized subtracted flux moves in $\leq\pm0.2$. H$\alpha$ presents the same situation achieving an error in EW determination as low as 0.09\%. For the case of Ca \textsc{ii} IRT-a $\lambda 4980$, $rms = 0.06$ and the error in determining EW is: $e(EW)\approx0.05$\%. The reference star is the old and relatively quiescent well-known Barnard's Star, with spectral type M3.5V and $T_\text{eff}$ = 3273 K and the spectrum provided by CARMENES DR1 with ID J17578+046. The right panel of Fig. \ref{fig:istarmod1} corresponds to the CARMENES target YZ Cet (J0115--169), also a nearby red dwarf star but with spectral type M4.5Ve. YZ Cet is a flare star, showing intermittent episodes of magnetic activity. This suggests a slightly lower level of chromospheric activity, as reflected in their levels of Ca \textsc{ii} IRT-a excess emission. However, being a nearby star, the S/N and the values of rms are the same order as in the previous case, and the same applies to the errors obtained in measuring EW. The reference star employed was GJ 1235, with spectral type M4.5V and $T_\text{eff} = 3059$ K, provided by CARMENES DR1 with ID J19216+208.

\subsection{Case: Spectral Subtraction of Spectroscopic Binaries SB2 \label{subsect:examplesSB2}}

Two systems have been chosen as examples of the application of \texttt{iSTARMOD} for spectral subtraction in binary stars.
The first one, GZ Leo (or 2RE J1101+223), is a chromospherically active binary system classified as K1Ve+K1Ve. It is a double-lined spectroscopic binary (SB2), with its components clearly separated, and has been the subject of several observational studies. Early high-resolution spectroscopic observations were obtained with FOCES \citep{pfeifferetalonFOCES} at the 2.2 m telescope at CAHA in 2002 by \cite{2009AJ....137.3965G}, who investigated its activity and orbital parameters, using \texttt{STARMOD}. One of these spectra has been reprocessed using \texttt{iSTARMOD} and the results are shown in the top row of Figure \ref{fig:binaries}. The weights were fixed to $w_1 = 0.508$ and $w_2= 0.492$ in the example for H$\alpha$, according to the fact of both components being of the same type, and resulting in a rms = 0.025. This translates for Ca \textsc{ii} IRT-a $\lambda{8498}$ to weights $w_1 = 0.572$ and $w_2= 0.428$, resulting in rms = 0.020. The error in EW determination is in both cases $e(EW) \approx 1$\%. As in the case for single stars, the reference stars employed, provided by FOCES, are the G9IV and K0V stars HD 92588 and HD 97004, with similar $T_\text{eff} \approx 5100$.

More recent observations with the CARMENES spectrograph, from CARMENES Data Release 1 (DR1; \citealt{datarelease1, caballerocarmenesinst}) have been included and shown in the bottom row of Figure \ref{fig:binaries}. The star is LP 395-8, with spectral type M3.0V+ and $T_\text{eff,1} = 3600$ K and $T_\text{eff,2} = 3300$ K (spectral type M3.5V). Their luminosity ratio $L_2/L_1 = 0.14 \pm 0.01$ in the VIS channel translates in the weights: $w_1 = 0.893$ and $w_2= 0.107$ for H$\alpha$ and $w_1 = 0.834$ and $w_2= 0.166$ for Ca \textsc{ii} IRT-a $\lambda{8498}$. The error in EW determination is in both cases $e(EW) \approx 2$\%. The reference stars are taken from CARMENES DR1 and referred to in \cite{taloretal}. They are HO Lib, with spectral type M3.0V and $T_\text{eff} = 3441 K$, and again Barnard's Star (M3.5V).
\section{Conclusions}\label{sect:section7conclusions}
\texttt{iSTARMOD} tool is a new implementation of the spectral subtraction technique. It is an upgraded and \texttt{Python}-coded version of the previous \texttt{STARMOD} code, with improved usability, modularity, and integration with modern data analysis workflows. This enhanced implementation of the code allows a more precise determination of radial and rotational projected velocities, as well as an automated calculation of the EWs in the study of single and binary chromospherically active stars. 
The code is publicly available \citep{Labarga2025_iSTARMOD} under an open-source license.

The code is also very useful to identify new lines with a significant chromospheric contribution, apart from well-known activity indicators, and to search for magnetically sensitive spectral lines, which are lines with detectable Zeeman broadening (see \citealt{2020sea..confE.168M}).

As a companion to the \texttt{iSTARMOD} code, a series of calibrations, allowing the determination of its corresponding fitted $\chi$-functions, have been discussed and provided for H$\alpha$, Ca \textsc{ii} H and K, He \textsc{i} D$_3$ + Na \textsc{i} D$_\text{1}$,D$_\text{2}$, Ca \textsc{ii} IRT, Pa$\delta$, Pa$\gamma$ + He \textsc{i} $\lambda 10830$ and P$\beta$. 

The application of both tools together will allow the calculation of flux--flux relationships \cite{Labarga2025} and temporal series, given the ability to process large spectral datasets or extensive stellar samples, and will be the subject of further studies. 
\begin{acknowledgments}
Part of the spectral subtraction examples in this work have been elaborated on usin spectra from the CARMENES project. CARMENES is an instrument at the Centro Astronómico Hispano en Andalucia (CAHA) in Calar Alto Observatory, operated jointly by the Junta de Andalucía and the Instituto de Astrofísica de Andalucía (CSIC). 
We used data from the CARMENES data archive at CAB (CSIC-INTA), made publicly available in CARMENES Data Release 1 (DR1). 

We acknowledge financial support from the Universidad Complutense de Madrid (UCM) and the Agencia Estatal de Investigaci\'on (AEI/10.13039/501100011033) of the Ministerio de Ciencia e Innovaci\'on and the ERDF ``A way of making Europe'' through project PID2022-137241NB-C4[4]

We appreciate our anonymous referee for helpful suggestions that greatly improved the quality of this paper.

\end{acknowledgments}
\begin{contribution}
FL was responsible for writing and submitting the manuscript.
DM came up with the initial research concept as PhD Thesis director and edited the manuscript.
\end{contribution}

\facilities{CAO:2.2m (FOCES),CAO:3.5m (CARMENES)}
\software{ARES is available at this webpage: \href{https://github.com/sousa-sag/ARES}{https://github.com/sousasag/ARES}.\\
Astropy \citep{astropy2018} is available at GitHub: \href{https://github.com/astropy/astropy}{https://github.com/astropy/astropy}.\\
Numpy \citep{numpy2020} is available at GitHub: \href{https://github.com/numpy/numpy}{https://github.com/numpy/numpy}.\\
Scipy \citep{scipy2020} is available at GitHub: \href{https://github.com/scipy/scipy}{https://github.com/scipy/scipy}.\\
TeXstudio -- A LaTeX Editor. Available at: https://www.texstudio.org/ (accessed May 20, 2025).}
\appendix
\section{Calibrations curves of log\texorpdfstring{$\chi$}{χ} for the most important chromospheric activity indicators \label{annexa}}
\begin{figure}[ht]
\centering
\includegraphics[width=0.3\linewidth]{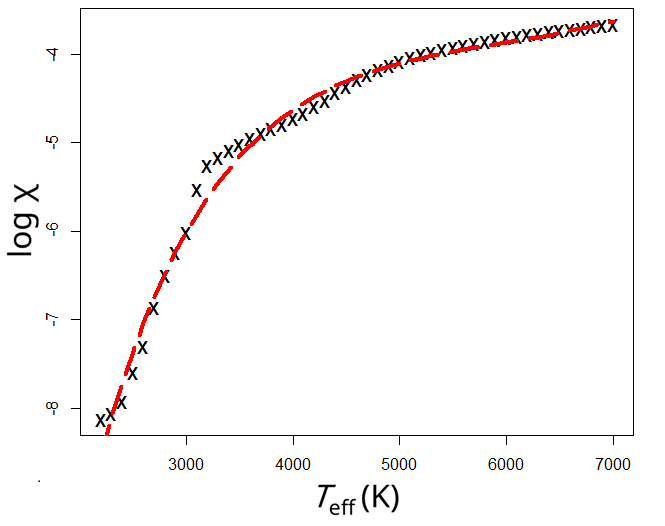}
\includegraphics[width=0.3\linewidth]{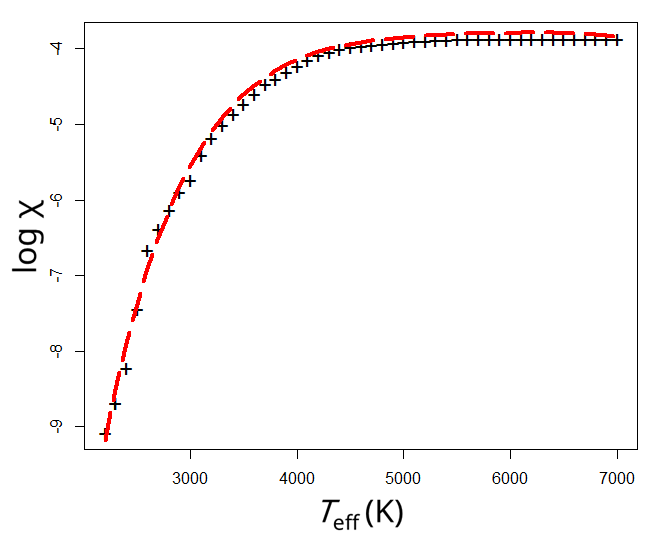}
\includegraphics[width=0.3\linewidth]{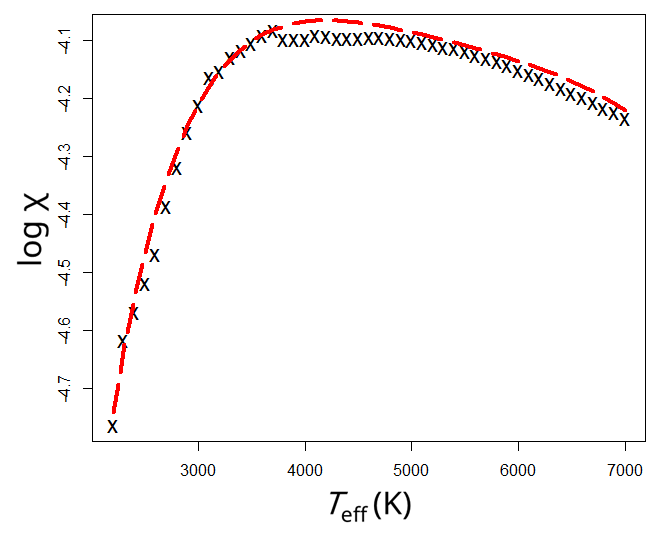}\\
\includegraphics[width=0.3\linewidth]{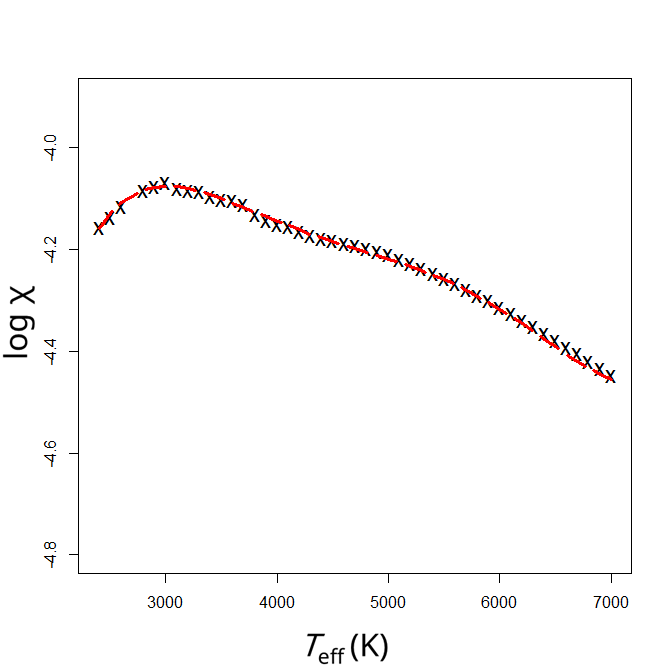}
\includegraphics[width=0.3\linewidth]{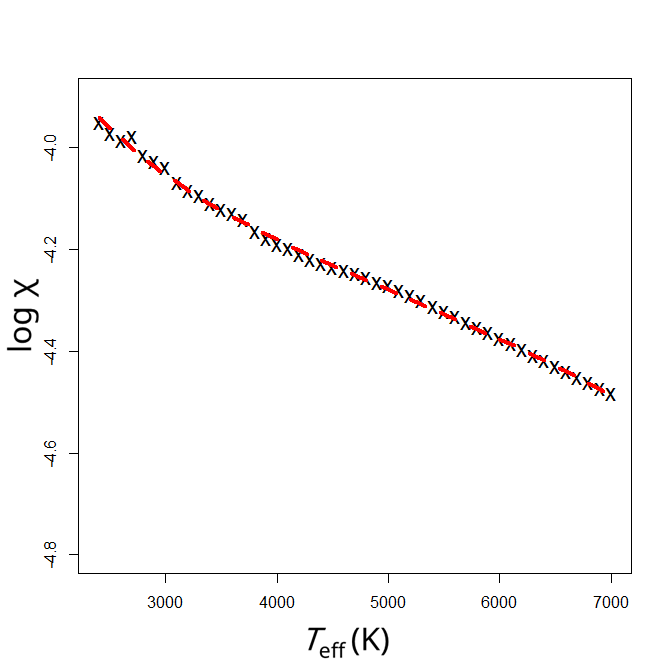}
\includegraphics[width=0.3\linewidth]{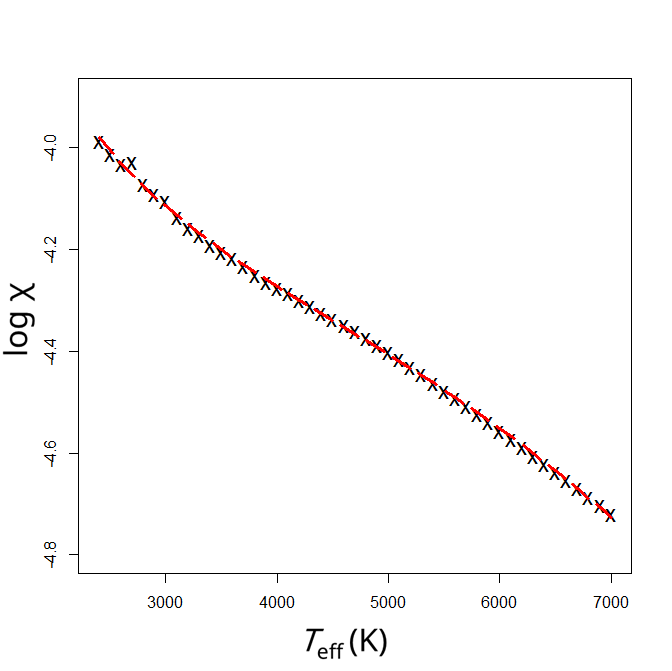}
\caption{First row, calibration for Ca \textsc{ii} H\&K doublet (\textit{Left Panel}), He \textsc{i} D$_3$ and also Na \textsc{i} D$_1$,D$_2$.  (\textit{Center Panel}), and Ca \textsc{ii} IRT lines (\textit{Right Panel}). Second row, the curves for Pa$\delta$ (\textit{Left Panel}), Pa$\gamma$ + He \textsc{i} $\lambda 10830$ (\textit{middle panel}) and Pa$\beta$ (\textit{right panel}).All in the range of effective temperatures [2200, 7000] K. The calibration curve for H$\alpha$ can be seen in the Figure \ref{fig:FigCalibChiHalpha}.
\label{fig:FigCalibChi}}
\end{figure}

The following Table \ref{table:t1} and Fig. \ref{fig:FigCalibChi} show the fit parameters of the $\chi$-functions mentioned in Section \ref{sect:section4chi}
\begin{splitdeluxetable}{lccccBlccc}
	\tabletypesize{\scriptsize}
	\tablewidth{0pt} 
	\tablecaption{Parameters of the fit for $\log \chi = C_1 + \alpha \log T_\text{eff} + P_5(T_\text{eff})$ for all studied lines.
		\label{table:t1}}
	\tablehead{
		\colhead{Coefficient } & \colhead{Ca \textsc{ii} H\&K}& \colhead{He \textsc{i} D$_3$ \& Na \textsc{i} D$_1$,D$_2$} & \colhead{H$\alpha$} & \colhead{Ca \textsc{ii} IRT} & \colhead{Coefficient} & \colhead{Paschen$\delta$} & \colhead{He \textsc{i} $\lambda10830$ \& Paschen$\gamma$} & \colhead{Paschen$\beta$}}
	\colnumbers
	\startdata
	$10^{-2}$ $C_1$ &  $-1.75\pm0.13$ & $-4.5\pm0.3$ & $-0.46\pm0.02$ & $-0.90\pm0.07$ & $10^{-2}$ $C_1$ & $-0.12\pm0.04$ & $0.045\pm0.009$ & $0.064\pm0.008$ \\
	$10^{-1}$ $\alpha$ & $5.4\pm0.5$ & $14.9\pm1.2$ & $1.281\pm0.006$ & $2.9\pm0.2$ & $10^{-1}$ $\alpha$& $0.0$ & $-0.27\pm0.03$ & $-0.34\pm0.03$ \\
	$10^{3}$ $\beta_1$ & $-7.8\pm1.0$ & $-33.3\pm4.0$ & $-1.01\pm0.06$ & $-7.1\pm0.8$ & $10^{3}$ $\beta_1$ & $8.5\pm0.5$ & $0.38\pm0.06$ & $0.51\pm0.05$ \\
	$10^{7}$ $\beta_2$ & $3.4\pm0.5$ & $28.8\pm4.7$ & $0$ & $6.6\pm0.9$ & $10^{7}$ $\beta_2$ & $-37.0\pm2.2$ &  $-0.24\pm0.03$ & $-0.35\pm0.03$ \\
	$10^{10}$ $\beta_3$ & $0$ & $-1.1\pm0.2$ & $0$ & $-0.26\pm0.05$ & $10^{10}$ $\beta_3$ & $7.8\pm0.5$ & $0$ & $0$ \\
	$10^{14}$ $\beta_4$ & $0$ & $0$ & $0$ & $0$ & $10^{14}$ $\beta_4$ & $-8.0\pm0.5$ & $0$ & $0$ \\
	$10^{18}$ $\beta_5$ & $0$ & $0$ & $0$ & $0$ & $10^{18}$ $\beta_5$ & $3.2\pm0.2$ & $0$ & $0$ \\
	&&&&&&&& \\
	Adj. Param. R\textsuperscript{2} & $0.9893$ & $0.9969$ & $0.9712$ & $0.9895$ & Adj. Param. R\textsuperscript{2} & $0.9986$ & $0.9995$ & $0.9995$ \\
	\enddata
\end{splitdeluxetable}
\clearpage
\newpage
\section{\texttt{iSTARMOD} Input Parameters and Configuration file \label{annexb}}
\small
As shown in Figure \ref{fig:workflow}, the first step is to provide the algorithm a very large set of parameters, which are specified in an input \texttt{inputParameters.sm} file with the syntax \texttt{KEYWORD = value}.

The parameters are grouped in the following sections, whose names are kept as in \texttt{STARMOD} for historical reasons and retro-compatibility:
\begin{itemize}
	\item \texttt{GENERAL INFORMATION}
	\begin{itemize}
		\item \texttt{IM\_PATH =v.\textbackslash path\textbackslash{}}	\# This is the path where the input FITS spectrum is located. Can be a relative path.
		\item \texttt{OBJ\_NAME = name\_of\_file.fits} 		\# Name of the input FITS file(s).
		If there are several files, wildcards can be used.
	\end{itemize}
	\item \texttt{OUTPUT SPECTRA}: All these files will be generated in the same path as defined by \texttt{IM\_PATH}
	\begin{itemize}
		\item \texttt{SYN\_SPEC = VALUE}  			\# write output spectra: only admits YES/NO
		\item \texttt{SYN\_NAME = name\_of\_synth}	    \# name of the synthetic spectrum (if SYN\_SPEC =YES)
		\item \texttt{SUB\_NAME = name\_of\_subtr}      \# name of subtracted spectrum (if SYN\_SPEC =YES) 
	\end{itemize}
	\item \texttt{INTERPOLATION PARAMETERS}: In this group, the number of executions of the least-squares iterative loop is defined, as well as the wavelength range in \si{\angstrom}, for the order to be analyzed, as well as the wavelength intervals to be excluded of the sum of the residuals for the least squares. The intervals are typically those affected by activity, where the lines present excess emission instead or absorption. If the radial velocities are large, it is prudent to also exclude the blue or red ends of the spectrum.
	\begin{itemize}
		\item \texttt{N\_ITER = 8}                 \# number of iterations: integer
		\item \texttt{PIX\_ZONE = XXXX YYYY  wvl}  \# 'pixel' range for the fit: 2 integers
		\item \texttt{PIX\_EXCL = ZZZ1 ZZZ2  wvl}  \# Blue/Red exclusion skip subrange: 2 integers (optional)
		\item \texttt{PIX\_EXCL = ZZZ3 ZZZ4  wvl}	 \# Line Zone skip subrange: 2 integers (optional)
		\item \texttt{PIX\_EXCL = ZZZ5 ZZZ6  Wvl}	 \# skip subrange: 2 integers (optional)
		\item \texttt{PIX\_EXCL = 		}		 \# skip subrange: 2 integers (optional)
		\item \texttt{PIX\_EXCL =      }           \# skip subrange: 2 integers (optional)
	\end{itemize}
	\item \texttt{PRIMARY STAR}. Reference star parameters. The radial velocity, $v sin i$, and weight values for the initial guesses can be fixed so that they do not vary across loop iterations, or allowed to vary freely from those initial guesses.
	\begin{itemize}
		\item \texttt{REF\_PATH = .\textbackslash path\textbackslash{}}	\# path (absolute or relative) of fits spectrum for primary star of reference
		\item \texttt{PRM\_NAME = name\_of\_file.fits }	\# primary file name
		\item \texttt{PRM\_RAD  = X.xxx var/fix}	\# doppler shift: float, keyword (var/fix)
		\item \texttt{PRM\_ROT  = Y.yyy var/fix}	\# doppler broadening: float, keyword (var/fix)
		\item \texttt{PRM\_WGT  = Z.zzz var/fix}	\# weight: float, keyword (var/fix).
	\end{itemize}
	\item \texttt{SECONDARY STAR}. The same as previous. In this case for the secondary star.
	\begin{itemize}
		\item \texttt{SEC\_PATH = .\textbackslash path\textbackslash{}}		\# path (absolute or relative) of fits spectrum for primary star of reference
		\item \texttt{SEC\_NAME = name\_of\_file.fits } \# secondary file name: Name /NONE
		\item \texttt{SEC\_RAD = X.xxx var/fix}	    \# doppler shift: float, keyword (var/fix)
		\item \texttt{SEC\_ROT = Y.yyy var/fix}      \# doppler broadening: float, keyword (var/fix)
		\item \texttt{SEC\_WGT = Z.zzz var/fix}      \# weight: float, keyword (var/fix).
	\end{itemize}
	\item \texttt{SPECTRA FORMAT}. To define the order of the input spectrum to analyze. The \texttt{LINE } parameter must be compliant with the specification of a configuration file \texttt{lambdas.dat}. If \texttt{NONE} the tool will not perform any calculation
	\begin{itemize}
		\item \texttt{MODE =  ech     } \# NOT USED
		\item \texttt{APERTURE =  46  } \# aperture number/ Order : integer
		\item \texttt{BAND = }         \# NOT USED
		\item \texttt{LINE = CaIRT-a } \# To specify the LINE for measuring the EW
	\end{itemize}
	\item \texttt{ALGORITHM \& VISUALISATION PARAMS}. To define the maximum value of the y-axis in each of the window of the created figure. The FINDMAX\_TOLERANCE restricts the wavelength interval where the emission maximum is to be searched.
	\begin{itemize}
		\item \texttt{MAXFLUXDISP\_OBJ = X.xx 	} \# Maximum y-value to display in the object+ref spectra window
		\item \texttt{MAXFLUXDISP\_SUB = X.xx 	} \# Maximum y-value to display in the subtracted spectrum window
		\item \texttt{FINDMAX\_TOLERANCE = 0.00x} \# to find the maximum of the subtracted spectra when calculating the integral 
	\end{itemize}
\end{itemize}
\normalsize

\newpage
\bibliography{updated_refs,sample7,starmod_references}{}

@ARTICLE{2025MNRAS.tmp..723L,
	author = {{Liu}, Fei and {Li}, Kai and {Gao}, Xiang and {Wang}, Jing-Yi and {Xu}, Xin and {Wang}, Yi-Fan and {Wu}, Cheng-Yu and {Li}, Mu-Zi-Mei and {Gao}, Xing and {Sun}, Guo-You},
	title = "{Extremely low mass ratio contact binaries {\textendash} II. The first photometric and spectroscopic investigations of six systems with orbital periods longer than 0.5 d}",
	journal = {\mnras},
	year = 2025,
	month = jun,
	volume = {540},
	number = {1},
	pages = {1290-1307},
	doi = {10.1093/mnras/staf763},
	archivePrefix = {arXiv},
	eprint = {2505.04029},
	primaryClass = {astro-ph.SR},
	adsurl = {https://ui.adsabs.harvard.edu/abs/2025MNRAS.540.1290L}
}

@ARTICLE{2025AJ....169..198C,
	author = {{Cao}, Dongtao and {Gu}, Shenghong},
	title = "{Further Investigation on Chromospheric Activity Patterns of the RS Canum Venaticorum Star V711 Tauri}",
	journal = {\aj},
	year = 2025,
	month = apr,
	volume = {169},
	number = {4},
	eid = {198},
	pages = {198},
	doi = {10.3847/1538-3881/adb5fe},
	adsurl = {https://ui.adsabs.harvard.edu/abs/2025AJ....169..198C}
}

@ARTICLE{2025AJ....169..139X,
	author = {{Xia}, Qiqi and {Wang}, Xiaofeng and {Li}, Kai and {Gao}, Xiang and {Guo}, Fangzhou and {Lin}, Jie and {Liu}, Cheng and {Mo}, Jun and {Peng}, Haowei and {Liu}, Qichun and {Xi}, Gaobo and {Yan}, Shengyu and {Jiang}, Xiaojun and {Zhang}, Jicheng and {Song}, Cui-Ying and {Shi}, Jianrong and {Ma}, Xiaoran and {Xiang}, Danfeng and {Li}, Wenxiong},
	title = "{Minute-cadence Observations of the LAMOST Fields with the TMTS. VI. Physical Parameters of Contact Binaries}",
	journal = {\aj},
	year = 2025,
	month = mar,
	volume = {169},
	number = {3},
	eid = {139},
	pages = {139},
	doi = {10.3847/1538-3881/ada7eb},
	archivePrefix = {arXiv},
	eprint = {2412.11545},
	primaryClass = {astro-ph.SR},
	adsurl = {https://ui.adsabs.harvard.edu/abs/2025AJ....169..139X}
}

@ARTICLE{2025AJ....169...85X,
	author = {{Xu}, Xin and {Li}, Kai and {Liu}, Fei and {Yan}, Qian-Xue and {Wang}, Yi-Fan and {Cui}, Xin-Yu and {Wang}, Jing-Yi and {Gao}, Xing and {Sun}, Guo-You and {Wu}, Cheng-Yu and {Li}, Mu-Zi-Mei},
	title = "{Photometric and Spectroscopic Investigations of Three Large Amplitude Contact Binaries}",
	journal = {\aj},
	year = 2025,
	month = feb,
	volume = {169},
	number = {2},
	eid = {85},
	pages = {85},
	doi = {10.3847/1538-3881/ad9a59},
	archivePrefix = {arXiv},
	eprint = {2412.00331},
	primaryClass = {astro-ph.SR},
	adsurl = {https://ui.adsabs.harvard.edu/abs/2025AJ....169...85X}
}

@ARTICLE{2025ApJ...979...69W,
	author = {{Wang}, Jing-Yi and {Li}, Kai and {Gao}, Xiang and {Guo}, Di-Fu and {Wang}, Li-Heng and {Gao}, Dong-Yang and {Li}, Ling-Zhi and {Guo}, Ya-Ni and {Gao}, Xing and {Sun}, Guo-You},
	title = "{Search for and Analysis of Eclipsing Binaries in the LAMOST Medium-resolution Survey Field. I. R.A.: 23$^{h}$01$^{m}$51$^{s}$, Decl.: +34{\textdegree}36'45''}",
	journal = {\apj},
	year = 2025,
	month = jan,
	volume = {979},
	number = {1},
	eid = {69},
	pages = {69},
	doi = {10.3847/1538-4357/ad9929},
	archivePrefix = {arXiv},
	eprint = {2412.00377},
	primaryClass = {astro-ph.SR},
	adsurl = {https://ui.adsabs.harvard.edu/abs/2025ApJ...979...69W}
}

@ARTICLE{2024ApJ...963...13C,
	author = {{Cao}, Dongtao and {Gu}, Shenghong},
	title = "{Red Asymmetry of H$_{ {\ensuremath{\alpha}} }$ Line Profiles during the Flares on the Active RS CVn-type Star II Pegasi}",
	journal = {\apj},
	year = 2024,
	month = mar,
	volume = {963},
	number = {1},
	eid = {13},
	pages = {13},
	doi = {10.3847/1538-4357/ad1928},
	archivePrefix = {arXiv},
	eprint = {2402.16336},
	primaryClass = {astro-ph.SR},
	adsurl = {https://ui.adsabs.harvard.edu/abs/2024ApJ...963...13C}
}

@ARTICLE{2024MNRAS.527.6406L,
	author = {{Liu}, Fei and {Li}, Kai and {Gao}, Xiang and {Guo}, Ya-Ni and {Li}, Ling-Zhi and {Liu}, Xin-Yi and {Li}, Ke-Xin and {Gao}, Xin-Yi and {Gao}, Xing and {Sun}, Guo-You and {Wang}, Xi and {Yin}, Shi-Peng},
	title = "{The first analysis of three long-period low mass-ratio contact binaries}",
	journal = {\mnras},
	year = 2024,
	month = jan,
	volume = {527},
	number = {3},
	pages = {6406-6418},
	doi = {10.1093/mnras/stad3591},
	adsurl = {https://ui.adsabs.harvard.edu/abs/2024MNRAS.527.6406L}
}

@ARTICLE{2023MNRAS.523.4146C,
	author = {{Cao}, Dongtao and {Gu}, Shenghong and {Wolter}, U. and {Mittag}, M. and {Schmitt}, J.~H.~M.~M. and {Gao}, Dongyang and {Hu}, Shaoming},
	title = "{Prominence detection and chromosphere feature on the prototype RS CVn of active binary systems}",
	journal = {\mnras},
	year = 2023,
	month = aug,
	volume = {523},
	number = {3},
	pages = {4146-4157},
	doi = {10.1093/mnras/stad1700},
	archivePrefix = {arXiv},
	eprint = {2401.02583},
	primaryClass = {astro-ph.SR},
	adsurl = {https://ui.adsabs.harvard.edu/abs/2023MNRAS.523.4146C}
}

@ARTICLE{2022AJ....164..202L,
	author = {{Li}, Kai and {Gao}, Xiang and {Liu}, Xin-Yi and {Gao}, Xing and {Li}, Ling-Zhi and {Chen}, Xu and {Sun}, Guo-You},
	title = "{Extremely Low Mass Ratio Contact Binaries. I. The First Photometric and Spectroscopic Investigations of Ten Systems}",
	journal = {\aj},
	year = 2022,
	month = nov,
	volume = {164},
	number = {5},
	eid = {202},
	pages = {202},
	doi = {10.3847/1538-3881/ac8ff2},
	archivePrefix = {arXiv},
	eprint = {2209.03653},
	primaryClass = {astro-ph.SR},
	adsurl = {https://ui.adsabs.harvard.edu/abs/2022AJ....164..202L}
}

@ARTICLE{2022ApJ...927...12P,
	author = {{Panchal}, A. and {Joshi}, Y.~C. and {De Cat}, Peter and {Tiwari}, S.~N.},
	title = "{Long-term Photometric and Low-resolution Spectroscopic Analysis of Five Contact Binaries}",
	journal = {\apj},
	year = 2022,
	month = mar,
	volume = {927},
	number = {1},
	eid = {12},
	pages = {12},
	doi = {10.3847/1538-4357/ac45fb},
	archivePrefix = {arXiv},
	eprint = {2112.12379},
	primaryClass = {astro-ph.SR},
	adsurl = {https://ui.adsabs.harvard.edu/abs/2022ApJ...927...12P}
}

@ARTICLE{2021AJ....161..221P,
	author = {{Panchal}, A. and {Joshi}, Y.~C.},
	title = "{Photometric and Spectroscopic Analysis of Four Contact Binaries}",
	journal = {\aj},
	year = 2021,
	month = may,
	volume = {161},
	number = {5},
	eid = {221},
	pages = {221},
	doi = {10.3847/1538-3881/abea0c},
	archivePrefix = {arXiv},
	eprint = {2102.13401},
	primaryClass = {astro-ph.SR},
	adsurl = {https://ui.adsabs.harvard.edu/abs/2021AJ....161..221P}
}

@article{alonsofloriano2019,
	author  = {{Alonso-Floriano}, F.~J. and {Snellen}, I.~A.~G. and {Czesla}, S. and {Bauer}, F.~F. and {Salz}, M. and {Lamp{\'o}n}, M. and {Lara}, L.~M. and {Nagel}, E. and {L{\'o}pez-Puertas}, M. and {Nortmann}, L. and {S{\'a}nchez-L{\'o}pez}, A. and {Sanz-Forcada}, J. and {Caballero}, J.~A. and {Reiners}, A. and {Ribas}, I. and {Quirrenbach}, A. and {Amado}, P.~J. and {Aceituno}, J. and {Anglada-Escud{\'e}}, G. and {B{\'e}jar}, V.~J.~S. and {Brinkm{\"o}ller}, M. and {Hatzes}, A.~P. and {Henning}, Th. and {Kaminski}, A. and {K{\"u}rster}, M. and {Labarga}, F. and {Montes}, D. and {Pall{\'e}}, E. and {Schmitt}, J.~H.~M.~M. and {Zapatero Osorio}, M.~R.},
	title = "{He I {\ensuremath{\lambda}} 10 830 {\r{A}} in the transmission spectrum of HD209458 b}",
	journal = {\aap},
	year = 2019,
	month = sep,
	volume = {629},
	eid = {A110},
	pages = {A110},
	doi = {10.1051/0004-6361/201935979},
	archivePrefix = {arXiv},
	eprint = {1907.13425},
	primaryClass = {astro-ph.EP},
	adsurl = {https://ui.adsabs.harvard.edu/abs/2019A&A...629A.110A}
}

@ARTICLE{2018Ap&SS.363..174Z,
	author = {{Zhang}, Li-Yun},
	title = "{Photospheric and chromospheric activity of the short period X-ray and Algol eclipsing binary UX CrB}",
	journal = {\apss},
	year = 2018,
	month = aug,
	volume = {363},
	number = {8},
	eid = {174},
	pages = {174},
	doi = {10.1007/s10509-018-3395-x},
	adsurl = {https://ui.adsabs.harvard.edu/abs/2018Ap&SS.363..174Z}
}

@ARTICLE{2015MNRAS.449.1380C,
	author = {{Cao}, Dongtao and {Gu}, Shenghong},
	title = "{Chromospheric activity and rotational modulation of the RS Canum Venaticorum binary V711 Tauri during 1998-2004}",
	journal = {\mnras},
	year = 2015,
	month = may,
	volume = {449},
	number = {2},
	pages = {1380-1390},
	doi = {10.1093/mnras/stv110},
	adsurl = {https://ui.adsabs.harvard.edu/abs/2015MNRAS.449.1380C}
}

@ARTICLE{2014NewA...32....1Z,
	author = {{Zhang}, Liyun and {Pi}, Qingfeng and {Zhu}, Zhongzhong and {Zhang}, Xiliang and {Li}, Zhongmu},
	title = "{Chromospheric activity on late-type star LQ Hya}",
	journal = {\na},
	year = 2014,
	month = oct,
	volume = {32},
	pages = {1-5},
	doi = {10.1016/j.newast.2014.02.010},
	adsurl = {https://ui.adsabs.harvard.edu/abs/2014NewA...32....1Z}
}

@ARTICLE{2014AJ....147...50P,
	author = {{Pi}, Qing-feng and {Zhang}, Li-Yun and {Li}, Zhong-mu and {Zhang}, Xi-liang},
	title = "{Magnetic Activity and Orbital Period Variation of the Short-period Eclipsing Binary DV Psc}",
	journal = {\aj},
	year = 2014,
	month = mar,
	volume = {147},
	number = {3},
	eid = {50},
	pages = {50},
	doi = {10.1088/0004-6256/147/3/50},
	adsurl = {https://ui.adsabs.harvard.edu/abs/2014AJ....147...50P}
}

@ARTICLE{2014NewA...27...81Z,
	author = {{Zhang}, Liyun and {Pi}, Qingfeng and {Yang}, Yuangui and {Li}, Zhongmu},
	title = "{Magnetic activity and orbital period variation of the eclipsing binary KV Gem}",
	journal = {\na},
	year = 2014,
	month = feb,
	volume = {27},
	pages = {81-94},
	doi = {10.1016/j.newast.2013.09.002},
	adsurl = {https://ui.adsabs.harvard.edu/abs/2014NewA...27...81Z}
}

@ARTICLE{2012A&A...538A.130C,
	author = {{Cao}, Dong-Tao and {Gu}, Sheng-Hong},
	title = "{New observations of chromospheric and prominence activity on the RS CVn-type binary SZ Piscium}",
	journal = {\aap},
	year = 2012,
	month = feb,
	volume = {538},
	eid = {A130},
	pages = {A130},
	doi = {10.1051/0004-6361/201118184},
	adsurl = {https://ui.adsabs.harvard.edu/abs/2012A&A...538A.130C}
}

@article{martinezarnaiz2011,
	author = {{Mart{\'\i}nez-Arn{\'a}iz}, R. and {L{\'o}pez-Santiago}, J. and {Crespo-Chac{\'o}n}, I. and {Montes}, D.},
	title = "{Effect of magnetic activity saturation in chromospheric flux-flux relationships}",
	journal = {\mnras},
	year = 2011,
	month = jul,
	volume = {414},
	number = {3},
	pages = {2629-2641},
	doi = {10.1111/j.1365-2966.2011.18584.x},
	archivePrefix = {arXiv},
	eprint = {1102.4506},
	primaryClass = {astro-ph.SR},
	adsurl = {https://ui.adsabs.harvard.edu/abs/2011MNRAS.414.2629M}
}

@INPROCEEDINGS{2020sea..confE.168M,
	author = {{Montes}, D. and {L{\'o}pez-Gallifa}, A. and {Labarga}, F. and {Caballero}, J.~A. and {Marfil}, E. and {Tabernero}, H.~M. and {Lafarga}, M. and {Jeffers}, S.~V. and {Ribas}, I. and {Reiners}, A. and {Quirrenbach}, A. and {Amado}, P.~J. and {CARMENES Consortium}},
	title = "{Identifying activity-sensitive spectral lines in the CARMENES VIS and NIR spectral range of M dwarfs}",
	booktitle = {XIV.0 Scientific Meeting (virtual) of the Spanish Astronomical Society},
	year = 2020,
	month = jul,
	eid = {168},
	pages = {168},
	url = {https://ui.adsabs.harvard.edu/abs/2020sea..confE.168M}
}

@ARTICLE{2010A&A...520A..79M,
	author = {{Mart{\'\i}nez-Arn{\'a}iz}, R. and {Maldonado}, J. and {Montes}, D. and {Eiroa}, C. and {Montesinos}, B.},
	title = "{Chromospheric activity and rotation of FGK stars in the solar vicinity. An estimation of the radial velocity jitter}",
	journal = {\aap},
	year = 2010,
	month = sep,
	volume = {520},
	eid = {A79},
	pages = {A79},
	doi = {10.1051/0004-6361/200913725},
	archivePrefix = {arXiv},
	eprint = {1002.4391},
	primaryClass = {astro-ph.SR},
	adsurl = {https://ui.adsabs.harvard.edu/abs/2010A&A...520A..79M}
}

@ARTICLE{2009AJ....137.3965G,
	author = {{G{\'a}lvez}, M.~C. and {Montes}, D. and {Fern{\'a}ndez-Figueroa}, M.~J. and {De Castro}, E. and {Cornide}, M.},
	title = "{Multiwavelength Optical Observations of Two Chromospherically Active Binary Systems: V789 Mon and GZ Leo}",
	journal = {\aj},
	year = 2009,
	month = apr,
	volume = {137},
	number = {4},
	pages = {3965-3975},
	doi = {10.1088/0004-6256/137/4/3965},
	archivePrefix = {arXiv},
	eprint = {0901.1634},
	primaryClass = {astro-ph.SR},
	adsurl = {https://ui.adsabs.harvard.edu/abs/2009AJ....137.3965G}
}

@ARTICLE{2007A&A...472..587G,
	author = {{G{\'a}lvez}, M.~C. and {Montes}, D. and {Fern{\'a}ndez-Figueroa}, M.~J. and {de Castro}, E. and {Cornide}, M.},
	title = "{Multiwavelength optical observations of chromospherically active binary systems. V. FF UMa (2RE J0933+624): a system with orbital period variation}",
	journal = {\aap},
	year = 2007,
	month = sep,
	volume = {472},
	number = {2},
	pages = {587-598},
	doi = {10.1051/0004-6361:20067015},
	archivePrefix = {arXiv},
	eprint = {0706.3665},
	primaryClass = {astro-ph},
	adsurl = {https://ui.adsabs.harvard.edu/abs/2007A&A...472..587G}
}

@ARTICLE{2006Ap&SS.304...59G,
	author = {{G{\'a}lvez}, M.~C. and {Montes}, D. and {Fern{\'a}ndez-Figueroa}, M.~J. and {L{\'o}pez-Santiago}, J.},
	title = "{Chromospheric Activity and Orbital Solution of Six New Late-type Spectroscopic Binary Systems}",
	journal = {\apss},
	year = 2006,
	month = aug,
	volume = {304},
	number = {1-4},
	pages = {59-61},
	doi = {10.1007/s10509-006-9074-3},
	adsurl = {https://ui.adsabs.harvard.edu/abs/2006Ap&SS.304...59G}
}

@INCOLLECTION{2004LNEA....1..119M,
	author = {{Montes}, D. and {Crespo-Chac{\'o}n}, I. and {G{\'a}lvez}, M.~C. and {Fern{\'a}ndez-Figueroa}, M.~J. and {L{\'o}pez-Santiago}, J. and {de Castro}, E. and {Cornide}, M. and {Hern{\'a}n-Obispo}, M.},
	title = "{Cool Stars: Chromospheric Activity, Rotation, Kinematic and Age}",
	booktitle = {Lecture Notes and Essays in Astrophysics},
	year = 2004,
	editor = {{Ulla}, Ana and {Manteiga}, Minia},
	publisher = {Real Sociedad Española de Física},
	volume = {1},
	pages = {119-132},
	adsurl = {https://ui.adsabs.harvard.edu/abs/2004LNEA....1..119M}
}

@ARTICLE{2003A&A...411..489L,
	author = {{L{\'o}pez-Santiago}, J. and {Montes}, D. and {Fern{\'a}ndez-Figueroa}, M.~J. and {Ramsey}, L.~W.},
	title = "{Rotational modulation of the photospheric and chromospheric activity in the young, single K2-dwarf PW And}",
	journal = {\aap},
	year = 2003,
	month = dec,
	volume = {411},
	pages = {489-502},
	doi = {10.1051/0004-6361:20031377},
	archivePrefix = {arXiv},
	eprint = {astro-ph/0309072},
	primaryClass = {astro-ph},
	adsurl = {https://ui.adsabs.harvard.edu/abs/2003A&A...411..489L}
}

@ARTICLE{2002A&A...389..524G,
	author = {{G{\'a}lvez}, M.~C. and {Montes}, D. and {Fern{\'a}ndez-Figueroa}, M.~J. and {L{\'o}pez-Santiago}, J. and {De Castro}, E. and {Cornide}, M.},
	title = "{Multiwavelength optical observations of chromospherically active binary systems. IV. The X-ray/EUV selected binary BK Psc (2RE J0039+103)}",
	journal = {\aap},
	year = 2002,
	month = jul,
	volume = {389},
	pages = {524-536},
	doi = {10.1051/0004-6361:20020644},
	archivePrefix = {arXiv},
	eprint = {astro-ph/0204490},
	primaryClass = {astro-ph},
	adsurl = {https://ui.adsabs.harvard.edu/abs/2002A&A...389..524G}
}

@article{montesonpwandLi,
	author = {{Montes}, D. and {L{\'o}pez-Santiago}, J. and {Fern{\'a}ndez-Figueroa}, M.~J. and {G{\'a}lvez}, M.~C.},
	title = "{Chromospheric activity, lithium and radial velocities of single late-type stars possible members of young moving groups}",
	journal = {\aap},
	year = 2001,
	month = dec,
	volume = {379},
	pages = {976-991},
	doi = {10.1051/0004-6361:20011385},
	archivePrefix = {arXiv},
	eprint = {astro-ph/0110066},
	primaryClass = {astro-ph},
	adsurl = {https://ui.adsabs.harvard.edu/abs/2001A&A...379..976M}
}

@ARTICLE{2001AJ....122.1954O,
	author = {{O'Neal}, Douglas and {Neff}, James E. and {Saar}, Steven H. and {Mines}, Jonathan K.},
	title = "{Hydroxyl 1.563 Micron Absorption from Starspots on Active Stars}",
	journal = {\aj},
	year = 2001,
	month = oct,
	volume = {122},
	number = {4},
	pages = {1954-1964},
	doi = {10.1086/323093},
	adsurl = {https://ui.adsabs.harvard.edu/abs/2001AJ....122.1954O}
}

@ARTICLE{1998ApJ...501L..73O,
	author = {{O'Neal}, Douglas and {Saar}, Steven H. and {Neff}, James E.},
	title = "{Spectroscopic Evidence for Nonuniform Starspot Properties on II Pegasi}",
	journal = {\apjl},
	year = 1998,
	month = jul,
	volume = {501},
	number = {1},
	pages = {L73-L76},
	doi = {10.1086/311441},
	adsurl = {https://ui.adsabs.harvard.edu/abs/1998ApJ...501L..73O}
}

@ARTICLE{1997AJ....113.1129O,
	author = {{O'Neal}, Douglas and {Neff}, James E.},
	title = "{OH 1.563 micron Absorption from Starspots on Active Stars}",
	journal = {\aj},
	year = 1997,
	month = mar,
	volume = {113},
	pages = {1129-1137},
	doi = {10.1086/118331},
	adsurl = {https://ui.adsabs.harvard.edu/abs/1997AJ....113.1129O}
}

@ARTICLE{1994AJ....107.1149H,
	author = {{Hall}, Jeffrey C. and {Ramsey}, Lawrence W.},
	title = "{Eclipse Observations of RS CVN Binaries II. A Parametric Model of Extended Matter}",
	journal = {\aj},
	year = 1994,
	month = mar,
	volume = {107},
	pages = {1149},
	doi = {10.1086/116927},
	adsurl = {https://ui.adsabs.harvard.edu/abs/1994AJ....107.1149H}
}

@ARTICLE{1992AJ....104.1942H,
	author = {{Hall}, Jeffrey C. and {Ramsey}, Lawrence W.},
	title = "{Eclipse Observations of RS CVn Binaries. I. A Survey for Extended Matter}",
	journal = {\aj},
	year = 1992,
	month = nov,
	volume = {104},
	pages = {1942},
	doi = {10.1086/116370},
	adsurl = {https://ui.adsabs.harvard.edu/abs/1992AJ....104.1942H}
}

@misc{Labarga2025_iSTARMOD,
	author       = {Labarga, F. and Montes, D.},
	title        = {iSTARMOD: a Python Code to Quantify Chromospheric Activity by Using Spectral Subtraction Technique},
	year         = {2025},
	publisher    = {iSTARMOD: a Python Code to Quantify Chromospheric Activity by Using Spectral Subtraction Technique (10.0). Zenodo},
	doi          = {10.5281/zenodo.17329154},
	url          = {https://doi.org/10.5281/zenodo.17329154}
}

@article{allard2012,
       author = {{Allard}, F. and {Homeier}, D. and {Freytag}, B.},
        title = "{Models of very-low-mass stars, brown dwarfs and exoplanets}",
      journal = {Philosophical Transactions of the Royal Society of London Series A},
         year = 2012,
        month = jun,
       volume = {370},
       number = {1968},
        pages = {2765-2777},
          doi = {10.1098/rsta.2011.0269},
archivePrefix = {arXiv},
       eprint = {1112.3591},
 primaryClass = {astro-ph.SR},
       adsurl = {https://ui.adsabs.harvard.edu/abs/2012RSPTA.370.2765A}
}

@article{astropy2018,
	author = {{Astropy Collaboration} and {Price-Whelan}, A.~M. and {Sip{\H{o}}cz}, B.~M. and {G{\"u}nther}, H.~M. and {Lim}, P.~L. and {Crawford}, S.~M. and {Conseil}, S. and {Shupe}, D.~L. and {Craig}, M.~W. and {Dencheva}, N. and {Ginsburg}, A. and {VanderPlas}, J.~T. and {Bradley}, L.~D. and {P{\'e}rez-Su{\'a}rez}, D. and {de Val-Borro}, M. and {Aldcroft}, T.~L. and {Cruz}, K.~L. and {Robitaille}, T.~P. and {Tollerud}, E.~J. and {Ardelean}, C. and {Babej}, T. and {Bach}, Y.~P. and {Bachetti}, M. and {Bakanov}, A.~V. and {Bamford}, S.~P. and {Barentsen}, G. and {Barmby}, P. and {Baumbach}, A. and {Berry}, K.~L. and {Biscani}, F. and {Boquien}, M. and {Bostroem}, K.~A. and {Bouma}, L.~G. and {Brammer}, G.~B. and {Bray}, E.~M. and {Breytenbach}, H. and {Buddelmeijer}, H. and {Burke}, D.~J. and {Calderone}, G. and {Cano Rodr{\'\i}guez}, J.~L. and {Cara}, M. and {Cardoso}, J.~V.~M. and {Cheedella}, S. and {Copin}, Y. and {Corrales}, L. and {Crichton}, D. and {D'Avella}, D. and {Deil}, C. and {Depagne}, {\'E}. and {Dietrich}, J.~P. and {Donath}, A. and {Droettboom}, M. and {Earl}, N. and {Erben}, T. and {Fabbro}, S. and {Ferreira}, L.~A. and {Finethy}, T. and {Fox}, R.~T. and {Garrison}, L.~H. and {Gibbons}, S.~L.~J. and {Goldstein}, D.~A. and {Gommers}, R. and {Greco}, J.~P. and {Greenfield}, P. and {Groener}, A.~M. and {Grollier}, F. and {Hagen}, A. and {Hirst}, P. and {Homeier}, D. and {Horton}, A.~J. and {Hosseinzadeh}, G. and {Hu}, L. and {Hunkeler}, J.~S. and {Ivezi{\'c}}, {\v{Z}}. and {Jain}, A. and {Jenness}, T. and {Kanarek}, G. and {Kendrew}, S. and {Kern}, N.~S. and {Kerzendorf}, W.~E. and {Khvalko}, A. and {King}, J. and {Kirkby}, D. and {Kulkarni}, A.~M. and {Kumar}, A. and {Lee}, A. and {Lenz}, D. and {Littlefair}, S.~P. and {Ma}, Z. and {Macleod}, D.~M. and {Mastropietro}, M. and {McCully}, C. and {Montagnac}, S. and {Morris}, B.~M. and {Mueller}, M. and {Mumford}, S.~J. and {Muna}, D. and {Murphy}, N.~A. and {Nelson}, S. and {Nguyen}, G.~H. and {Ninan}, J.~P. and {N{\"o}the}, M. and {Ogaz}, S. and {Oh}, S. and {Parejko}, J.~K. and {Parley}, N. and {Pascual}, S. and {Patil}, R. and {Patil}, A.~A. and {Plunkett}, A.~L. and {Prochaska}, J.~X. and {Rastogi}, T. and {Reddy Janga}, V. and {Sabater}, J. and {Sakurikar}, P. and {Seifert}, M. and {Sherbert}, L.~E. and {Sherwood-Taylor}, H. and {Shih}, A.~Y. and {Sick}, J. and {Silbiger}, M.~T. and {Singanamalla}, S. and {Singer}, L.~P. and {Sladen}, P.~H. and {Sooley}, K.~A. and {Sornarajah}, S. and {Streicher}, O. and {Teuben}, P. and {Thomas}, S.~W. and {Tremblay}, G.~R. and {Turner}, J.~E.~H. and {Terr{\'o}n}, V. and {van Kerkwijk}, M.~H. and {de la Vega}, A. and {Watkins}, L.~L. and {Weaver}, B.~A. and {Whitmore}, J.~B. and {Woillez}, J. and {Zabalza}, V. and {Astropy Contributors}},
    title = "{The Astropy Project: Building an Open-science Project and Status of the v2.0 Core Package}",
    journal = {\aj},
    year = 2018,
    month = sep,
    volume = {156},
    number = {3},
    eid = {123},
    pages = {123},
    doi = {10.3847/1538-3881/aabc4f},
	archivePrefix = {arXiv},
    eprint = {1801.02634},
 	primaryClass = {astro-ph.IM},
    adsurl = {https://ui.adsabs.harvard.edu/abs/2018AJ....156..123A}
}

@ARTICLE{baharonPWAnd,
	author = {{Bahar}, Engin and {{\c{S}}enavc{\i}}, Hakan V. and {I{\c{s}}{\i}k}, Emre and {Hussain}, Gaitee A.~J. and {Kochukhov}, Oleg and {Montes}, David and {Xiang}, Yue},
	title = "{First Chromospheric Activity and Doppler Imaging Study of PW And Using a New Doppler Imaging Code: SpotDIPy}",
	journal = {\apj},
	year = 2024,
	month = jan,
	volume = {960},
	number = {1},
	eid = {60},
	pages = {60},
	doi = {10.3847/1538-4357/ad055d},
	archivePrefix = {arXiv},
	eprint = {2310.14865},
	primaryClass = {astro-ph.SR},
	adsurl = {https://ui.adsabs.harvard.edu/abs/2024ApJ...960...60B}
}

@article{baraffe2015,
       	author = {{Baraffe}, Isabelle and {Homeier}, Derek and {Allard}, France and {Chabrier}, Gilles},
       	title = "{New evolutionary models for pre-main sequence and main sequence low-mass stars down to the hydrogen-burning limit}",
       	journal = {\aap},
    	year = 2015,
        month = may,
       	volume = {577},
        eid = {A42},
        pages = {A42},
        doi = {10.1051/0004-6361/201425481},
		archivePrefix = {arXiv},
       	eprint = {1503.04107},
 		primaryClass = {astro-ph.SR},
       	adsurl = {https://ui.adsabs.harvard.edu/abs/2015A&A...577A..42B}
}

@article{barden1,
	author = {{Barden}, S.~C.},
    title = "{A study of short-period RS Canum Venaticorum and W Ursae Majoris binary systems : the global nature of H alpha.}",
    journal = {\apj},
    year = 1985,
    month = aug,
    volume = {295},
    pages = {162-170},
    doi = {10.1086/163361},
    adsurl = {https://ui.adsabs.harvard.edu/abs/1985ApJ...295..162B}
}

@article{caballerocarmenesinst,
	author = {{Caballero}, Jos{\'e} A. and {Seifert}, Walter and {Quirrenbach}, Andreas and {Amado}, Pedro J. and {Ribas}, Ignasi and {Reiners}, Ansgar},
	title = "{CARMENES as an Instrument for Exoplanet Research}",
	journal = {arXiv e-prints},
	year = 2025,
	month = mar,
	eid = {arXiv:2503.05501},
	pages = {arXiv:2503.05501},
	doi = {10.48550/arXiv.2503.05501},
	archivePrefix = {arXiv},
	eprint = {2503.05501},
	primaryClass = {astro-ph.EP},
	adsurl = {https://ui.adsabs.harvard.edu/abs/2025arXiv250305501C}
}

@PROCEEDINGS{1988IAUS..132.....C,
	title = "{The impact of very high S/N spectroscopy on stellar physics: proceedings of the 132nd Symposium of the International Astronomical Union held in Paris, France, June 29-July 3, 1987.}",
	booktitle = {The Impact of Very High S/N Spectroscopy on Stellar Physics},
    year = 1988,
	editor = {{Cayrel de Strobel}, G. and {Spite}, Monique},
	series = {IAU Symposium},
	volume = {132},
 	month = jan,
	url = {https://ui.adsabs.harvard.edu/abs/1988IAUS..132.....C}
}

@ARTICLE{cayrel2004,
       author = {{Cayrel}, R. and {Depagne}, E. and {Spite}, M. and {Hill}, V. and {Spite}, F. and {Fran{\c{c}}ois}, P. and {Plez}, B. and {Beers}, T. and {Primas}, F. and {Andersen}, J. and {Barbuy}, B. and {Bonifacio}, P. and {Molaro}, P. and {Nordstr{\"o}m}, B.},
        title = "{First stars V - Abundance patterns from C to Zn and supernova yields in the early Galaxy}",
      journal = {\aap},
         year = 2004,
        month = mar,
       volume = {416},
        pages = {1117-1138},
          doi = {10.1051/0004-6361:20034074},
archivePrefix = {arXiv},
       eprint = {astro-ph/0311082},
 primaryClass = {astro-ph},
       adsurl = {https://ui.adsabs.harvard.edu/abs/2004A&A...416.1117C}
}

@ARTICLE{douglasAltChi1,
       author = {{Douglas}, S.~T. and {Ag{\"u}eros}, M.~A. and {Covey}, K.~R. and {Bowsher}, E.~C. and {Bochanski}, J.~J. and {Cargile}, P.~A. and {Kraus}, A. and {Law}, N.~M. and {Lemonias}, J.~J. and {Arce}, H.~G. and {Fierroz}, D.~F. and {Kundert}, A.},
        title = "{The Factory and the Beehive. II. Activity and Rotation in Praesepe and the Hyades}",
      journal = {\apj},
         year = 2014,
        month = nov,
       volume = {795},
       number = {2},
          eid = {161},
        pages = {161},
          doi = {10.1088/0004-637X/795/2/161},
archivePrefix = {arXiv},
       eprint = {1409.7603},
 primaryClass = {astro-ph.SR},
       adsurl = {https://ui.adsabs.harvard.edu/abs/2014ApJ...795..161D}
}

@article{FrascaCatalanoHalphaforbinaries,
	author = {{Frasca}, A. and {Catalano}, S.},
	title = "{H{\ensuremath{\alpha}} survey of late-type active binaries.}",
	journal = {\aap},
	year = 1994,
	month = apr,
	volume = {284},
	pages = {883-899},
	url = {https://ui.adsabs.harvard.edu/abs/1994A&A...284..883F}
}

@article{frascaetalonrotfit2003,
	author = {{Frasca}, A. and {Alcal{\'a}}, J.~M. and {Covino}, E. and {Catalano}, S. and {Marilli}, E. and {Paladino}, R.},
	title = "{Further identification of ROSAT all-sky survey sources in Orion}",
	journal = {\aap},
	year = 2003,
	month = jul,
	volume = {405},
	pages = {149-163},
	doi = {10.1051/0004-6361:20030644},
	adsurl = {https://ui.adsabs.harvard.edu/abs/2003A&A...405..149F}
}

@article{frascaetalonrotfit2006,
	author = {{Frasca}, A. and {Guillout}, P. and {Marilli}, E. and {Freire Ferrero}, R. and {Biazzo}, K. and {Klutsch}, A.},
	title = "{Newly discovered active binaries in the RasTyc sample of stellar X-ray sources. I. Orbital and physical parameters of six new binaries}",
	journal = {\aap},
	year = 2006,
	month = jul,
	volume = {454},
	number = {1},
	pages = {301-309},
	doi = {10.1051/0004-6361:20054573},
	adsurl = {https://ui.adsabs.harvard.edu/abs/2006A&A...454..301F}
}

@article{frascaetal2018,
	author = {{Frasca}, A. and {Guillout}, P. and {Klutsch}, A. and {Ferrero}, R. Freire and {Marilli}, E. and {Biazzo}, K. and {Gandolfi}, D. and {Montes}, D.},
	title = "{A spectroscopic survey of the youngest field stars in the solar neighborhood . II. The optically faint sample}",
	journal = {\aap},
	year = 2018,
	month = may,
	volume = {612},
	eid = {A96},
	pages = {A96},
	doi = {10.1051/0004-6361/201732028},
	archivePrefix = {arXiv},
	eprint = {1801.00671},
	primaryClass = {astro-ph.SR},
	adsurl = {https://ui.adsabs.harvard.edu/abs/2018A&A...612A..96F}
}

@article{fuhrmeisterHeI10830,
	author = {{Fuhrmeister}, B. and {Czesla}, S. and {Hildebrandt}, L. and {Nagel}, E. and {Schmitt}, J.~H.~M.~M. and {Jeffers}, S.~V. and {Caballero}, J.~A. and {Hintz}, D. and {Johnson}, E.~N. and {Sch{\"o}fer}, P. and {Zechmeister}, M. and {Reiners}, A. and {Ribas}, I. and {Amado}, P.~J. and {Quirrenbach}, A. and {Nortmann}, L. and {Bauer}, F.~F. and {B{\'e}jar}, V.~J.~S. and {Cort{\'e}s-Contreras}, M. and {Dreizler}, S. and {Galad{\'\i}-Enr{\'\i}quez}, D. and {Hatzes}, A.~P. and {Kaminski}, A. and {K{\"u}rster}, M. and {Lafarga}, M. and {Montes}, D.},
	title = "{The CARMENES search for exoplanets around M dwarfs. Variability of the He I line at 10 830 {\r{A}}}",
    journal = {\aap},
    year = 2020,
    month = aug,
   	volume = {640},
	eid = {A52},
   	pages = {A52},
   	doi = {10.1051/0004-6361/202038279},
	archivePrefix = {arXiv},
   	eprint = {2006.09372},
 	primaryClass = {astro-ph.SR},
   	adsurl = {https://ui.adsabs.harvard.edu/abs/2020A&A...640A..52F}
}

@ARTICLE{otherchi,
       author = {{Garc{\'\i}a Soto}, Aylin and {Duvvuri}, Girish M. and {Newton}, Elisabeth R. and {Howard}, Ward S. and {N{\'u}{\~n}ez}, Alejandro and {Douglas}, Stephanie T.},
        title = "{Short-term Balmer Line Emission Variability in M Dwarfs}",
      journal = {\apj},
         year = 2025,
        month = apr,
       volume = {982},
       number = {2},
          eid = {98},
        pages = {98},
          doi = {10.3847/1538-4357/adb615},
archivePrefix = {arXiv},
       eprint = {2502.02568},
 primaryClass = {astro-ph.SR},
       adsurl = {https://ui.adsabs.harvard.edu/abs/2025ApJ...982...98G}
}

@BOOK{graybible,
    author = {{Gray}, David F.},
    title = "{The observation and analysis of stellar photospheres}",
    year = 2022,
    publisher = {Cambridge University Press},
    doi = {10.1017/9781009082136},
    adsurl = {https://ui.adsabs.harvard.edu/abs/2022oasp.book.....G}
}

@article{guilloutonfurtherrotfit,
	author = {{Guillout}, P. and {Klutsch}, A. and {Frasca}, A. and {Freire Ferrero}, R. and {Marilli}, E. and {Mignemi}, G. and {Biazzo}, K. and {Bouvier}, J. and {Monier}, R. and {Motch}, C. and {Sterzik}, M.},
	title = "{A spectroscopic survey of the youngest field stars in the solar neighbourhood. I. The optically bright sample}",
	journal = {\aap},
	year = 2009,
	month = sep,
	volume = {504},
	number = {3},
	pages = {829-843},
	doi = {10.1051/0004-6361/200811313},
	archivePrefix = {arXiv},
	eprint = {0907.1157},
	primaryClass = {astro-ph.SR},
	adsurl = {https://ui.adsabs.harvard.edu/abs/2009A&A...504..829G}
}

@article{numpy2020,
	author = {{Harris}, Charles R. and {Millman}, K. Jarrod and {van der Walt}, St{\'e}fan J. and {Gommers}, Ralf and {Virtanen}, Pauli and {Cournapeau}, David and {Wieser}, Eric and {Taylor}, Julian and {Berg}, Sebastian and {Smith}, Nathaniel J. and {Kern}, Robert and {Picus}, Matti and {Hoyer}, Stephan and {van Kerkwijk}, Marten H. and {Brett}, Matthew and {Haldane}, Allan and {del R{\'\i}o}, Jaime Fern{\'a}ndez and {Wiebe}, Mark and {Peterson}, Pearu and {G{\'e}rard-Marchant}, Pierre and {Sheppard}, Kevin and {Reddy}, Tyler and {Weckesser}, Warren and {Abbasi}, Hameer and {Gohlke}, Christoph and {Oliphant}, Travis E.},
	title = "{Array programming with NumPy}",
	journal = {\nat},
	year = 2020,
	month = sep,
	volume = {585},
	number = {7825},
	pages = {357-362},
	doi = {10.1038/s41586-020-2649-2},
	archivePrefix = {arXiv},
	eprint = {2006.10256},
	primaryClass = {cs.MS},
	adsurl = {https://ui.adsabs.harvard.edu/abs/2020Natur.585..357H}
}

@ARTICLE{herbig1985onHalphaemission,
	author = {{Herbig}, G.~H.},
	title = "{Chromospheric H alpha emission in F8-G3 dwarfs and its connection with the T Tauri stars.}",
	journal = {\apj},
	year = 1985,
	month = feb,
	volume = {289},
	pages = {269-278},
	doi = {10.1086/162887},
	adsurl = {https://ui.adsabs.harvard.edu/abs/1985ApJ...289..269H}
}

@article{hintzCaii,
  author = {Hintz, D. and Fuhrmeister, B. and Czesla, S. et al.},
  title = {, },
  year = {2019},
  journal = {A\&A 623, A136},
  doi = {10.1051/0004-6361/201834788}
}

@ARTICLE{huenemoerderonTiO,
	author = {{Huenemoerder}, David P. and {Ramsey}, Lawrence W. and {Buzasi}, Derek L.},
	title = "{Titanium Oxide Variations in II Pegasi}",
	journal = {\aj},
	year = 1989,
	month = dec,
	volume = {98},
	pages = {2264},
	doi = {10.1086/115295},
	adsurl = {https://ui.adsabs.harvard.edu/abs/1989AJ.....98.2264H}
}

@ARTICLE{husser2013,
       author = {{Husser}, T. -O. and {Wende-von Berg}, S. and {Dreizler}, S. and {Homeier}, D. and {Reiners}, A. and {Barman}, T. and {Hauschildt}, P.~H.},
        title = "{A new extensive library of PHOENIX stellar atmospheres and synthetic spectra}",
      journal = {\aap},
         year = 2013,
        month = may,
       volume = {553},
          eid = {A6},
        pages = {A6},
          doi = {10.1051/0004-6361/201219058},
archivePrefix = {arXiv},
       eprint = {1303.5632},
 primaryClass = {astro-ph.SR},
       adsurl = {https://ui.adsabs.harvard.edu/abs/2013A&A...553A...6H}
}

@article{kumaretfares2023,
	author = {{Kumar}, M. and {Fares}, R.},
	title = "{A study of the magnetic activity and variability of GJ 436}",
	journal = {\mnras},
	year = 2023,
	month = jan,
	volume = {518},
	number = {2},
	pages = {3147-3163},
	doi = {10.1093/mnras/stac2766},
	archivePrefix = {arXiv},
	eprint = {2209.11258},
	primaryClass = {astro-ph.SR},
	adsurl = {https://ui.adsabs.harvard.edu/abs/2023MNRAS.518.3147K}
}

@phdthesis{Labarga2025,
	author       = {Labarga, F.},
	title        = {PhD Thesis},
	school       = {Universidad Complutense de Madrid (UCM)},
	year         = {2025},
	note         = {In preparation}
}

@INPROCEEDINGS{2023hsa..conf..272L,
	author = {{Labarga}, F. and {Montes}, D. and {Duque-Arribas}, C. and {Lopez-Gallifa}, A. and {Caballero}, J.~A. and {Jeffers}, S.~V. and {Reiners}, A. and {Ribas}, I. and {Quirrenbach}, A. and {Amado}, P.~J.},
	title = "{Analysis of chromospheric flux-flux relationships of M Dwarfs using visible and near-infrared CARMENES spectra}",
	booktitle = {Highlights on Spanish Astrophysics XI},
	year = 2023,
	editor = {{Manteiga}, M. and {Bellot}, L. and {Benavidez}, P. and {de Lorenzo-C{\'a}ceres}, A. and {Fuente}, M.~A. and {Mart{\'\i}nez}, M.~J. and {V{\'a}zquez Acosta}, M. and {Dafonte}, C.},
	month = may,
	pages = {272},
	doi = {10.5281/zenodo.7668217},
	url = {https://ui.adsabs.harvard.edu/abs/2023hsa..conf..272L}
}

@INPROCEEDINGS{istarmodatsea,
	author = {{Labarga}, F. and {Montes}, D.},
	title = "{iSTARMOD: a Python code to quantify the chromospheric activity of FGKM stars using the Spectral istarmodatseaSubtraction Technique}",
	booktitle = {XIV.0 Scientific Meeting (virtual) of the Spanish Astronomical Society},
	year = 2020,
	month = jul,
	eid = {153},
	pages = {153},
	url = {https://ui.adsabs.harvard.edu/abs/2020sea..confE.153L}
}

@article{linskyetayres,
       author = {{Linsky}, J.~L. and {Ayres}, T.~R.},
        title = "{Stellar model chromospheres. VI. Empirical estimates of the chromospheric radiative losses of late-type stars.}",
      journal = {\apj},
         year = 1978,
        month = mar,
       volume = {220},
        pages = {619-628},
          doi = {10.1086/155945},
       adsurl = {https://ui.adsabs.harvard.edu/abs/1978ApJ...220..619L}
}

@article{linskyetal,
	author = {{Linsky}, J.~L. and {Worden}, S.~P. and {McClintock}, W. and {Robertson}, R.~M.},
	title = "{Stellar model chromospheres. X. High-resolution, absolute flux profiles of the Ca II H and K lines in stars of spectral types F0 - M2.}",
	journal = {\apjs},
	year = 1979,
	month = sep,
	volume = {41},
	pages = {47-74},
	doi = {10.1086/190607},
	adsurl = {https://ui.adsabs.harvard.edu/abs/1979ApJS...41...47L}
}

@article{lopezsantiago2010,
	author = {{L{\'o}pez-Santiago}, J. and {Montes}, D. and {G{\'a}lvez-Ortiz}, M.~C. and {Crespo-Chac{\'o}n}, I. and {Mart{\'\i}nez-Arn{\'a}iz}, R.~M. and {Fern{\'a}ndez-Figueroa}, M.~J. and {de Castro}, E. and {Cornide}, M.},
    title = "{A high-resolution spectroscopic survey of late-type stars: chromospheric activity, rotation, kinematics, and age}",
    journal = {\aap},
    year = 2010,
    month = may,
    volume = {514},
    eid = {A97},
    pages = {A97},
    doi = {10.1051/0004-6361/200913437},
	archivePrefix = {arXiv},
    eprint = {1002.1663},
 	primaryClass = {astro-ph.SR},
adsurl = {https://ui.adsabs.harvard.edu/abs/2010A&A...514A..97L}
}

@article{lucyongravitydarkening1967,
	author = {{Lucy}, L.~B.},
	title = "{Gravity-Darkening for Stars with Convective Envelopes}",
	journal = {\zap},
	year = 1967,
	month = jan,
	volume = {65},
	pages = {89},
	url = {https://ui.adsabs.harvard.edu/abs/1967ZA.....65...89L}
}

@article{marfil2021,
	author = {{Marfil}, E. and {Tabernero}, H.~M. and {Montes}, D. and {Caballero}, J.~A. and {L{\'a}zaro}, F.~J. and {Gonz{\'a}lez Hern{\'a}ndez}, J.~I. and {Nagel}, E. and {Passegger}, V.~M. and {Schweitzer}, A. and {Ribas}, I. and {Reiners}, A. and {Quirrenbach}, A. and {Amado}, P.~J. and {Cifuentes}, C. and {Cort{\'e}s-Contreras}, M. and {Dreizler}, S. and {Duque-Arribas}, C. and {Galad{\'\i}-Enr{\'\i}quez}, D. and {Henning}, Th. and {Jeffers}, S.~V. and {Kaminski}, A. and {K{\"u}rster}, M. and {Lafarga}, M. and {L{\'o}pez-Gallifa}, {\'A}. and {Morales}, J.~C. and {Shan}, Y. and {Zechmeister}, M.},
    title = "{The CARMENES search for exoplanets around M dwarfs. Stellar atmospheric parameters of target stars with SteParSyn}",
    journal = {\aap},
    year = 2021,
    month = dec,
    volume = {656},
    eid = {A162},
    pages = {A162},
    doi = {10.1051/0004-6361/202141980},
	archivePrefix = {arXiv},
    eprint = {2110.07329},
 	primaryClass = {astro-ph.SR},
 adsurl = {https://ui.adsabs.harvard.edu/abs/2021A&A...656A.162M}
}

@article{marvinonTeff,
	author = {{Marvin}, C.~J. and {Reiners}, A. and {Anglada-Escud{\'e}}, G. and {Jeffers}, S.~V. and {Boro Saikia}, S.},
	title = "{Absolute Ca II H \& K and H-alpha flux measurements of low-mass stars: Extending R'$_{HK}$ to M dwarfs}",
	journal = {\aap},
	year = 2023,
	month = mar,
	volume = {671},
	eid = {A162},
	pages = {A162},
	doi = {10.1051/0004-6361/201937306},
	archivePrefix = {arXiv},
	eprint = {2302.00056},
	primaryClass = {astro-ph.SR},
	adsurl = {https://ui.adsabs.harvard.edu/abs/2023A&A...671A.162M}
}

@ARTICLE{montes1995a,
    author = {{Montes}, D. and {Fernandez-Figueroa}, M.~J. and {de Castro}, E. and {Cornide}, M.},
    title = "{Excess H{\ensuremath{\alpha}} emission in chromospherically active binaries.}",
    journal = {\aap},
    year = 1995,
    month = feb,
    volume = {294},
    pages = {165-176},
    url = {https://ui.adsabs.harvard.edu/abs/1995A&A...294..165M}
}

@ARTICLE{montes1995b,
       author = {{Montes}, D. and {de Castro}, E. and {Fernandez-Figueroa}, M.~J. and {Cornide}, M.},
        title = "{Application of the spectral subtraction technique to the CA II H \& K and Hɛ lines in a sample of chromospherically active binaries.}",
      journal = {\aaps},
         year = 1995,
        month = dec,
       volume = {114},
        pages = {287},
       url = {https://ui.adsabs.harvard.edu/abs/1995A&AS..114..287M}
}

@ARTICLE{montes2000,
       author = {{Montes}, D. and {Fern{\'a}ndez-Figueroa}, M.~J. and {De Castro}, E. and {Cornide}, M. and {Latorre}, A. and {Sanz-Forcada}, J.},
        title = "{Multiwavelength optical observations of chromospherically active binary systems. III. High resolution echelle spectra from Ca II H \& K to Ca II IRT}",
      journal = {\aaps},
         year = 2000,
        month = oct,
       volume = {146},
        pages = {103-140},
          doi = {10.1051/aas:2000359},
archivePrefix = {arXiv},
       eprint = {astro-ph/0007041},
 primaryClass = {astro-ph},
       adsurl = {https://ui.adsabs.harvard.edu/abs/2000A&AS..146..103M}
}

@article{douglasAltChi2,
       author = {{N{\'u}{\~n}ez}, Alejandro and {Ag{\"u}eros}, Marcel A. and {Curtis}, Jason L. and {Covey}, Kevin R. and {Douglas}, Stephanie T. and {Chu}, Sabine R. and {DeLaurentiis}, Stanislav and {Wang}, Minzhi (Luna) and {Drake}, Jeremy J.},
        title = "{The Factory and the Beehive. V. Chromospheric and Coronal Activity and Its Dependence on Rotation in Praesepe and the Hyades}",
      journal = {\apj},
         year = 2024,
        month = feb,
       volume = {962},
       number = {1},
          eid = {12},
        pages = {12},
          doi = {10.3847/1538-4357/ad117e},
archivePrefix = {arXiv},
       eprint = {2311.18690},
 primaryClass = {astro-ph.SR},
       adsurl = {https://ui.adsabs.harvard.edu/abs/2024ApJ...962...12N}
}

@ARTICLE{pfeifferetalonFOCES,
	author = {{Pfeiffer}, M.~J. and {Frank}, C. and {Baumueller}, D. and {Fuhrmann}, K. and {Gehren}, T.},
	title = "{FOCES - a fibre optics Cassegrain Echelle spectrograph}",
	journal = {\aaps},
	year = 1998,
	month = jun,
	volume = {130},
	pages = {381-393},
	doi = {10.1051/aas:1998231},
	adsurl = {https://ui.adsabs.harvard.edu/abs/1998A&AS..130..381P}
}

@INPROCEEDINGS{quirrenbach1,
       author = {{Quirrenbach}, A. and {Amado}, P.~J. and {Ribas}, I. and {Reiners}, A. and {Caballero}, J.~A. and {Seifert}, W. and {Aceituno}, J. and {Azzaro}, M. and {Baroch}, D. and {Barrado}, D. and {Bauer}, F. and {Becerril}, S. and {B{\`e}jar}, V.~J.~S. and {Ben{\'\i}tez}, D. and {Brinkm{\"o}ller}, M. and {Cardona Guill{\'e}n}, C. and {Cifuentes}, C. and {Colom{\'e}}, J. and {Cort{\'e}s-Contreras}, M. and {Czesla}, S. and {Dreizler}, S. and {Fr{\"o}lich}, K. and {Fuhrmeister}, B. and {Galad{\'\i}-Enr{\'\i}quez}, D. and {Gonz{\'a}lez Hern{\'a}ndez}, J.~I. and {Gonz{\'a}lez Peinado}, R. and {Guenther}, E.~W. and {de Guindos}, E. and {Hagen}, H. -J. and {Hatzes}, A.~P. and {Hauschildt}, P.~H. and {Helmling}, J. and {Henning}, Th. and {Herbort}, O. and {Hern{\'a}ndez Casta{\~n}o}, L. and {Herrero}, E. and {Hintz}, D. and {Jeffers}, S.~V. and {Johnson}, E.~N. and {de Juan}, E. and {Kaminski}, A. and {Klahr}, H. and {K{\"u}rster}, M. and {Lafarga}, M. and {Sairam}, L. and {Lamp{\'o}n}, M. and {Lara}, L.~M. and {Launhardt}, R. and {L{\'o}pez del Fresno}, M. and {L{\'o}pez-Puertas}, M. and {Luque}, R. and {Mandel}, H. and {Marfil}, E.~G. and {Mart{\'\i}n}, E.~L. and {Mart{\'\i}n-Ruiz}, S. and {Mathar}, R.~J. and {Montes}, D. and {Morales}, J.~C. and {Nagel}, E. and {Nortmann}, L. and {Nowak}, G. and {Pall{\'e}}, E. and {Passegger}, V. -M. and {Pavlov}, A. and {Pedraz}, S. and {P{\'e}rez-Medialdea}, D. and {Perger}, M. and {Rebolo}, R. and {Reffert}, S. and {Rodr{\'\i}guez}, E. and {Rodr{\'\i}guez L{\'o}pez}, C. and {Rosich}, A. and {Sabotta}, S. and {Sadegi}, S. and {Salz}, M. and {S{\'a}nchez-L{\'o}pez}, A. and {Sanz-Forcada}, J. and {Sarkis}, P. and {Sch{\"a}fer}, S. and {Schiller}, J. and {Schmitt}, J.~H.~M.~M. and {Sch{\"o}fer}, P. and {Schweitzer}, A. and {Shulyak}, D. and {Solano}, E. and {Stahl}, O. and {Tala Pinto}, M. and {Trifonov}, T. and {Zapatero Osorio}, M.~R. and {Yan}, F. and {Zechmeister}, M. and {Abell{\'a}n}, F.~J. and {Abril}, M. and {Alonso-Floriano}, F.~J. and {Ammler-von Eiff}, M. and {Anglada-Escud{\'e}}, G. and {Anwand-Heerwart}, H. and {Arroyo-Torres}, B. and {Berdi{\~n}as}, Z.~M. and {Bergondy}, G. and {Bl{\"u}mcke}, M. and {del Burgo}, C. and {Cano}, J. and {Carro}, J. and {C{\'a}rdenas}, M.~C. and {Casal}, E. and {Claret}, A. and {D{\'\i}ez-Alonso}, E. and {Doellinger}, M. and {Dorda}, R. and {Feiz}, C. and {Fern{\'a}ndez}, M. and {Ferro}, I.~M. and {Gaisn{\'e}}, G. and {Gallardo}, I. and {G{\'a}lvez-Ortiz}, M.~C. and {Garc{\'\i}a-Piquer}, A. and {Garc{\'\i}a-Vargas}, M.~L. and {Garrido}, R. and {Gesa}, L. and {G{\'o}mez Galera}, V. and {Gonz{\'a}lez-{\'A}lvarez}, E. and {Gonz{\'a}lez-Cuesta}, L. and {Grohnert}, S. and {Gr{\"o}zinger}, U. and {Gu{\`a}rdia}, J. and {Guijarro}, A. and {Hedrosa}, R.~P. and {Hermann}, D. and {Hermelo}, I. and {Hern{\'a}ndez Arab{\'\i}}, R. and {Hern{\'a}ndez Hernando}, F. and {Hidalgo}, D. and {Holgado}, G. and {Huber}, A. and {Huber}, K. and {Huke}, P. and {Kehr}, M. and {Kim}, M. and {Klein}, R. and {Kl{\"u}ter}, J. and {Klutsch}, A. and {Labarga}, F. and {Labiche}, N. and {Lamert}, A. and {Laun}, W. and {L{\'a}zaro}, F.~J. and {Lemke}, U. and {Lenzen}, R. and {Llamas}, M. and {Lizon}, J. -L. and {Lodieu}, N. and {L{\'o}pez Gonz{\'a}lez}, M.~J. and {L{\'o}pez-Morales}, M. and {L{\'o}pez Salas}, J.~F. and {L{\'o}pez-Santiago}, J. and {Mag{\'a}n Madinabeitia}, H. and {Mall}, U. and {Mancini}, L. and {Mar{\'\i}n Molina}, J.~A. and {Mart{\'\i}nez-Rodr{\'\i}guez}, H. and {Maroto Fern{\'a}ndez}, D. and {Marvin}, C.~J. and {Mirabet}, E. and {Moreno-Raya}, M.~E. and {Moya}, A. and {Mundt}, R. and {Naranjo}, V. and {Panduro}, J. and {Pascual}, J. and {P{\'e}rez-Calpena}, A. and {Perryman}, M.~A.~C. and {Pluto}, M. and {Ram{\'o}n}, A. and {Redondo}, P. and {Reinhart}, S. and {Rhode}, P. and {Rix}, H. -W. and {Rodler}, F. and {Rohloff}, R. -R. and {S{\'a}nchez-Blanco}, E. and {S{\'a}nchez Carrasco}, M.~A. and {Sarmiento}, L.~F. and {Schmidt}, C. and {Storz}, C. and {Strachan}, J.~B.~P. and {St{\"u}rmer}, J. and {Su{\'a}rez}, J.~C. and {Tabernero}, H.~M. and {Tal-Or}, L. and {Tulloch}, S.~M. and {Ulbrich}, R. -G. and {Veredas}, G. and {Vico Linares}, J.~L. and {Vidal-Dasilva}, M. and {Vilardell}, F. and {Wagner}, K. and {Winkler}, J. and {Wolthoff}, V. and {Xu}, W.},
        title = "{CARMENES: high-resolution spectra and precise radial velocities in the red and infrared}",
    booktitle = {Ground-based and Airborne Instrumentation for Astronomy VII},
         year = 2018,
       editor = {{Evans}, Christopher J. and {Simard}, Luc and {Takami}, Hideki},
       series = {Society of Photo-Optical Instrumentation Engineers (SPIE) Conference Series},
       volume = {10702},
        month = jul,
          eid = {107020W},
        pages = {107020W},
          doi = {10.1117/12.2313689},
       adsurl = {https://ui.adsabs.harvard.edu/abs/2018SPIE10702E..0WQ}
}

@INPROCEEDINGS{quirrenbach2,
	author = {{Quirrenbach}, Andreas and {CARMENES Consortium} and {Amado}, P.~J. and {Ribas}, I. and {Reiners}, A. and {Caballero}, J.~A. and {Aceituno}, J. and {Alacid}, J.~M. and {Alonso-Floriano}, F.~J. and {Anglada-Escud{\'e}}, G. and {Azzaro}, M. and {Baroch}, D. and {Bauer}, F.~F. and {Becerril}, S. and {B{\'e}jar}, V.~J.~S. and {Bluhm}, P. and {Calvo Ortega}, R. and {Cardona Guill{\'e}n}, C. and {Casasayas-Barris}, N. and {Chaturvedi}, P. and {Cifuentes}, C. and {Colom{\'e}}, J. and {Conte}, D. and {Cort{\'e}s-Contreras}, M. and {Czesla}, S. and {D{\'\i}ez-Alonso}, E. and {Dom{\'\i}nguez Fern{\'a}ndez}, A.~J. and {Dreizler}, S. and {Duque-Arribas}, C. and {Espinoza}, N. and {Fuhrmeister}, B. and {Galad{\'\i}-Enr{\'\i}quez}, D. and {Gar{\textasciiacute}a Quintana}, E. and {Gonz{\'a}lez-Alvare}, E. and {Gonz{\'a}lez Cuesta}, z. L. and {Gonz{\'a}lez Hern{\'a}ndez}, J.~I. and {Guenther}, E.~W. and {de Guindos}, E. and {Hatzes}, A.~P. and {Henning}, T. and {Herbort}, O. and {Herrero}, E. and {Hintz}, D. and {Iglesias-P{\'a}ra}, J. and {Jeffers}, S.~V. and {Johnson}, E.~N. and {de Juan}, E. and {Kaminski}, A. and {Kemmer}, J. and {Khaimova}, J. and {Khalafinejad}, S. and {Klahr}, H. and {Kossakowski}, D. and {Kreidberg}, L. and {K{\"u}rster}, M. and {Labarga}, F. and {Lafarga}, M. and {Lamp{\'o}n}, M. and {Lara}, L.~M. and {Lillo-Box}, J. and {Lodieu}, N. and {L{\'o}pez Gallifa}, A. and {L{\'o}pez Gonz{\'a}lez}, M.~J. and {L{\'o}pez-Puertas}, M. and {Luque}, R. and {Marfil}, E. and {Mart{\'\i}n-Ruiz}, S. and {Matth{\'e}}, C. and {Molaverdikhani}, K. and {Montes}, D. and {Morales}, J.~C. and {Morales-Calder{\'o}on}, M. and {Nagel}, E. and {Nortmann}, L. and {Nowak}, G. and {Ofir}, A. and {Oshaghi}, M. and {Pall{\'e}}, E. and {Passegger}, V.~M. and {Pavlov}, A. and {Pedraz}, S. and {Perdelwitz}, V. and {Perger}, M. and {Reffert}, S. and {Revilla}, D. and {Rodr{\'\i}guez}, E. and {Rodr{\'\i}guez L{\'o}pez}, C. and {Sabotta}, S. and {Sadegi}, S. and {Sairam}, L. and {Salz}, M. and {S{\'a}nchez-L{\'o}pez}, A. and {Sanz-Forcada}, J. and {Sarkis}, P. and {Sch{\"a}fer}, S. and {Schiller}, J. and {Schlecker}, M. and {Schmitt}, J.~H.~M.~M. and {Sch{\"o}fer}, P. and {Schweitzer}, A. and {Seiferta}, W. and {Shan}, Y. and {Shulyak}, D. and {Skrzypinski}, S.~L. and {Solano}, E. and {Soto}, M.~G. and {Stahl}, O. and {Stangret}, M. and {Stock}, S.~A. and {Strachan}, J.~B.~P. and {Stuber}, T. and {St{\"u}rmer}, J. and {Tabernero}, H.~M. and {Tal-Or}, L. and {Tala-Pinto}, M. and {Trifonov}, T. and {Vanaverbeke}, S. and {Yan}, F. and {Zapatero Osorio}, M.~R. and {Zechmeister}, M.},
	title = "{The CARMENES M-dwarf planet survey}",
	booktitle = {Ground-based and Airborne Instrumentation for Astronomy VIII},
	year = 2020,
	editor = {{Evans}, Christopher J. and {Bryant}, Julia J. and {Motohara}, Kentaro},
	series = {Society of Photo-Optical Instrumentation Engineers (SPIE) Conference Series},
	volume = {11447},
	month = dec,
	eid = {114473C},
	pages = {114473C},
	doi = {10.1117/12.2561380},
	adsurl = {https://ui.adsabs.harvard.edu/abs/2020SPIE11447E..3CQ}
}

@ARTICLE{reinersbasri,
       author = {{Reiners}, A. and {Basri}, G.},
        title = "{Chromospheric Activity, Rotation, and Rotational Braking in M and L Dwarfs}",
      journal = {\apj},
         year = 2008,
        month = sep,
       volume = {684},
       number = {2},
        pages = {1390-1403},
          doi = {10.1086/590073},
archivePrefix = {arXiv},
       eprint = {0805.1059},
 primaryClass = {astro-ph},
       adsurl = {https://ui.adsabs.harvard.edu/abs/2008ApJ...684.1390R}
}

@article{datarelease1,
       author = {{Ribas}, I. and {Reiners}, A. and {Zechmeister}, M. and {Caballero}, J.~A. and {Morales}, J.~C. and {Sabotta}, S. and {Baroch}, D. and {Amado}, P.~J. and {Quirrenbach}, A. and {Abril}, M. and {Aceituno}, J. and {Anglada-Escud{\'e}}, G. and {Azzaro}, M. and {Barrado}, D. and {B{\'e}jar}, V.~J.~S. and {Ben{\'\i}tez de Haro}, D. and {Bergond}, G. and {Bluhm}, P. and {Calvo Ortega}, R. and {Cardona Guill{\'e}n}, C. and {Chaturvedi}, P. and {Cifuentes}, C. and {Colom{\'e}}, J. and {Cont}, D. and {Cort{\'e}s-Contreras}, M. and {Czesla}, S. and {D{\'\i}ez-Alonso}, E. and {Dreizler}, S. and {Duque-Arribas}, C. and {Espinoza}, N. and {Fern{\'a}ndez}, M. and {Fuhrmeister}, B. and {Galad{\'\i}-Enr{\'\i}quez}, D. and {Garc{\'\i}a-L{\'o}pez}, A. and {Gonz{\'a}lez-{\'A}lvarez}, E. and {Gonz{\'a}lez Hern{\'a}ndez}, J.~I. and {Guenther}, E.~W. and {de Guindos}, E. and {Hatzes}, A.~P. and {Henning}, Th. and {Herrero}, E. and {Hintz}, D. and {Huelmo}, {\'A}. L. and {Jeffers}, S.~V. and {Johnson}, E.~N. and {de Juan}, E. and {Kaminski}, A. and {Kemmer}, J. and {Khaimova}, J. and {Khalafinejad}, S. and {Kossakowski}, D. and {K{\"u}rster}, M. and {Labarga}, F. and {Lafarga}, M. and {Lalitha}, S. and {Lamp{\'o}n}, M. and {Lillo-Box}, J. and {Lodieu}, N. and {L{\'o}pez Gonz{\'a}lez}, M.~J. and {L{\'o}pez-Puertas}, M. and {Luque}, R. and {Mag{\'a}n}, H. and {Mancini}, L. and {Marfil}, E. and {Mart{\'\i}n}, E.~L. and {Mart{\'\i}n-Ruiz}, S. and {Molaverdikhani}, K. and {Montes}, D. and {Nagel}, E. and {Nortmann}, L. and {Nowak}, G. and {Pall{\'e}}, E. and {Passegger}, V.~M. and {Pavlov}, A. and {Pedraz}, S. and {Perdelwitz}, V. and {Perger}, M. and {Ram{\'o}n-Ballesta}, A. and {Reffert}, S. and {Revilla}, D. and {Rodr{\'\i}guez}, E. and {Rodr{\'\i}guez-L{\'o}pez}, C. and {Sadegi}, S. and {S{\'a}nchez Carrasco}, M. {\'A}. and {S{\'a}nchez-L{\'o}pez}, A. and {Sanz-Forcada}, J. and {Sch{\"a}fer}, S. and {Schlecker}, M. and {Schmitt}, J.~H.~M.~M. and {Sch{\"o}fer}, P. and {Schweitzer}, A. and {Seifert}, W. and {Shan}, Y. and {Skrzypinski}, S.~L. and {Solano}, E. and {Stahl}, O. and {Stangret}, M. and {Stock}, S. and {St{\"u}rmer}, J. and {Tabernero}, H.~M. and {Tal-Or}, L. and {Trifonov}, T. and {Vanaverbeke}, S. and {Yan}, F. and {Zapatero Osorio}, M.~R.},
        title = "{The CARMENES search for exoplanets around M dwarfs. Guaranteed time observations Data Release 1 (2016-2020)}",
      journal = {\aap},
         year = 2023,
        month = feb,
       volume = {670},
          eid = {A139},
        pages = {A139},
          doi = {10.1051/0004-6361/202244879},
archivePrefix = {arXiv},
       eprint = {2302.10528},
 primaryClass = {astro-ph.EP},
       adsurl = {https://ui.adsabs.harvard.edu/abs/2023A&A...670A.139R}
}

@ARTICLE{solaronvsinigravitydarkening,
	author = {{Solar}, Mart{\'\i}n and {Arcos}, Catalina and {Cur{\'e}}, Michel and {Levenhagen}, Ronaldo S. and {Araya}, Ignacio},
	title = "{Automatic algorithm to obtain v sin i values via Fourier transform in the BeSOS database}",
	journal = {\mnras},
	year = 2022,
	month = apr,
	volume = {511},
	number = {3},
	pages = {4404-4416},
	doi = {10.1093/mnras/stac202},
	archivePrefix = {arXiv},
	eprint = {2201.08695},
	primaryClass = {astro-ph.SR},
	adsurl = {https://ui.adsabs.harvard.edu/abs/2022MNRAS.511.4404S}
}

@ARTICLE{schofer2019,
	author = {{Sch{\"o}fer}, P. and {Jeffers}, S.~V. and {Reiners}, A. and {Shulyak}, D. and {Fuhrmeister}, B. and {Johnson}, E.~N. and {Zechmeister}, M. and {Ribas}, I. and {Quirrenbach}, A. and {Amado}, P.~J. and {Caballero}, J.~A. and {Anglada-Escud{\'e}}, G. and {Bauer}, F.~F. and {B{\'e}jar}, V.~J.~S. and {Cort{\'e}s-Contreras}, M. and {Dreizler}, S. and {Guenther}, E.~W. and {Kaminski}, A. and {K{\"u}rster}, M. and {Lafarga}, M. and {Montes}, D. and {Morales}, J.~C. and {Pedraz}, S. and {Tal-Or}, L.},
	title = "{The CARMENES search for exoplanets around M dwarfs. Activity indicators at visible and near-infrared wavelengths}",
	journal = {\aap},
	year = 2019,
	month = mar,
	volume = {623},
	eid = {A44},
	pages = {A44},
	doi = {10.1051/0004-6361/201834114},
	archivePrefix = {arXiv},
	eprint = {1901.08861},
	primaryClass = {astro-ph.SR},
	adsurl = {https://ui.adsabs.harvard.edu/abs/2019A&A...623A..44S}
}

@ARTICLE{ares1,
       author = {{Sousa}, S.~G. and {Santos}, N.~C. and {Israelian}, G. and {Mayor}, M. and {Monteiro}, M.~J.~P.~F.~G.},
        title = "{A new code for automatic determination of equivalent widths: Automatic Routine for line Equivalent widths in stellar Spectra (ARES)}",
      journal = {\aap},
         year = 2007,
        month = jul,
       volume = {469},
       number = {2},
        pages = {783-791},
          doi = {10.1051/0004-6361:20077288},
archivePrefix = {arXiv},
       eprint = {astro-ph/0703696},
 primaryClass = {astro-ph},
       adsurl = {https://ui.adsabs.harvard.edu/abs/2007A&A...469..783S}
}

@ARTICLE{ares2,
       author = {{Sousa}, S.~G. and {Santos}, N.~C. and {Adibekyan}, V. and {Delgado-Mena}, E. and {Israelian}, G.},
        title = "{ARES v2: new features and improved performance}",
      journal = {\aap},
         year = 2015,
        month = may,
       volume = {577},
          eid = {A67},
        pages = {A67},
          doi = {10.1051/0004-6361/201425463},
archivePrefix = {arXiv},
       eprint = {1504.02725},
 primaryClass = {astro-ph.IM},
       adsurl = {https://ui.adsabs.harvard.edu/abs/2015A&A...577A..67S}
}

@article{taloretal,
	author = {{Tal-Or}, L. and {Zechmeister}, M. and {Reiners}, A. and {Jeffers}, S.~V. and {Sch{\"o}fer}, P. and {Quirrenbach}, A. and {Amado}, P.~J. and {Ribas}, I. and {Caballero}, J.~A. and {Aceituno}, J. and {Bauer}, F.~F. and {B{\'e}jar}, V.~J.~S. and {Czesla}, S. and {Dreizler}, S. and {Fuhrmeister}, B. and {Hatzes}, A.~P. and {Johnson}, E.~N. and {K{\"u}rster}, M. and {Lafarga}, M. and {Montes}, D. and {Morales}, J.~C. and {Reffert}, S. and {Sadegi}, S. and {Seifert}, W. and {Shulyak}, D.},
	title = "{The CARMENES search for exoplanets around M dwarfs. Radial-velocity variations of active stars in visual-channel spectra}",
	journal = {\aap},
	year = 2018,
	month = jun,
	volume = {614},
	eid = {A122},
	pages = {A122},
	doi = {10.1051/0004-6361/201732362},
	archivePrefix = {arXiv},
	eprint = {1803.02338},
	primaryClass = {astro-ph.SR},
	adsurl = {https://ui.adsabs.harvard.edu/abs/2018A&A...614A.122T}
}

@ARTICLE{scipy2020,
    author = {{Virtanen}, Pauli and {Gommers}, Ralf and {Oliphant}, Travis E. and {Haberland}, Matt and {Reddy}, Tyler and {Cournapeau}, David and {Burovski}, Evgeni and {Peterson}, Pearu and {Weckesser}, Warren and {Bright}, Jonathan and {van der Walt}, St{\'e}fan J. and {Brett}, Matthew and {Wilson}, Joshua and {Millman}, K. Jarrod and {Mayorov}, Nikolay and {Nelson}, Andrew R.~J. and {Jones}, Eric and {Kern}, Robert and {Larson}, Eric and {Carey}, C.~J. and {Polat}, {\.I}lhan and {Feng}, Yu and {Moore}, Eric W. and {VanderPlas}, Jake and {Laxalde}, Denis and {Perktold}, Josef and {Cimrman}, Robert and {Henriksen}, Ian and {Quintero}, E.~A. and {Harris}, Charles R. and {Archibald}, Anne M. and {Ribeiro}, Ant{\^o}nio H. and {Pedregosa}, Fabian and {van Mulbregt}, Paul and {SciPy 1. 0 Contributors}},
    title = "{SciPy 1.0: fundamental algorithms for scientific computing in Python}",
    journal = {Nature Methods},
    year = 2020,
    month = feb,
    volume = {17},
    pages = {261-272},
    doi = {10.1038/s41592-019-0686-2},
    archivePrefix = {arXiv},
    eprint = {1907.10121},
    primaryClass = {cs.MS},
    adsurl = {https://ui.adsabs.harvard.edu/abs/2020NatMe..17..261V}
    }

@ARTICLE{vaughan1978,
	author = {{Vaughan}, A.~H. and {Preston}, G.~W. and {Wilson}, O.~C.},
	title = "{Flux measurements of Ca II and K emission.}",
	journal = {\pasp},
	year = 1978,
	month = jun,
	volume = {90},
	pages = {267-274},
	doi = {10.1086/130324},
	adsurl = {https://ui.adsabs.harvard.edu/abs/1978PASP...90..267V}
}

@ARTICLE{voigtfunc,
       author = {{McLean}, A.~B. and {Mitchell}, C.~E.~J. and {Swanston}, D.~M.},
        title = "{Implementation of an efficient analytical approximation to the Voigt function for photoemission lineshape analysis}",
      journal = {Journal of Electron Spectroscopy and Related Phenomena},
         year = 1994,
        month = jan,
       volume = {69},
       number = {2},
        pages = {125-132},
          doi = {10.1016/0368-2048(94)02189-7},
       adsurl = {https://ui.adsabs.harvard.edu/abs/1994JESRP..69..125M}
}

@ARTICLE{walkowicz1,
       author = {{Walkowicz}, Lucianne M. and {Hawley}, Suzanne L. and {West}, Andrew A.},
        title = "{The {\ensuremath{\chi}} Factor: Determining the Strength of Activity in Low-Mass Dwarfs}",
      journal = {\pasp},
         year = 2004,
        month = nov,
       volume = {116},
       number = {826},
        pages = {1105-1110},
          doi = {10.1086/426792},
archivePrefix = {arXiv},
       eprint = {astro-ph/0410422},
 primaryClass = {astro-ph},
       adsurl = {https://ui.adsabs.harvard.edu/abs/2004PASP..116.1105W}
}

@ARTICLE{Gray1984,
	author       = {{Gray}, D.~F.},
	title        = "{The observation and analysis of stellar photospheric velocity fields. I - Effects of macroturbulence and rotation on line profiles}",
	journal      = {Astrophysical Journal},
	year         = 1984,
	volume       = 281,
	pages        = {719--727},
	doi          = {10.1086/162149}
}

@ARTICLE{Gray2010empiricaldec,
	author = {{Gray}, David F.},
	title = "{Empirical Decoding of the Shapes of Spectral-Line Bisectors}",
	journal = {\apj},
	year = 2010,
	month = feb,
	volume = {710},
	number = {2},
	pages = {1003-1008},
	doi = {10.1088/0004-637X/710/2/1003},
	adsurl = {https://ui.adsabs.harvard.edu/abs/2010ApJ...710.1003G}
}

@ARTICLE{Doyle2014,
	author       = {{Doyle}, A.~P. and {Davies}, G.~R. and {Smoker}, J.~V. and {Smalley}, B.},
	title        = "{Macroturbulence in slowly rotating solar-type stars}",
	journal      = {Monthly Notices of the Royal Astronomical Society},
	year         = 2014,
	volume       = 444,
	number       = 3,
	pages        = {3592--3602},
	doi          = {10.1093/mnras/stu1692}
}

@ARTICLE{Tsantaki2018,
	author       = {{Tsantaki}, M. and {Sousa}, S.~G. and {Adibekyan}, V.~Zh. and {Santos}, N.~C. and {Mortier}, A. and {Delgado Mena}, E.},
	title        = "{Macroturbulent and rotational velocities for FGK dwarfs and subgiants: calibration and implications}",
	journal      = {Monthly Notices of the Royal Astronomical Society},
	year         = 2018,
	volume       = 473,
	number       = 4,
	pages        = {5066--5078},
	doi          = {10.1093/mnras/stx2719}
}

@ARTICLE{2022A&A...666A.120G,
	author = {{Gilmore}, G. and {Randich}, S. and {Worley}, C.~C. and {Hourihane}, A. and {Gonneau}, A. and {Sacco}, G.~G. and {Lewis}, J.~R. and {Magrini}, L. and {Fran{\c{c}}ois}, P. and {Jeffries}, R.~D. and {Koposov}, S.~E. and {Bragaglia}, A. and {Alfaro}, E.~J. and {Allende Prieto}, C. and {Blomme}, R. and {Korn}, A.~J. and {Lanzafame}, A.~C. and {Pancino}, E. and {Recio-Blanco}, A. and {Smiljanic}, R. and {Van Eck}, S. and {Zwitter}, T. and {Bensby}, T. and {Flaccomio}, E. and {Irwin}, M.~J. and {Franciosini}, E. and {Morbidelli}, L. and {Damiani}, F. and {Bonito}, R. and {Friel}, E.~D. and {Vink}, J.~S. and {Prisinzano}, L. and {Abbas}, U. and {Hatzidimitriou}, D. and {Held}, E.~V. and {Jordi}, C. and {Paunzen}, E. and {Spagna}, A. and {Jackson}, R.~J. and {Ma{\'\i}z Apell{\'a}niz}, J. and {Asplund}, M. and {Bonifacio}, P. and {Feltzing}, S. and {Binney}, J. and {Drew}, J. and {Ferguson}, A.~M.~N. and {Micela}, G. and {Negueruela}, I. and {Prusti}, T. and {Rix}, H. -W. and {Vallenari}, A. and {Bergemann}, M. and {Casey}, A.~R. and {de Laverny}, P. and {Frasca}, A. and {Hill}, V. and {Lind}, K. and {Sbordone}, L. and {Sousa}, S.~G. and {Adibekyan}, V. and {Caffau}, E. and {Daflon}, S. and {Feuillet}, D.~K. and {Gebran}, M. and {Gonzalez Hernandez}, J.~I. and {Guiglion}, G. and {Herrero}, A. and {Lobel}, A. and {Merle}, T. and {Mikolaitis}, {\v{S}}. and {Montes}, D. and {Morel}, T. and {Ruchti}, G. and {Soubiran}, C. and {Tabernero}, H.~M. and {Tautvai{\v{s}}ien{\.{e}}}, G. and {Traven}, G. and {Valentini}, M. and {Van der Swaelmen}, M. and {Villanova}, S. and {Viscasillas V{\'a}zquez}, C. and {Bayo}, A. and {Biazzo}, K. and {Carraro}, G. and {Edvardsson}, B. and {Heiter}, U. and {Jofr{\'e}}, P. and {Marconi}, G. and {Martayan}, C. and {Masseron}, T. and {Monaco}, L. and {Walton}, N.~A. and {Zaggia}, S. and {Aguirre B{\o}rsen-Koch}, V. and {Alves}, J. and {Balaguer-Nunez}, L. and {Barklem}, P.~S. and {Barrado}, D. and {Bellazzini}, M. and {Berlanas}, S.~R. and {Binks}, A.~S. and {Bressan}, A. and {Capuzzo-Dolcetta}, R. and {Casagrande}, L. and {Casamiquela}, L. and {Collins}, R.~S. and {D'Orazi}, V. and {Dantas}, M.~L.~L. and {Debattista}, V.~P. and {Delgado-Mena}, E. and {Di Marcantonio}, P. and {Drazdauskas}, A. and {Evans}, N.~W. and {Famaey}, B. and {Franchini}, M. and {Fr{\'e}mat}, Y. and {Fu}, X. and {Geisler}, D. and {Gerhard}, O. and {Gonz{\'a}lez Solares}, E.~A. and {Grebel}, E.~K. and {Guti{\'e}rrez Albarr{\'a}n}, M.~L. and {Jim{\'e}nez-Esteban}, F. and {J{\"o}nsson}, H. and {Khachaturyants}, T. and {Kordopatis}, G. and {Kos}, J. and {Lagarde}, N. and {Ludwig}, H. -G. and {Mahy}, L. and {Mapelli}, M. and {Marfil}, E. and {Martell}, S.~L. and {Messina}, S. and {Miglio}, A. and {Minchev}, I. and {Moitinho}, A. and {Montalban}, J. and {Monteiro}, M.~J.~P.~F.~G. and {Morossi}, C. and {Mowlavi}, N. and {Mucciarelli}, A. and {Murphy}, D.~N.~A. and {Nardetto}, N. and {Ortolani}, S. and {Paletou}, F. and {Palou{\v{s}}}, J. and {Pickering}, J.~C. and {Quirrenbach}, A. and {Re Fiorentin}, P. and {Read}, J.~I. and {Romano}, D. and {Ryde}, N. and {Sanna}, N. and {Santos}, W. and {Seabroke}, G.~M. and {Spina}, L. and {Steinmetz}, M. and {Stonkut{\'e}}, E. and {Sutorius}, E. and {Th{\'e}venin}, F. and {Tosi}, M. and {Tsantaki}, M. and {Wright}, N. and {Wyse}, R.~F.~G. and {Zoccali}, M. and {Zorec}, J. and {Zucker}, D.~B.},
	title = "{The Gaia-ESO Public Spectroscopic Survey: Motivation, implementation, GIRAFFE data processing, analysis, and final data products}",
	journal = {\aap},
	year = 2022,
	month = oct,
	volume = {666},
	eid = {A120},
	pages = {A120},
	doi = {10.1051/0004-6361/202243134},
	archivePrefix = {arXiv},
	eprint = {2208.05432},
	primaryClass = {astro-ph.SR},
	adsurl = {https://ui.adsabs.harvard.edu/abs/2022A&A...666A.120G}
}
\bibliographystyle{aasjournal}

\end{document}